\documentclass[prd,amsmath,amssymb]{revtex4}

\usepackage{graphicx}
\usepackage{dcolumn}
\usepackage{bm}

\begin{document}

\title{Atmospheric Muon and Neutrino Flux From 3-Dimensional Simulation}

\author{Yong Liu}
\email{yongliu@fnal.gov}
\thanks{Present address: MS 309, FermiLab, P.O.Box 500, Batavia, IL 60510, USA.}
\affiliation{Institut des Sciences Nucl\'eaires, 53 av. des Martyrs, 38026 Grenoble cedex, France}

\author{L. Derome}
\email{derome@isn.in2p3.fr}
\affiliation{Institut des Sciences Nucl\'eaires, 53 av. des Martyrs, 38026 Grenoble cedex, France}

\author{M. Bu\'enerd}
\thanks{Corresponding author}\email{buenerd@isn.in2p3.fr}
\affiliation{Institut des Sciences Nucl\'eaires, 53 av. des Martyrs, 38026 Grenoble cedex, France}
\begin{abstract}

The atmospheric muon and neutrino flux have been simulated using the same approach which 
successfully accounted for the recent secondary proton, electron and positron flux measurements 
in orbit by the AMS experiment. For the muon flux, a good agreement is obtained with the CAPRICE 
and HEAT data for altitudes ranging from sea level up to about 38~km. The general features of the 
calculated atmospheric neutrino flux are reported and discussed. The flux obtained at the 
Super-Kamiokande experiment location are reported and compared with other calculations. For low 
neutrino energies the flux obtained is significantly smaller than that used in the data analysis of 
underground experiment. The simulation results for the SOUDAN experiment site are also reported.\\
\end{abstract}
\pacs{96.40.Tv, 95.85.Ry, 14.60.Pq}

\maketitle

\section{Introduction}

With the rapidly increasing amount and statistical significance of the data collected by 
underground neutrino detectors \cite{superk,soudan}, the precise calculation of the 
atmospheric neutrino flux is highly desirable, since the interpretation of the 
data relies on the calculated flux as a fundamental input. In the contained events 
analysis, the ratio of ratios, namely the ratio of the observed number of muon events to the 
number of observed electron events with the corresponding ratio of the same numbers from 
simulation calculations, is not sensitive to the absolute neutrino flux. However, in the 
analysis of upward going muon events, the absolute flux need to be known accurately in 
addition to the detailed features of the flux such as the zenithal and azimuthal angle 
distribution \cite{frati,fogli}. The availability of calculations of proven validity 
would discard all possible doubt on the conclusions.
\par
In the past 20 years, various methods including analytical, semianalytical \cite{bn89,pl93}, 
kinematical \cite{pk94} and Monte Carlo techniques \cite{bs89,hk90}, have been applied to 
calculate the atmospheric neutrino flux \cite{vv80,bn89,bs89,hk90,HO95,lk90}.
However, the accuracy of the results was limited by the accuracy on both the cosmic ray (CR) 
abundance, and the production cross section of the parent particle ($\pi^\pm, K^\pm, 
K^0_L, K^0_S$) in the collisions of CRs with atmospheric nuclei over the relevant energy 
range \cite{tk20,re20}, used in these calculations. These two quantities being basic inputs to 
the calculations. 
The analysis and the comparisons made in \cite{GA96} indicated that the particular choice
of the primary spectra and the strong interaction model, as well as the earth geomagnetic 
field could result in significant differences in the calculated neutrino flux. 

The new AMS \cite{ams00} and BESS \cite{bess00} measurements of the CR proton flux consistently
agree to within $5~\%$. The values reported are lower than those used in the previous works by 
about 20 to 25 \% \cite{honda1}. Besides, some analysis \cite{ms98} showed that the 
pion production cross sections given by the general purpose softwares based on theoretical 
models, may largely depart from experimental data. 

Although the values of the flavor ratio reported in previous works are very close to each
other, the absolute values of the flux obtained in these works fail to be consistent with
each other within satisfying precision \cite{GA96}. In particular, these results 
seem not to predict quantitatively the observed East-West (EW) asymmetry of the neutrino 
induced events \cite{ew111,ew122}
which is expected to originate from the EW asymmetry of the geomagnetic cut off (GC) 
on the incoming CR - mainly protons and helium particles here - momentum, due to the 
earth magnetic field. 
Any calculation, to be reliable, must reproduce the experimental EW asymmetry since flavor 
oscillations do not change the direction of motion. 
This requirement can be taken as a testing ground for the various approaches of the neutrino
flux calculations.

With the occurrence of the new measurements of the primary CR flux, together with the 
continuous improvement of the cross section calculations, most approaches have been 
updated \cite{honda1,honda2,naumov01,lap1,gai01}, and new calculations have been proposed
\cite{gflm,tknw,plya,ldbd,wentz}. In parallel, experimental groups also made 
greater efforts aiming to understand the neutrino production process in the atmosphere 
\cite{cbdv,capri,mass99} with the measurements on the muon flux at various altitudes, 
together with those of the primary proton and helium flux. These new data provided a sounder 
testing ground for the reliability of the numerical approaches to the particle production 
and to the particle dynamics and kinematics in the atmosphere and in the earth geomagnetic 
field. 

The work reported here is an extension of a research program which primary motivations were 
to interpret the AMS01 measurements of the charged particle flux in near earth orbit 
\cite{ams00,ams02}. To this purpose, an event generator describing the CR induced cascade in 
the atmosphere, particle propagation in geomagnetic field, and interactions with the medium, 
has been developed and successfully used to reproduce the proton \cite{DERP}, electron-positron 
\cite{DERL} and helium 3 \cite{DERH} flux data measured by AMS and their relevant dependence on 
the geomagnetic coordinates. Since the $e^{\pm}$ generator of the program was basically the 
same as needed to generate the muon and neutrino flux, the code could be rather straightforwardly 
extended to describe the latter and to address the important issue of the atmospheric neutrino 
flux. 
It should be emphasized that the present approach allows an evaluation of the neutrino flux
induced by both the primary CR flux and the atmospheric secondary flux, which have been tested 
sensitively and independently on other observables, conferring to the results a character of sound 
reliability. The above issues will be discussed in details below. This approach is also being 
successfully applied to the calculation of the atmospheric secondary flux of antinucleons 
\cite{hdb03}.

The article reports on the calculation of the atmospheric muon and neutrino flux. The paper 
is organized as follows. In section \ref{METH}, the method and models used in the 
calculations are introduced. Section \ref{MUS} is devoted to the results on the muon flux. 
In section \ref{GLOB}, the properties of the calculated neutrino flux averaged on the 
whole terrestrial sphere are discussed, while the properties of the local neutrino flux 
calculated at the location of the Super-Kamiokande detector are discussed in section 
\ref{SK}. In the following section \ref{SOUD} the neutrino flux at the Soudan detector 
location is discussed along the same lines. Summary and conclusions are given in section 
\ref{CONC}. 

\vspace{0.5cm}
\section{METHOD AND MODELS}\label{METH}
The method of calculation and the models used in the simulation program are described in 
\cite{DERP,DERL}. The calculation proceeds by means of a full 3D-simulation program. 
The main features of this approach are listed below for the reader's convenience, with 
emphasis on some particular points relevant to the neutrino and muon flux.
\begin{enumerate}
\item { CR generation:}
Incident Cosmic Rays are generated on a virtual sphere chosen at some distance from the 
earth as discussed below. Events are generated randomly and uniformly on this sphere. In 
order to get an isotropic flux at any point inside the volume of the virtual sphere, the 
differential element of the zenith angle distribution of the particle direction generated 
on the sphere must be proportional to $\cos\theta_z\ d(\cos\theta_z)$, $\theta_z$ being the 
zenithal angle of the particle \cite{HO95,honda2} (see also \cite{lap1}). The particle 
($A$ and $Z$) and its momentum are then generated according to the CR abundances and spectra 
discussed below. 
Generating random events on a virtual sphere far enough from the earth, i.e., at a distance 
where the geomagnetic field effects are negligible, so that the CR flux is isotropic, would 
imply a very large distance from earth and thus the generation of a tremendous number of 
events, most of them being useless for the simulation purpose since they would not reach the
earth-atmosphere system.
Instead of this time consuming direct method, the generation sphere is chosen close to 
earth, at a 2000~km altitude. The geomagnetic cut-off is then applied by back-tracing the 
particle trajectory in the geomagnetic field. In this procedure, only those particles 
reaching a backtracing distance of 10 Earth radii are kept in the sample. Flux conservation 
along any allowed particle path in the geomagnetic field is ensured by Liouville's theorem 
application \cite{VA61}.
\item { CR abundances:}
For the incident CR proton and helium flux, functional forms fitted to the 1998 AMS 
measurements were used \cite{ams00,ams02} . For other periods of the solar cycle than those 
of the measurements, the incident cosmic flux are corrected for the different solar 
modulation effects using a simple force law approximation \cite{SOLMOD}.
\begin{table}
\begin{ruledtabular}
\begin{tabular}{cccccc}
 Particle type          & $p$   & $He$ & $CNO$ & $Ne-S$ & $Fe$   \\  
\hline
$<\phi(<\!z\!>)>$       & 106   &  16   & 1.1  &  0.34 &  0.08   \\ 
$<\mu(\pi^-)>$          & 0.3   & 0.63 & 1.23  & 1.53  & 1.86   \\
fractional $\pi^-$ flux & 0.72  & 0.22 & 0.032 & 0.011 & 0.0035 \\ 
\end{tabular}
\caption{\it Mean flux $\phi(z)$ $(m^2\ s\ sr)^{-1}$, $\pi^-$ multiplicity $<\!\mu(\pi)\!>$, 
of the components of the CR flux, and their product normalized to 1, for the conditions 
described in the text.
\label{ABUND}
}
\end{ruledtabular}
\end{table}
The main components of the incident CR flux are usually divided in the following most 
abundant groups of elements: $p, He, CNO, Ne-S$, and $Fe$. The contribution of each group 
to the neutrino flux scales roughly with the product $\phi(<\!z\!>)\mu(\pi)$ with $\phi(\!<\!z>)$ 
the CR abundance of the element or group of elements with mean electric charge $<\!z\!>$, 
and $\mu(\pi)$ the average multiplicities of pions produced in the CR collisions with 
atmospheric nuclei. This is approximate since secondary protons induced by the $He$ 
component also contributing to this production are not taken into account in this product.
Table~\ref{ABUND} shows the relative abundances of the estimated pion 
flux induced by the various components of the CR flux. They have been evaluated using the 
AMS measurements for $p$ and $He$ and those from \cite{SI83} for the heavier elements with 
the spectral index from \cite{WI98}. The flux have been calculated for particle rigidities 
above a geomagnetic cutoff of 10~GV. The pion multiplicities were taken from 
\cite{AG81,AG84} at 4.2~GeV/c per nucleon. The sum of the contributions from the $A>4$ flux 
components amounts to about 6\%, while they are 72\% for $p$ and 22\% for $He$. This choice 
for the GC maximizes the $He$ (and heavier elements) over $p$ fraction, which would be 
smaller for a lower cutoff (the $He$ fraction would be around 10\% for no cutoff). 
The $A>4$ components of the CR flux were not taken into account in the present calculations. 
They will be included in the further developments of the code.
\item { Particle propagation:}
Each particle is propagated in the geomagnetic field and interacts with nuclei of the local 
atmospheric density. The specific ionization energy loss is computed for each step along 
the trajectory. The model used to describe the atmospheric density was taken from references 
\cite{agfm1}.

Every secondary particles are processed the same way as their parent particle, leading to 
the generation of atmospheric cascades.
\item { Secondary particle production:}
Nucleons, pions and kaons are produced with their respective cross 
sections. For these particles the Kalinovsky-Mokhov-Nikitin (KMN) parametrization of the 
inclusive cross sections \cite{kmnn} was used. For pions the KMN parameters have been fit 
on a wide range of $p+A\rightarrow \pi^{\pm,0}$ data between 1.38 and 400~GeV/c incident 
momenta \cite{ms98,ldl11}. 
Leptons - electrons, muons, and neutrinos - are produced in the decay chains of mesons, 
mainly pions and kaons, hadronically produced on atmospheric nuclei $A$ in processes of 
type: $CR+A\rightarrow M+X$, $M\rightarrow\mu +\nu$, $\mu\rightarrow e+\nu +\bar{\nu}$. 
See below for details.
It must be emphasized here, that the constraint to reproduce simultaneously the lepton 
population measurements at various altitudes and geographical coordinates, is extremely 
strong, and that a successful result would ensure a high level of reliability of the 
theoretical grounds and of the numerical approach to the problem.  
\item { $\mu$ decay:}
For the decay of muons, the spectra of the products ($\nu, \bar{\nu}, e^{\pm}$) are 
generated according to the Fermi theory. 
\item { $K$ decay:}
For kaon decay, the Dalitz plot distribution given in \cite{pdg00} is used.
\item { $\mu$ polarization effects:}
Muon polarization was taken into account for pion decay and for the $K^\pm \rightarrow 
\mu^\pm \nu_{\mu}(\bar{\nu_{\mu}})$ channel \cite{hk90,HO95,sh57}. 
\item { CR Energy range:}
The kinetic energy range of incident CRs covered in the simulation is $[0.2, 2000]$~GeV, 
from around the pion production threshold up to a value where the corresponding neutrino 
flux produced becomes negligible.
\item { Processing:}
Each particle history, trajectory parameters and kinematics, is traced and recorded by the 
program. The event file was then analyzed separately to generate the various distributions 
of interest.
\item { $\nu$ detection:}
For the neutrino flux calculation, a virtual detection sphere is defined at the 
Super-Kamiokande detector altitude. 
\item{ Flux Normalisation}
The differential particle flux produced in the simulation 
($\textrm{m}^{-2}\cdot \textrm{s}^{-1}\cdot \textrm{sr}^{-1}\cdot \textrm{GeV}^{-1}$) were calculated by dividing the number of
particles detected by the (local) surface of the particle collection area, solid angle, 
energy bin size, and equivalent sampling time of the CR flux. 
The latter was obtained from the total event number generated in the simulation run(s)
divided by the surface of the generation sphere times the integrated ($\cos\theta_z$ weighted) 
solid angle ($\pi$) times the energy integrated flux. 

To increase the statistics for the energy regions of low particle flux, the primary spectra 
were separated into a set of energy intervals such as $[0.2, 50]$, $[50, 100]$, ..., $[450, 
500]$, $[500, 600]$, ..., $[900, 1000]$, $[1000, 1250]$, ..., $[1750,2000]$~GeV, the 
simulation program was run separately for each primary energy bin, and appropriately 
renormalized later when the full sample was constituted.

The corresponding exposure times ranged from $2.1 \times 10^{-12}$s for the lowest energy 
bin to $4.7 \times 10^{-10}$s for the highest energy bin, corresponding to a total of 
the order of 10$^8$ events generated in the sample. 
\end{enumerate}
These calculations as those reported previously \cite{DERP,DERL,DERH} include no 
adjustable parameter.

\section{Atmospheric muon spectra \label{MUS}}

As it was mentioned above, atmospheric muons are produced in the decay chain of pions and kaons 
produced in the collisions, i.e., in the same reaction chain as neutrinos. A successful 
reproduction of the muon flux in the atmosphere is thus a further support to the reliability of 
the neutrino flux calculated in the same framework, physical conditions and computational 
environment. 
This is true as well for the electron-positron flux. For the latter, the present approach has 
allowed to successfully reproduce the AMS measurements of the $e^\pm$ flux and of their 
latitude dependence, in near earth orbit recently \cite{DERL} (see \cite{DERLICRC} for the 
$e^\pm$ in the atmosphere). The results on the muon flux are presented in this section.
  
It must be emphasized first that the muon experimental observables, namely the energy 
dependence of the flux at various altitudes, are not available for neutrinos. Therefore 
they offer a test ground for flux calculations which is more sensitive to the various 
elements of the calculations than the neutrino observables discussed below. For this 
reason they deserve to be investigated with care.

\subsection{Negative muons}

\begin{figure*}
\begin{tabular}{cc}
\includegraphics[width=9.6cm]{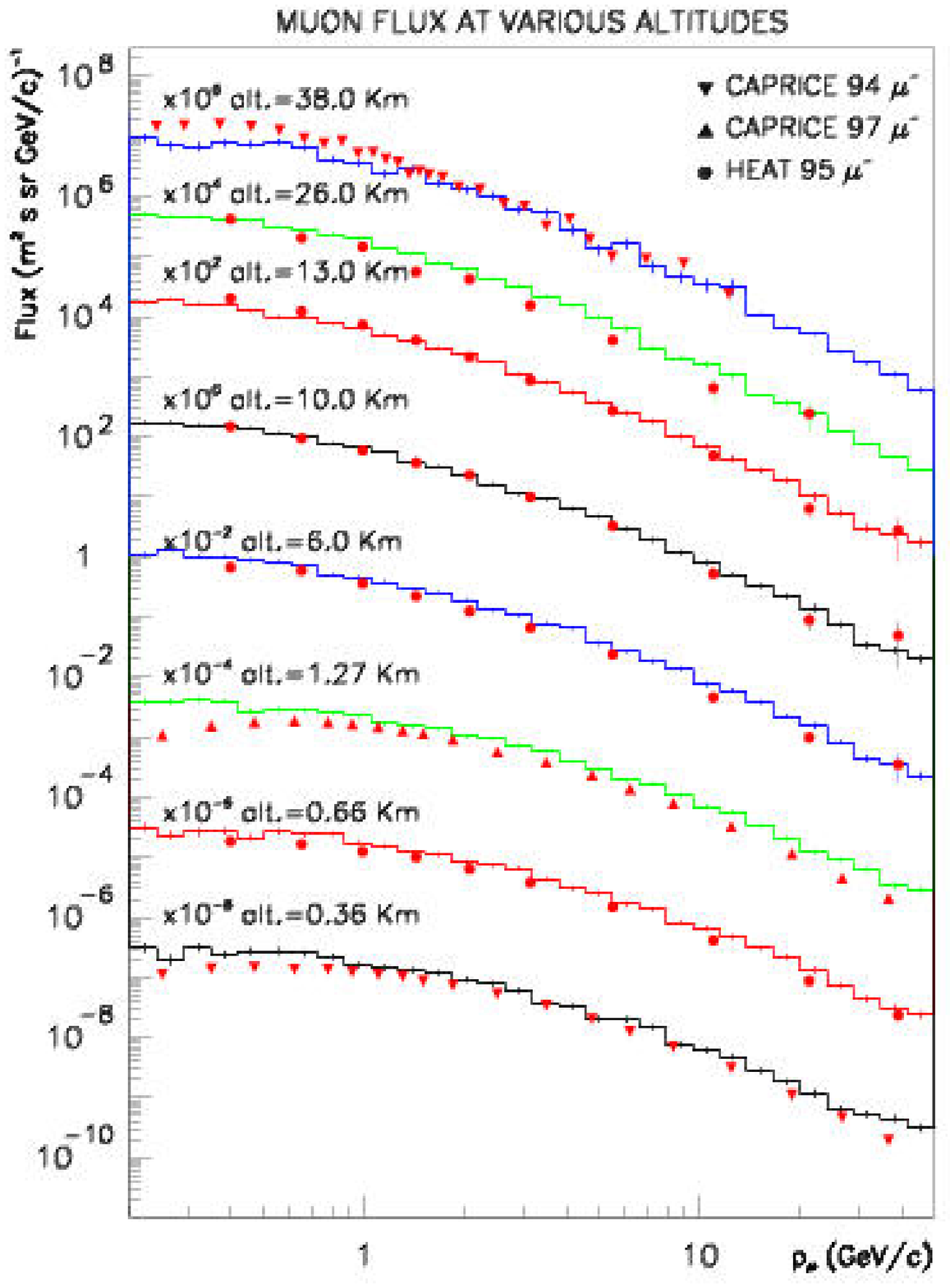}&\hspace{-1cm}
\includegraphics[width=9.6cm]{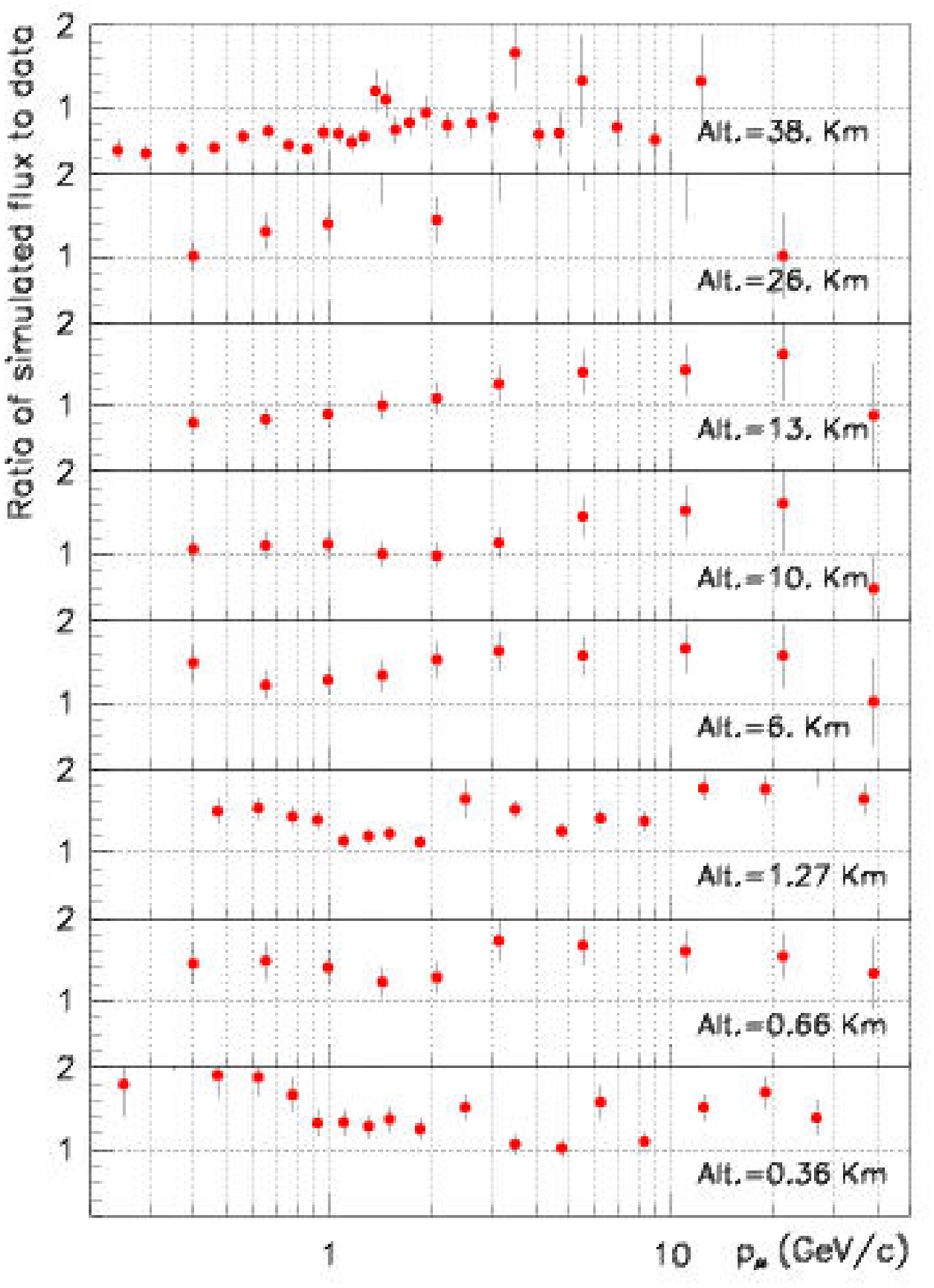}
\end{tabular}
\caption{Left: Simulation results (histograms) for the negative muon flux at various
altitudes in the atmosphere, compared to measurements (full circles
and triangles), from sea level up to about 38~km. See text for
details. Right: Ratio of simulated flux to the measurement for each
data point.\label{muonfr}}
\end{figure*}

Figure~\ref{muonfr} shows the calculated muon flux compared to the data measured by the 
CAPRICE and HEAT experiments \cite{capri,cbdv} at various altitudes (left), and the ratios 
of the simulated to measured values for each data point (right). The agreement between 
simulation results and data is quite good through the energy range investigated (0.5-50)~GeV 
for all altitudes, from about the sea level to 38~km. It is especially good over the region 
from 10km to 26km altitude where a large fraction of the neutrinos detected by underground 
detectors are produced (see below). A significant difference between simulation 
results and data is observed however at 38~km altitude for muon energies below 1~GeV. 
This might result from an underestimated production of low energy pions for CR incident energy 
below about 10~GeV, since the pion production cross section in this energy region is poorly
constrained by the very little data available. 
Note however that relatively large uncertainties and variations of the measured values, 
are observed for the low energy muon data \cite{phan}.

\begin{figure}
\includegraphics[width=9.5cm]{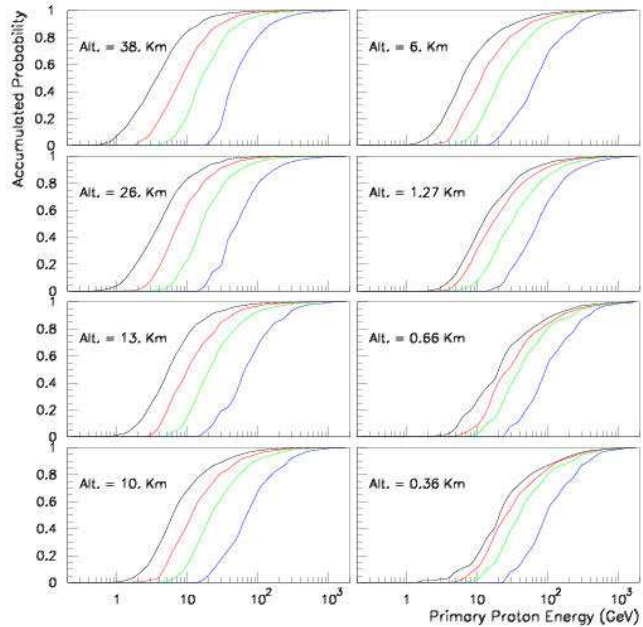}
\caption{CR proton kinetic energy distributions of the cumulative probability 
for producing downgoing muons in the energy bins of 0.2-0.4, 0.8-1.2,
2.0-4.0 and 8.0-12.0~GeV (from the left to the right), with zenith
($\theta_{zenith} \le 15^o$), for the altitudes indicated.
\label{muonp}}
\end{figure}

The energy distribution of the cumulated probability for the incident CR protons to produce 
downward-going muons close to zenith in different energy bins and for different altitudes,
is illustrated in figure~\ref{muonp}. The corresponding mean values of the primary proton energy are 
tabulated in table~\ref{pmuon}.

\begin{table}
\begin{ruledtabular}
\begin{tabular}{c|cccccccc}
 \multicolumn{1}{c|}{ } & \multicolumn{8}{c}{Altitude (km)}\\
 $E_\mu$ (GeV) & 38 & 26 & 13 & 10 & 6 &
1.27 & 0.66 & 0.36  \\  \hline
0.2-0.4  & 8.3 & 9.0 & 14.0 & 18.6 & 35.6 & 66.1 & 63.6 & 68.1 \\ 
0.8-1.2  & 17.5 & 16.6 & 25.0 & 30.8 & 46.8 & 75.9 & 84.9 & 82.0 \\ 
2.0-4.0  & 33.0 & 34.1 & 46.9 & 53.7 & 74.7 & 100.3 & 102.9 & 101.9 \\ 
8.0-12.  & 93.2  & 97.9 & 127.2 & 136.3 & 142.3 & 186.8 & 189.5 & 190.3 \\ 
\end{tabular}
\caption{ Mean incident CR proton energy (GeV) producing near vertical ($\theta_{zenith} 
\le 15^o$) downgoing muons in the specified energy bins and altitudes. 
\label{pmuon}}
\end{ruledtabular}
\end{table}

From figure~\ref{muonp}, it is easy to see that the cumulative probability curves shift 
gradually toward the high energy region with the decreasing altitude, while the distances 
between the curves are compressed. This can be clearly seen as well from table~\ref{pmuon}. 
At 38~km altitude, the mean energies of primary protons producing 0.2-0.4~GeV muons and 
producing 8.0-12.0~GeV muons differ by a factor of about 11, this factor becomes about 7 at 
10km, and 3 at around sea level. 

Figure~\ref{muonp}, table~\ref{pmuon} and the following figure~\ref{prisec} and figure~\ref{highrank}
show that, 
\begin{enumerate}
\item at the highest float altitude,
the simulated muon flux is most sensitive to the pion production cross section used in the 
simulation and hence, the higher altitude muon flux data can thus be taken as a good 
test of this cross section. 
\item in contrast, at around sea level, 
the low energy muon flux should be more sensitive to the proton/neutron 
production cross section because on the average:
\begin{enumerate}
\item The low energy muon path length is roughly on the km scale, and the muons measured 
at sea level are thus produced at relatively low altitude.

\item Above 10~km, about 5 collisions have occurred. In these collisions, more secondary protons/neutrons 
are produced, that can contribute largely to muons parent meson production.
\end{enumerate}\end{enumerate}

\subsection{Secondary proton contribution}

\begin{figure}
\includegraphics[width=9.5cm]{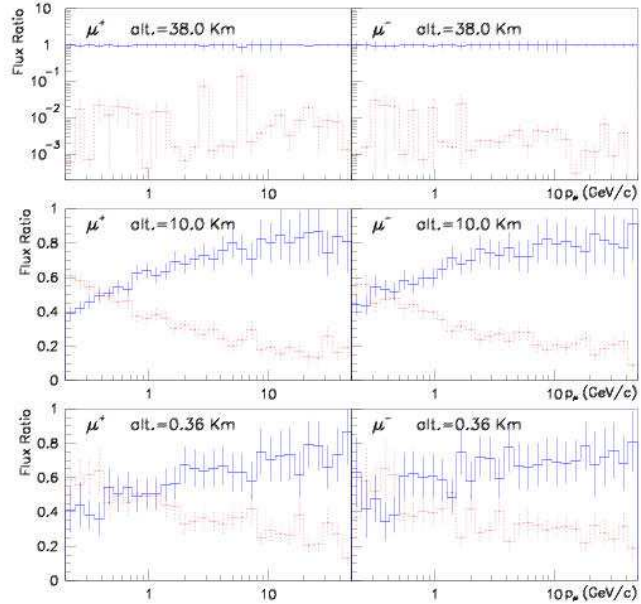}
\caption{Flux ratios for muons induced by the primary CR (proton + helium) flux (solid 
line), and by atmospheric secondaries (dotted lines), respectively, to the total flux, for three 
detection altitudes. The total flux are those from figure~\ref{muonfr}. \label{prisec}}
\end{figure}

The ratios of the muon flux originating from primary CRs and from secondaries respectively, 
to the total flux, are shown on figure~\ref{prisec}. The flux are nearly vertical and the 
corresponding normalization area are the same as those shown in figure~\ref{muonfr} for the 
three altitudes. 

This figure shows as it could be expected from simple collision rank considerations, that in 
the whole muon energy range, at the highest float altitude, more than 98$\%$ muon flux  originate 
from primary CRs. At intermediate altitudes, the flux induced by secondaries decreases 
continuously with the energy from about $60 \%$ at 0.2~GeV down to about $20 \%$ at 10~GeV, while
at around sea level, the same trend is observed with a slower decrease with energy.

\subsection{Positive muons}

\begin{figure*}
\begin{tabular}{cc}
\includegraphics[width=9.6cm]{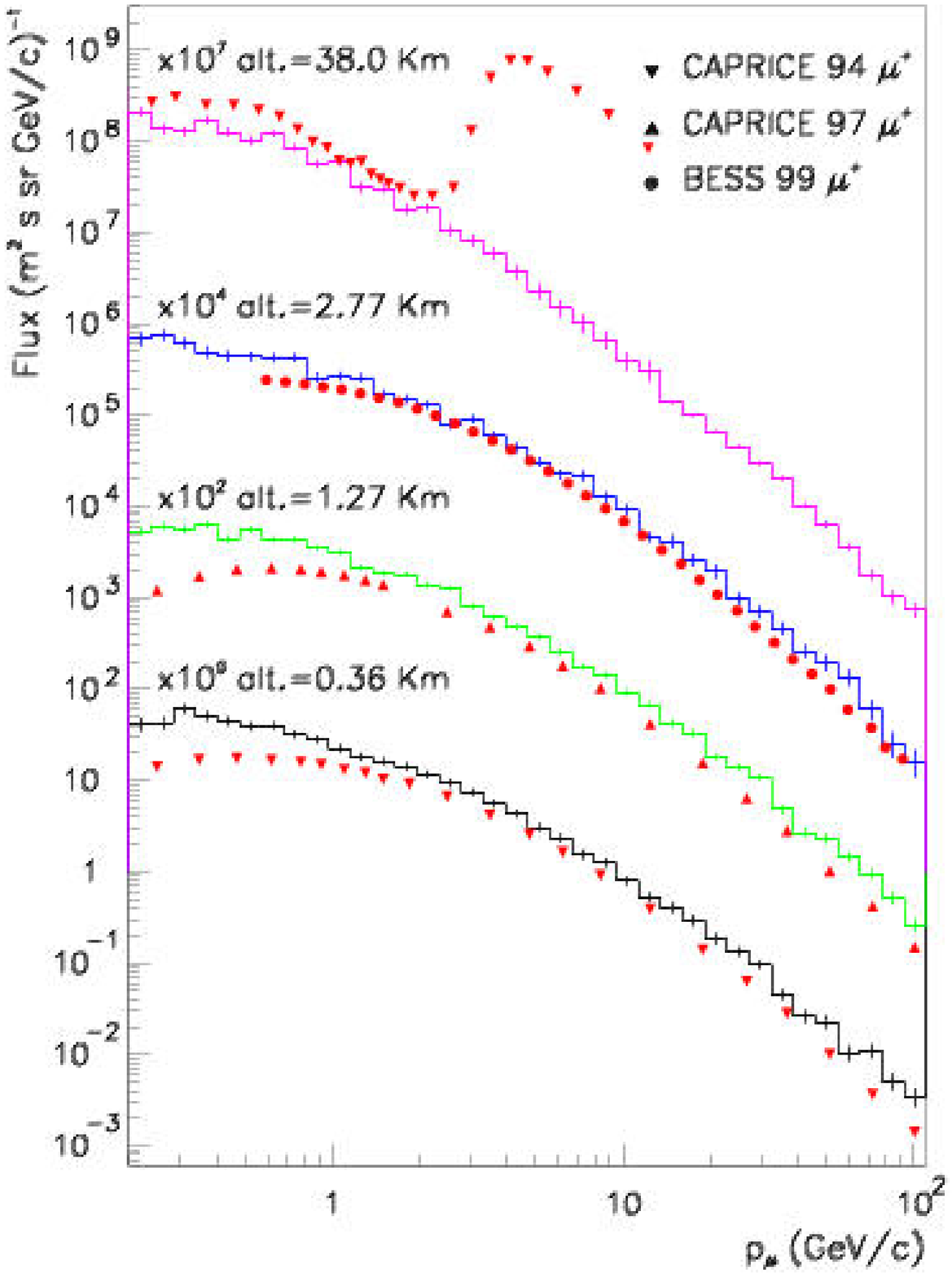}&\hspace{-1cm}
\includegraphics[width=9.6cm]{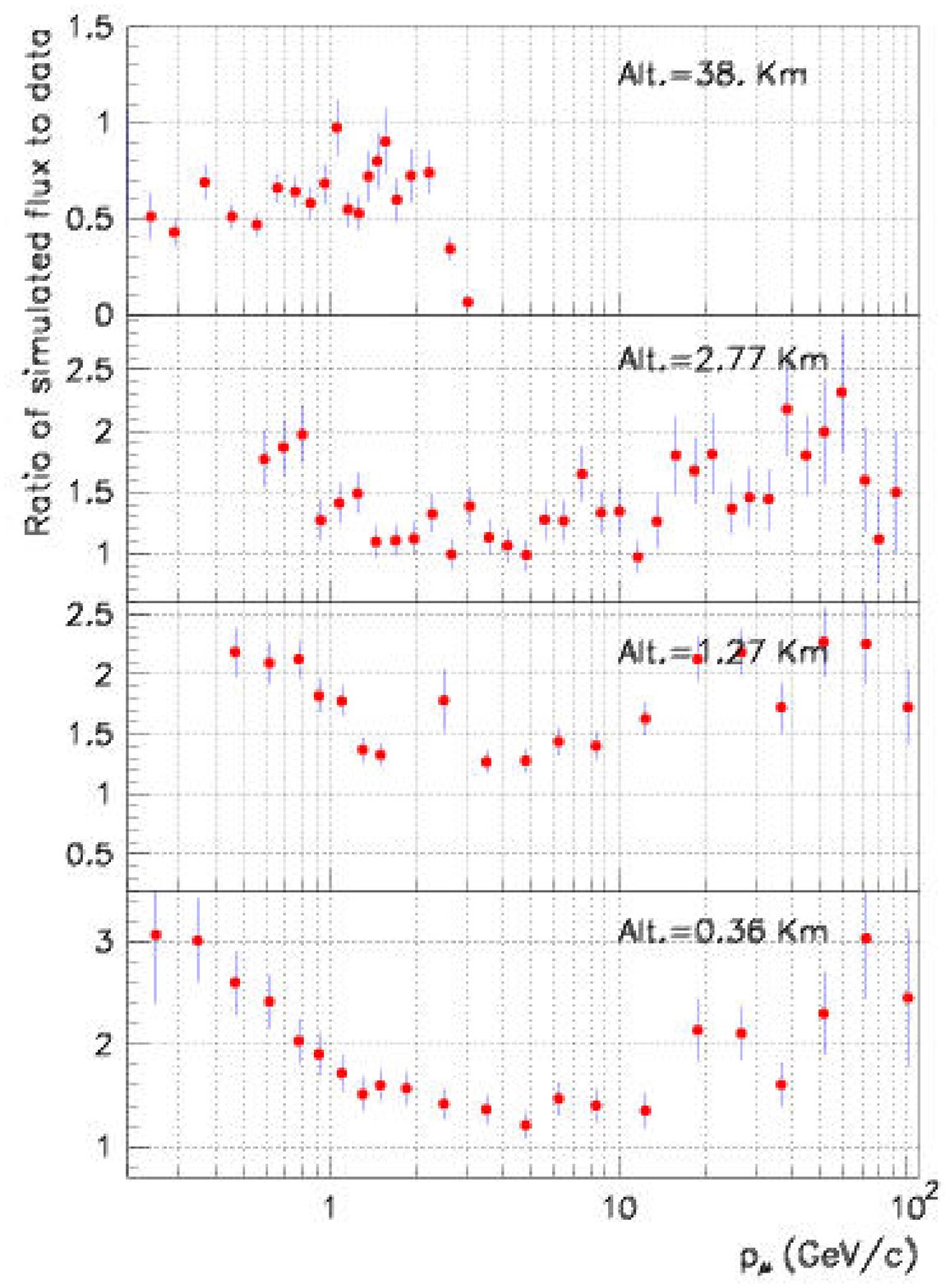}
\end{tabular}
\caption{Left: Simulation results (histograms) for the positive muon flux at various altitudes 
in the atmosphere, compared with measurements (full circles and triangles), 
from sea level up to 38~km, from refs\cite{capri,bess02}. See text for details. 
Right: Ratio of simulated flux to the measurement for each data point.
\label{muonpfr}}
\end{figure*}

Figure~\ref{muonpfr} shows the calculated $\mu^+$ flux compared to the measured data from the 
CAPRICE and BESS experiments.
The agreement with the BESS 99 data at mountain altitude \cite{bess02} is very good, but in the 
low (below 1.0~GeV) and high (beyond 10.0~GeV) energy range, the departure is obvious. This may 
result from the
overestimation of the secondary proton/neutron production cross section at low energy and
pion/kaon production cross section at high energy.

\section{Properties of neutrino spectra}\label{GLOB}

In this section, the characteristics of the simulated neutrino spectra, averaged over the 
whole detection surface, for three bins of geomagnetic latitude and integrated over $4~\pi$ 
solid angle, are described in details. The detection sphere is defined at 0.372~km, the 
altitude of the Super-Kamiokande detector. Some features having important implications for 
underground experiments, are emphasized. 
 
\subsection{Absolute flux and flavor ratios}\label{GLFL}
   
\begin{figure*}
\begin{tabular}{ccc}
\includegraphics[width=6.6cm]{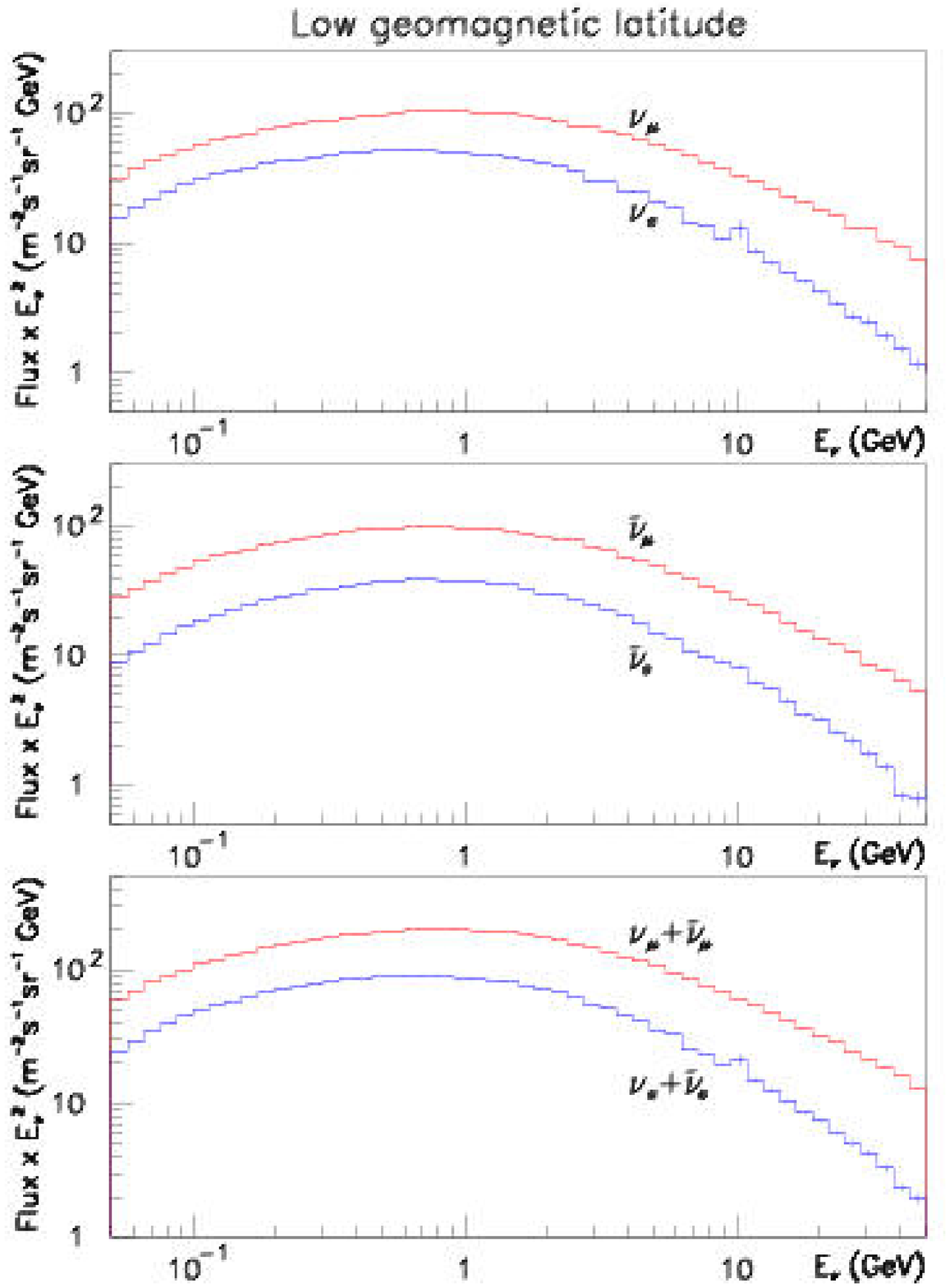}&\hspace{-1cm}
\includegraphics[width=6.6cm]{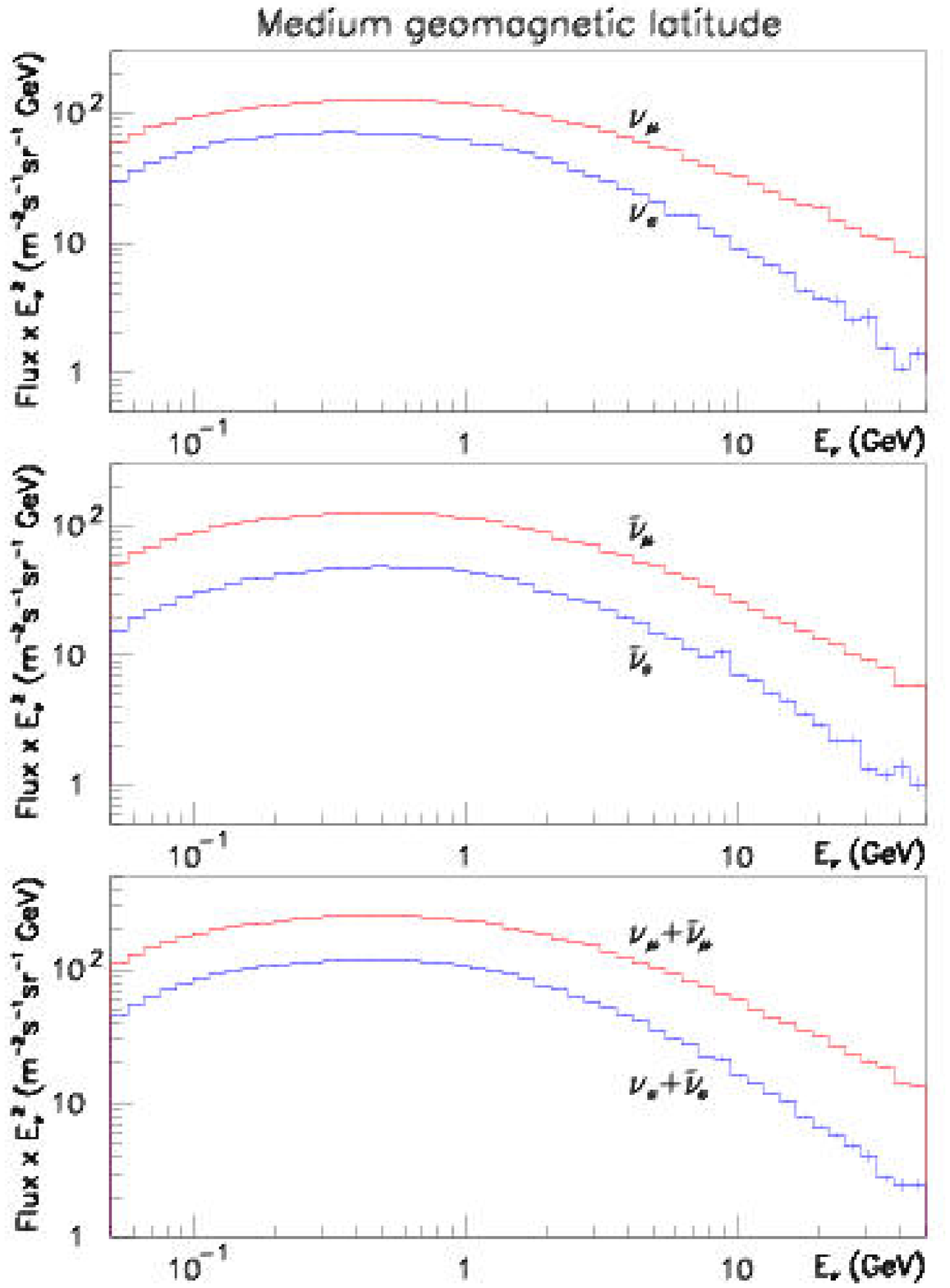}&\hspace{-1cm}
\includegraphics[width=6.6cm]{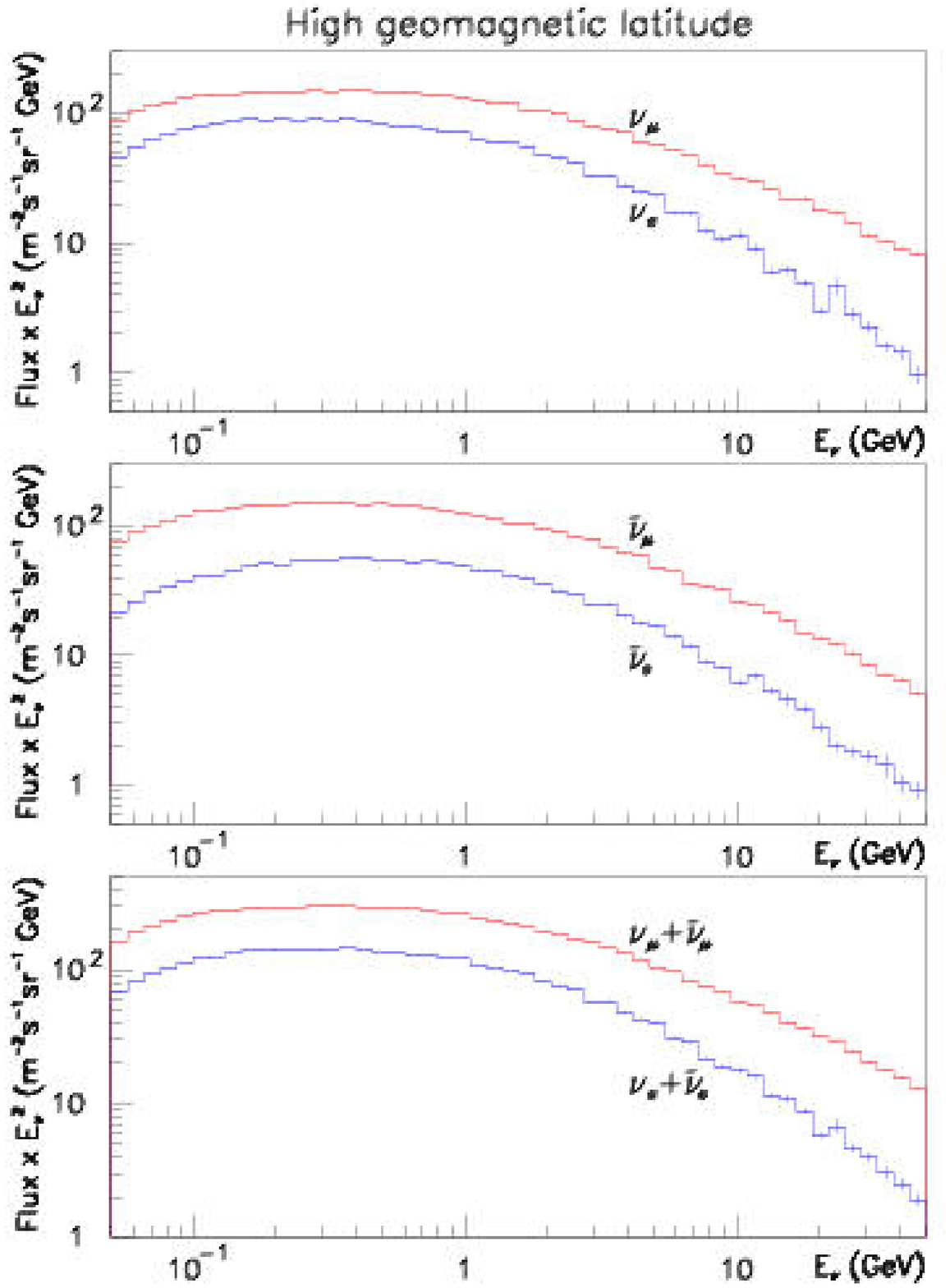}
\end{tabular}
\caption{Neutrino flux in three bins of geomagnetic latitude 0-0.5, 0.5-1, and 1-$\pi$/2 
radians, averaged over 4 $\pi$ solid angle. The error bars given in the figure and all the 
following figures are statistical unless specified otherwise.
\label{flux3}}
\end{figure*}

\begin{figure*}
\begin{tabular}{ccc}
\includegraphics[width=6.6cm]{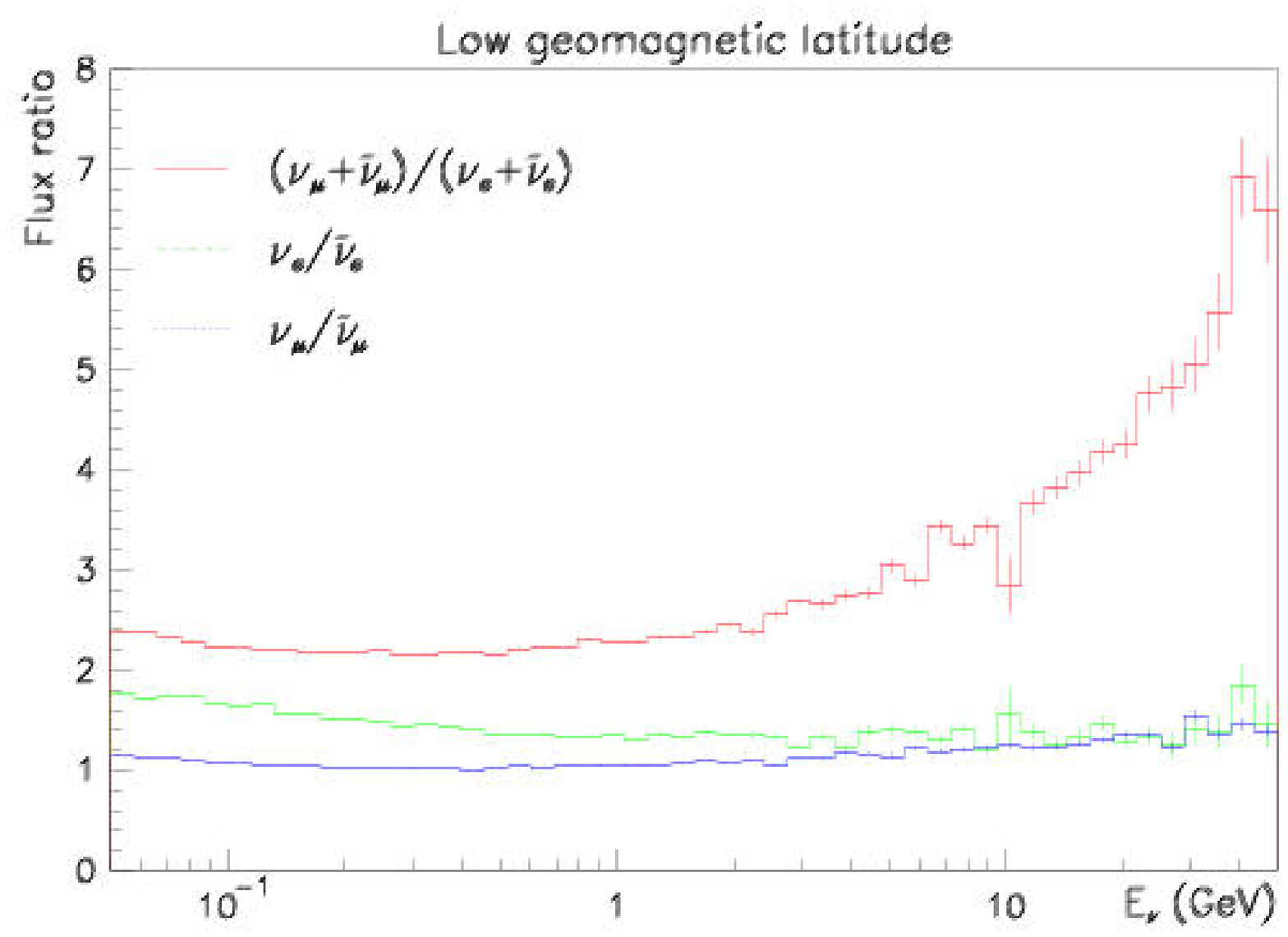}&\hspace{-1cm}
\includegraphics[width=6.6cm]{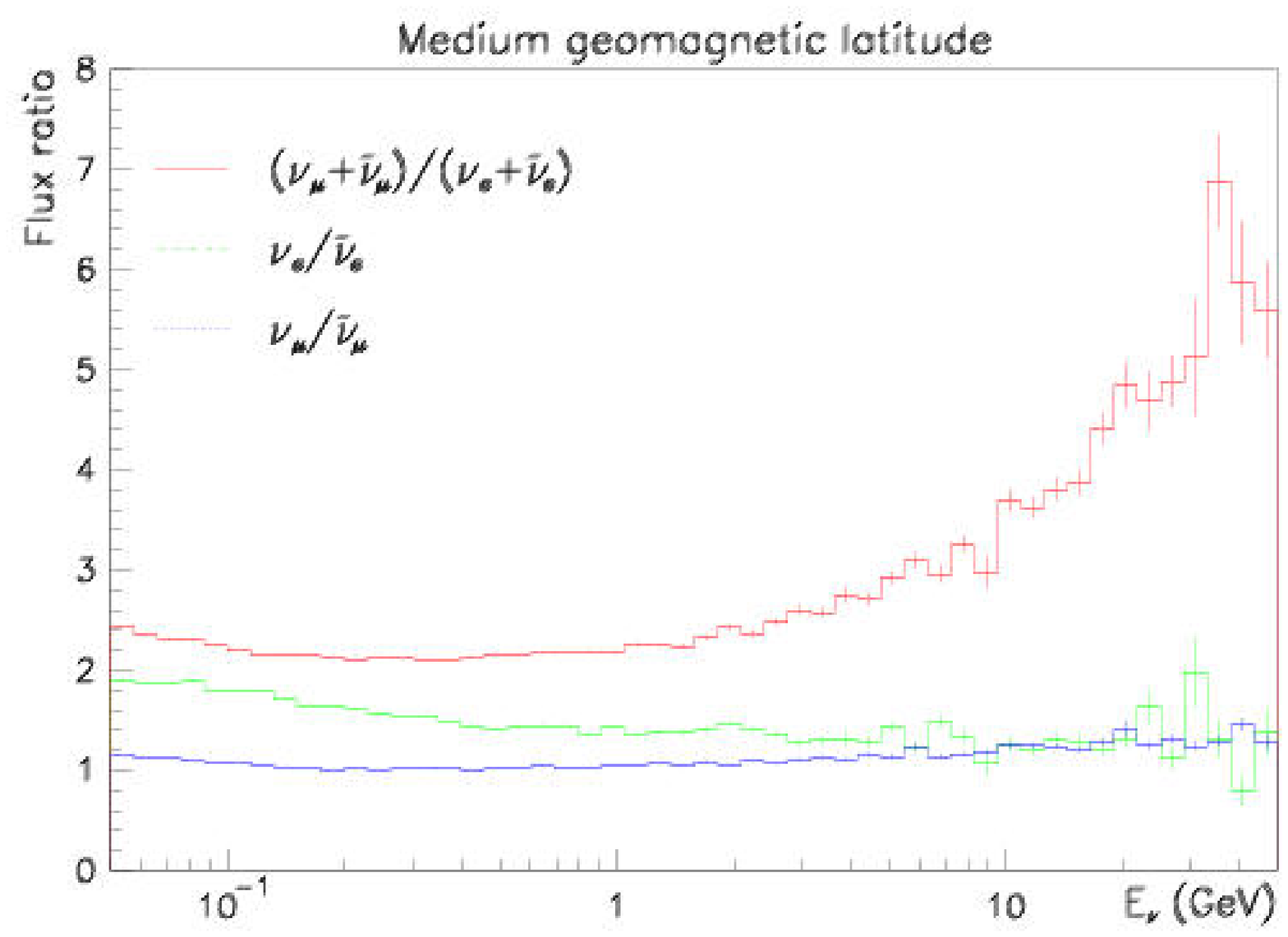}&\hspace{-1cm}
\includegraphics[width=6.6cm]{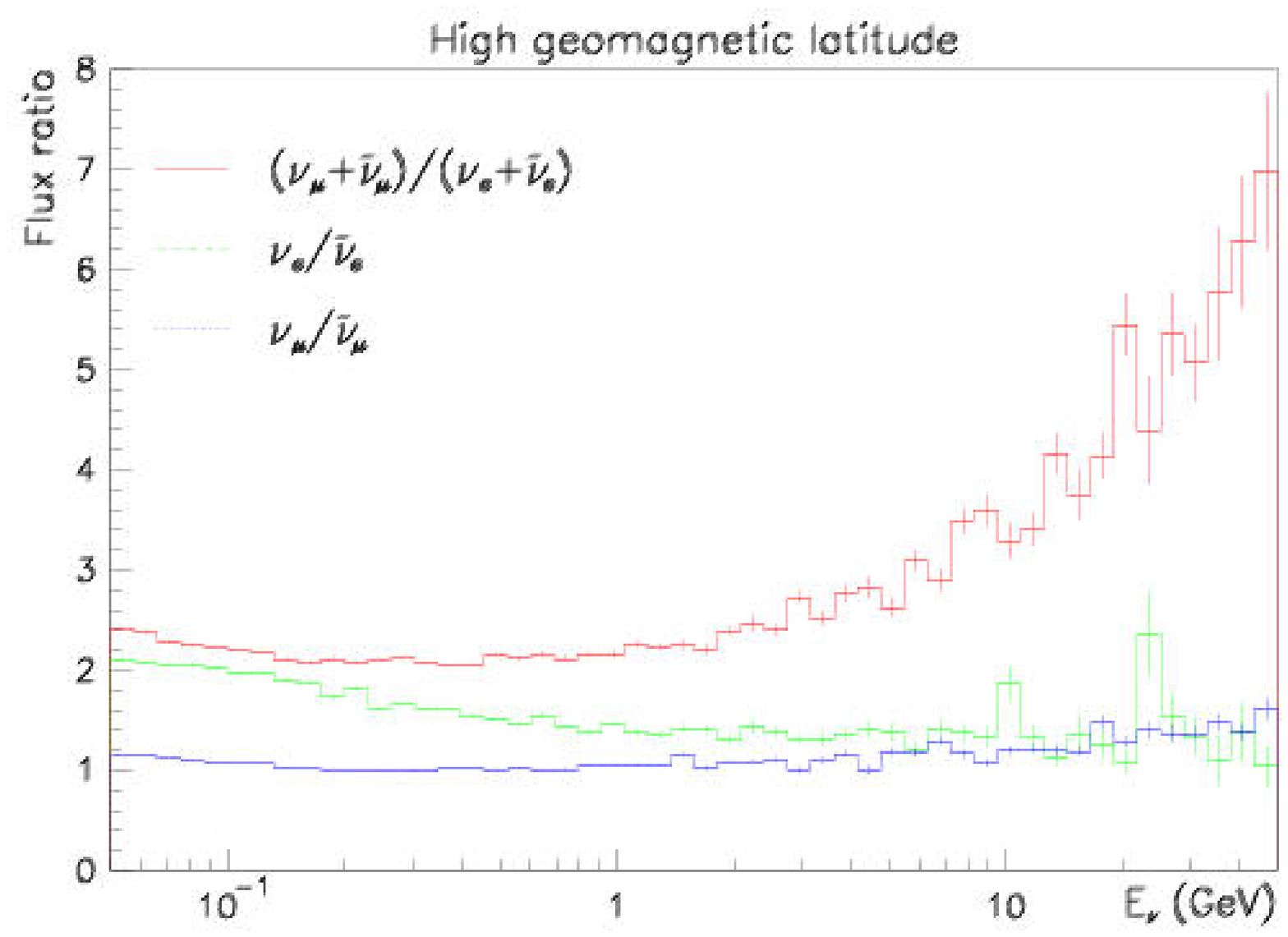}
\end{tabular}
\caption{Individual $\nu/\bar{\nu}$ and sum 
$(\nu_{\mu}+\bar{\nu_{\mu}})/(\nu_e+\bar{\nu_e}$) flux ratios, as in figure~\ref{flux3}. 
\label{fluxratio3}}
\end{figure*}

The calculated flux are shown in figure~\ref{flux3} for three bins of geomagnetic latitude.
The corresponding ratios of flux are shown on figure~\ref{fluxratio3}, where it is seen that 
the flux increases with the geomagnetic latitude in the low energy ($<$ few GeV) range. 
At 0.1~GeV, the difference between the flux at low and high latitudes can be more than a factor 
of 2. However, this difference decreases with the increasing particle energy, vanishing beyond 
a few GeV. This reflects the effect of the geomagnetic field. Indeed, the mean primary particle
energy associated to 1~GeV neutrino production, is about 60~GeV (see below), a value at which 
the effects of the geomagnetic cutoff (GC) are small.

Figure~\ref{fluxratio3} shows that the flavor ratios for low, middle and high latitudes
are very similar to each other. The $\nu_\mu/\bar{\nu}_\mu$ is flat through energy range, 
with its value near about 1.2. 
$\nu_e/\bar{\nu}_e$ decreases steadily from nearly 2 to about 1.2 - overlap with 
$\nu_\mu/{\bar{\nu}_\mu}$ at high energy end, 
while $(\nu_\mu+{\bar{\nu}_\mu})/(\nu_e+{\bar{\nu}_e})$ 
rises up steeply above about 3~GeV. Large difference for $\nu_e/{\bar{\nu}_e}$ from the 
previous works is seen here (See \cite{hk90,HO95} for comparisons). 

These features can be easily understood in terms of the productions of $\pi^+ (K^+)$ and 
$\pi^- (K^-)$, and of the muon kinematics. The ratio $\nu_e/{\bar{\nu}_e}$ is principally 
controlled by the ratio of $\pi^+ (K^+)$ to $\pi^- (K^-)$. The ratio 
$(\nu_\mu+{\bar{\nu}_\mu})/(\nu_e+{\bar{\nu}_e})$ is sensitive to the muon decay kinematics. 
The variation of $\nu_\mu/{\bar 
\nu_\mu}$ with energy is thus determined by both $\pi^+/\pi^- (K^+/K^-)$ production ratios 
and the muon decay kinematics. Because of the Lorentz time dilation, more higher 
energy muons do not decay while crossing the atmosphere. Hence, the $\nu_e ({\bar{\nu}_e})$ 
production (resulting from muon decay) is lesser than for $\nu_\mu ({\bar{\nu}_\mu})$, since
a $\nu_\mu$ has been produced in association with the $\mu$ in the $\pi$ decay, resulting 
in a $(\nu_\mu+{\bar{\nu}_\mu})/(\nu_e+{\bar{\nu}_e})$ ratio rising up with the neutrino 
energy.

Here, two points should be noted. First, the $\nu_e/{\bar{\nu}_e}$ ratio at low energy should 
be expected to be larger since the largest production ratio of $\pi^+/\pi^-$ is at low incident
CR energy (see \cite{coch} for example). 
Second, the $\nu_e/{\bar{\nu}_e}$ ratio should decrease to overlap with
$\nu_\mu/{\bar{\nu}_\mu}$ at high energy with their values approaching to 1.0. 
This is because 
at high incident primary energy, the production ratio of $\pi^+/\pi^-$ and $K^+/K^-$ approaches 
to 1.0 \cite{bogg}, 
and most of the high energy muons from which the high energy $\nu_e$ and ${\bar{\nu}_e}$ 
come will not decay before reaching the detection sphere. So, the $\nu_\mu$ flux is dominated 
by the $\pi^+$ decay while the ${\bar{\nu}_\mu}$ flux is dominant-ed by $\pi^-$ decay. And hence, 
at very high energy, 
$$\nu_e / {\bar{\nu}_e} \approx \nu_\mu / {\bar{\nu}_\mu} \approx \pi^+(K^+)/\pi^-(K^-).$$
Such characteristics can be seen clearly in figure~\ref{fluxratio3}.

\subsection{Contribution from the incident CR $He$ flux}\label{CRHE}

\begin{figure*}
\begin{tabular}{cc}
\includegraphics[width=9.6cm]{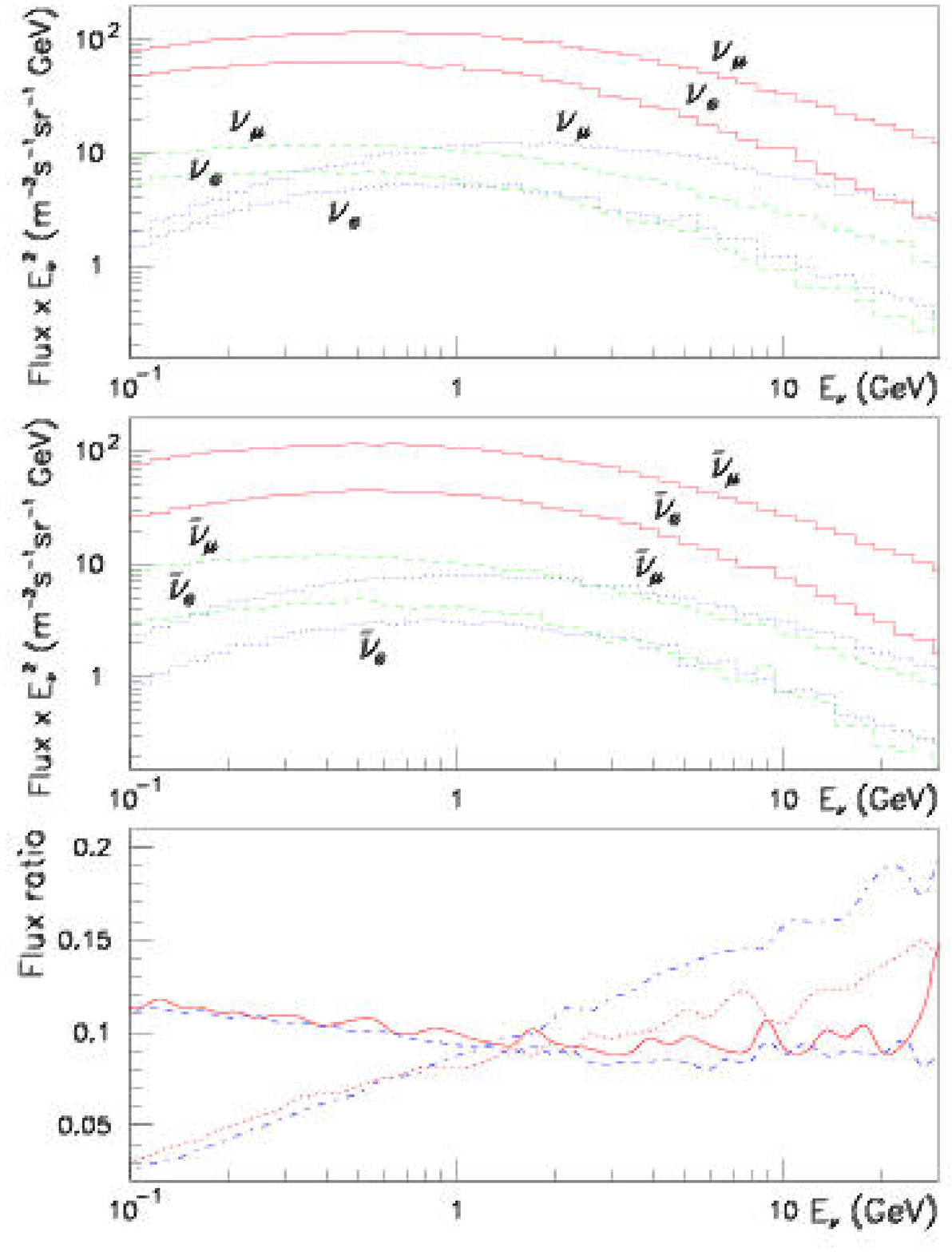}&\hspace{-1cm}
\includegraphics[width=9.6cm]{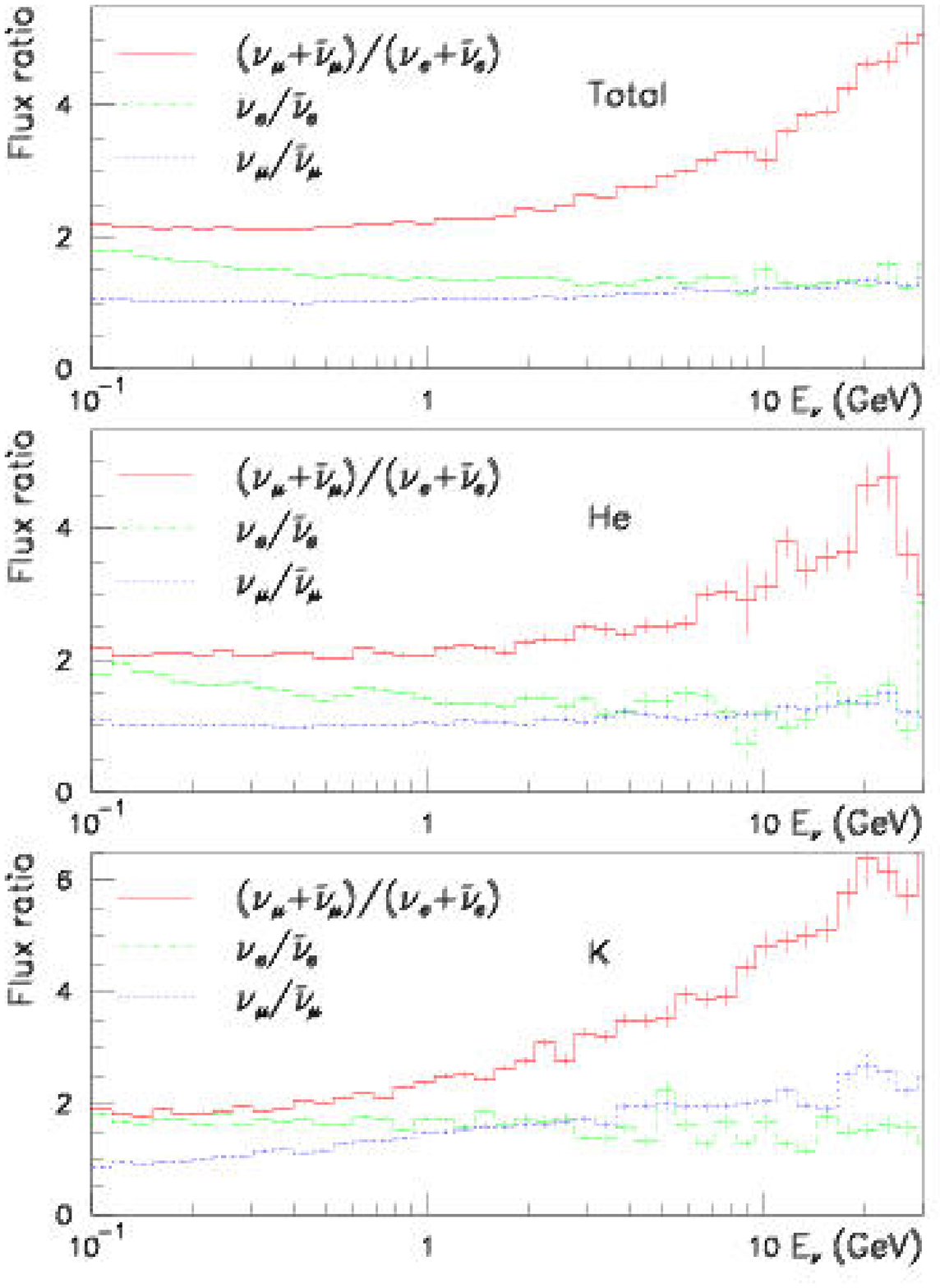}
\end{tabular}
\caption{Left up and middle: Energy spectra of the CR Helium flux 
contribution (dashed lines) and total Kaon decay contribution (dotted
lines), compared to the total flux (solid lines), for the $\nu$ and
$\bar{\nu}$, neutrino flavors from top to bottom as indicated in the
figure respectively. Left bottom: Ratios of CR Helium induced
($\nu_\mu + \bar{\nu_\mu}$) flux to total ($\nu_\mu + \bar{\nu_\mu}$)
flux (solid line), CR Helium induced ($\nu_e + \bar{\nu_e}$) flux to
total ($\nu_e + \bar{\nu_e}$) flux (dashed line), kaon decay induced
($\nu_\mu + \bar{\nu_\mu}$) flux to total ($\nu_\mu + \bar{\nu_\mu}$)
flux (dashdot line), and kaon decay induced ($\nu_e + \bar{\nu_e}$)
flux to total ($\nu_e + \bar{\nu_e}$) flux (dotted line).  Right:
Flavor ratios for the same three contributions (Total, CR He induced,
and Kaon decay, from top to bottom respectively), averaged over 4
$\pi$ solid angle and integrated over the whole detection sphere. See
text.\label{hekt}}
\end{figure*}

The calculated total neutrino flux is compared on left
figure~\ref{hekt} with the contribution induced by the CR Helium
($He$) flux and with the total kaon decay contribution. The
corresponding flavor ratios are also shown on the same figure (right
panel), where it is seen that the contribution of the CR $He$ flux to
the total neutrino flux is about $10 \%$ through the whole energy
range.

This contribution is constrained by both the CR $He$ spectral
abundance and the $He$ induced pion multiplicity in nuclear
collisions. Around 20~GeV per nucleon incident energy, where the
neutrino production around 1~GeV is maximum (see discussion below and
\cite{GA96}), the CR $He$ flux is about $5 \%$ of the $p$ flux, this
fraction decreasing slowly with the increasing primary
energy. Experimentally, the pion multiplicity increases with the mass
of the projectile according to a known law \cite{kmnn}. For the
$\alpha + C$ reaction at 4~GeV/nucleon, it is about 3 times the $p +
C$ value at the same incident energy~\cite{AG84}. These numbers are in
qualitative agreement with the $10 \%$ contribution to the neutrino
flux found to arise from the $He$ flux, assuming that the proton
induced helium contribution is small. This result is also consistent
with the previous work on the AMS lepton data by the authors~\cite{DERL}.

The difference between the contributions of CR $He$ to the
$(\nu_e+{\bar{\nu}_e})$ and $(\nu_\mu+{\bar{\nu}_\mu})$ flux ratios is
very small as seen on lower left figure~\ref{hekt}. For the flavor
ratio of the $He$ flux contributions shown on middle right
figure~\ref{hekt}, they have a similar energy dependence as the total
flux, as it could be expected (top).

\subsection{Contribution from secondary Kaons}\label{SECK}

At low energies the neutrino flux mostly results from $\pi$ decay in account of the dominance 
of the pion production. The Kaon decay contribution increases however with the increasing 
incident energy and the K production cross section, and finally dominates 
the production for very high energy neutrinos. This is shown on left figure~\ref{hekt} 
where the kaon decay contribution to the total neutrino flux (bottom left) is seen to 
increase continuously from about 2-3\% at 0.1~GeV up to $\approx$15\% for 
$(\nu_e+{\bar\nu_e})$, and $\approx$20\% for $(\nu_\mu+{\bar\nu_\mu})$, above 20~GeV.
This results is in agreement with \cite{hk90,HO95}.

The energy dependence of the flavor ratio for the kaon decay
contribution to the neutrino flux is a little different from the
averaged value (bottom right panel on figure~\ref{hekt}). The ratio
$\nu_e/{\bar{\nu}_e}$ decreases slowly with the increasing energy from
1.7 at 0.1~GeV down to about 1 at 10~GeV, while $\nu_\mu/{\bar
\nu_\mu}$ increases continuously from 0.9 at 0.1~GeV up to about 2.1
at 10~GeV. In contrast with the individual flavor ratios, the
$(\nu_\mu+{\bar{\nu}_\mu})/(\nu_e+{\bar{\nu}_e})$ ratio grows rapidly
from 1.8 at 0.1~GeV up to about 5 at 10~GeV. These differences
principally result from the $K^+$ to $K^-$ production ratio being
larger than for $\pi^+$ to $\pi^-$ \cite{acfs}.

Another noticeable feature of the kaon decay contribution is that, below 0.4~GeV neutrino
energy the ratio $(\nu_\mu+{\bar{\nu}_\mu})/(\nu_e+{\bar{\nu}_e})$ is less than 2. This can be 
understood from the $K_{3e\nu}$ decay mode having always larger a branching ratio than the 
$K_{3 \mu \nu}$ mode. In particular, for $K_L^0$, the  $K_{3e\nu}$ channel is the dominant 
decay mode~\cite{pdg00}.

\subsection{Contribution of atmospheric secondaries to the neutrino flux\label{ATSEC}}

\begin{figure*}
\includegraphics[width=15cm]{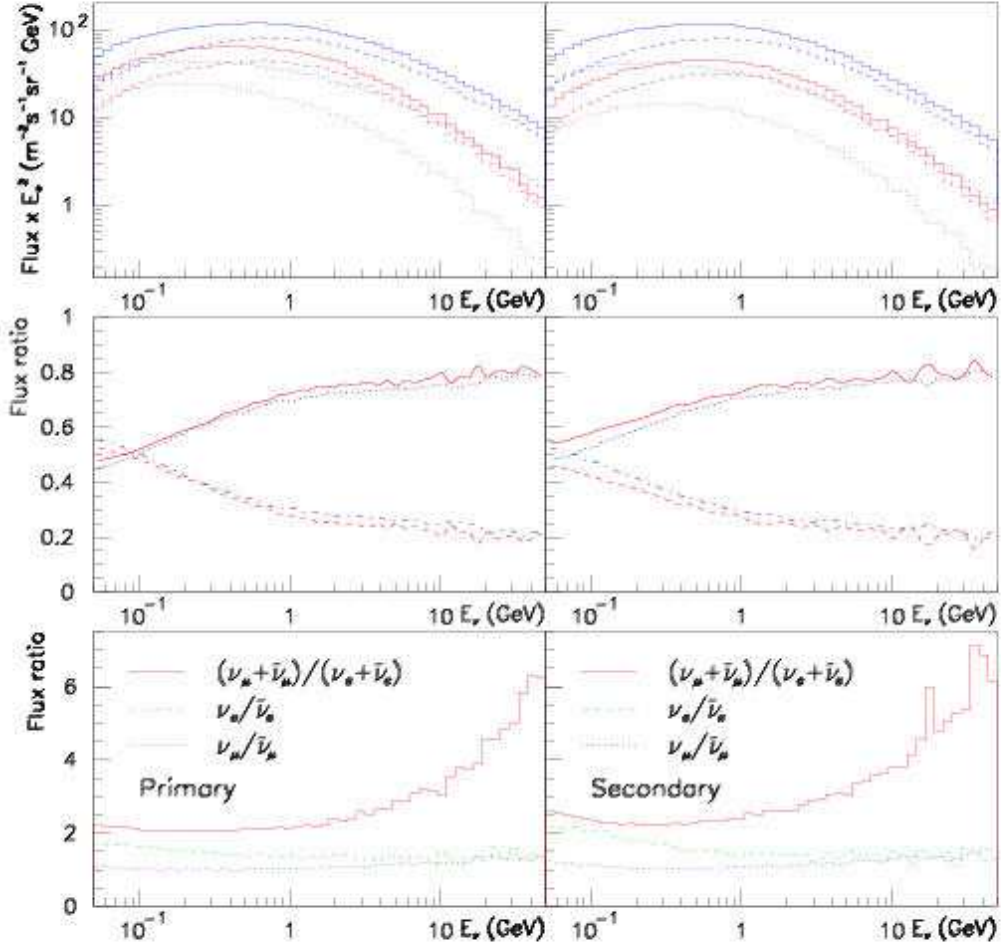}
\caption{Total neutrino flux, primary CR proton and helium directly induced neutrino flux, 
and the secondary induced neutrino flux, averaged over whole detection
sphere and over $4 \pi$ solid angle. Top left: CR Primary (dashed
lines, up: $\nu_\mu$, down: $\nu_e$), secondary (dotted lines, up:
$\nu_\mu$, down: $\nu_e$) contributions for $\nu_e$ and $\nu_\mu$, and
total $\nu_e$ and $\nu_\mu$ (solid lines, up: $\nu_\mu$, down:
$\nu_e$).  Top right: CR Primary (dashed lines, up: $\bar{\nu_\mu}$,
down: $\bar{\nu_e}$), secondary (dotted lines, up: $\bar{\nu_\mu}$,
down: $\bar{\nu_e}$) contributions for $\bar{\nu_e}$ and
$\bar{\nu_\mu}$, and total $\bar{\nu_e}$ and $\bar{\nu_\mu}$ (solid
lines, up: $\bar{\nu_\mu}$, down: $\bar{\nu_e}$).  Middle: Ratios of
Primary contributed flux to total flux and Secondary contributed flux
to total flux, for $\nu$ (left) and $\bar\nu$ (right). Dotted line:
ratio of primary directly induced $\nu_\mu$ ($\bar{\nu_\mu}$) to total
$\nu_\mu$ ($\bar{\nu_\mu}$). Solid line: ratio of primary directly
induced $\nu_e$ ($\bar{\nu_e}$) to total $\nu_e$ ($\bar{\nu_e}$).
Dashdot line: ratio of secondary induced $\nu_\mu$ ($\bar{\nu_\mu}$)
to total $\nu_\mu$ ($\bar{\nu_\mu}$). Dashed line: ratio of secondary
induced $\nu_e$ ($\bar{\nu_e}$) to total $\nu_e$
($\bar{\nu_e}$). Bottom: Flux ratios $\nu_e/{\bar\nu_e}$,
$\nu_\mu/{\bar\nu_\mu}$, and $(\nu_\mu+{\bar{\nu}_\mu})/(\nu_e+{\bar
\nu_e})$ for the Primary directly induced (left panel) and Secondary
induced (right panel) components. 
\label{muonsec}}
\end{figure*}

The contribution of the atmospheric secondary flux is compared on
figure~\ref{muonsec} with that of the primary component. It is seen
that, at 0.1~GeV, about 50~$\%$ of the neutrinos come from secondary
proton/neutron induced reactions, this fraction decreasing to about
20$\%$ at 50~GeV.

The ratio of the secondary contribution, $\nu_e/{\bar{\nu}_e}$ 
is larger than for the primary contribution. This is because of the
larger flux of low energy secondaries due to the larger $\pi^+/\pi^-$
production ratio at low incident energy.

\begin{figure}
\includegraphics[width=8cm]{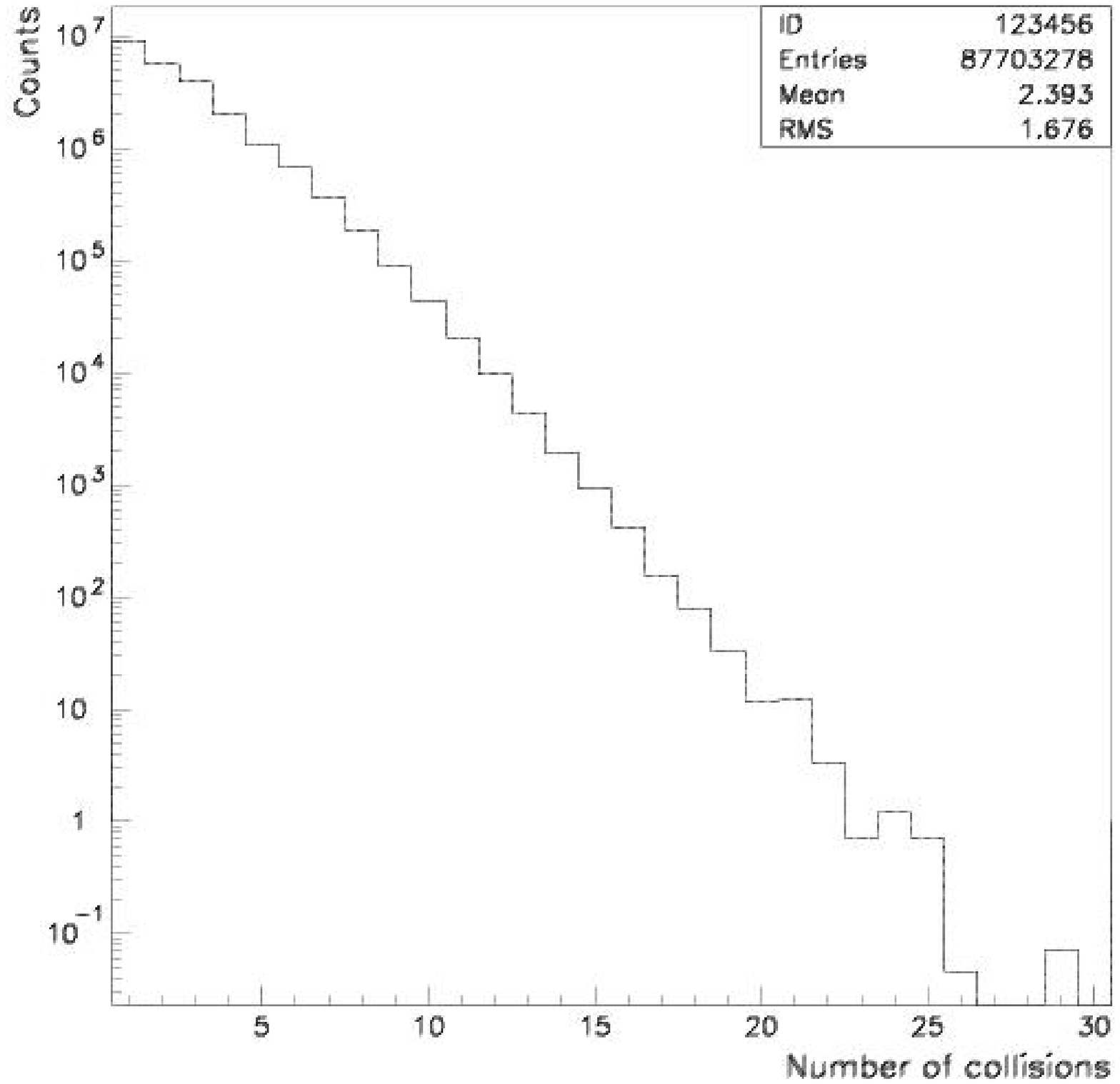}
\caption{Rank distribution of the neutrino producing collisions in the atmosphere. 
See text for details.\label{rank}}
\end{figure}

The distribution of the rank in the cascade of the neutrino producing
collision is plotted on figure~\ref{rank} \cite{DERP}.  The rank is
defined as the number of collisions taking place before the neutrino
is produced. It is 1 when the neutrino is produced from the incident
cosmic ray interaction. The figure shows that, on the average, about
2.4 collisions occur for the parent pion or kaon to be produced.

\subsection{Zenith angular distribution and geomagnetic latitude dependence of the 
neutrino flux\label{GLOBZEN}}

\begin{figure}[!htbp]
\begin{tabular}{c}
\includegraphics[width=9.5cm]{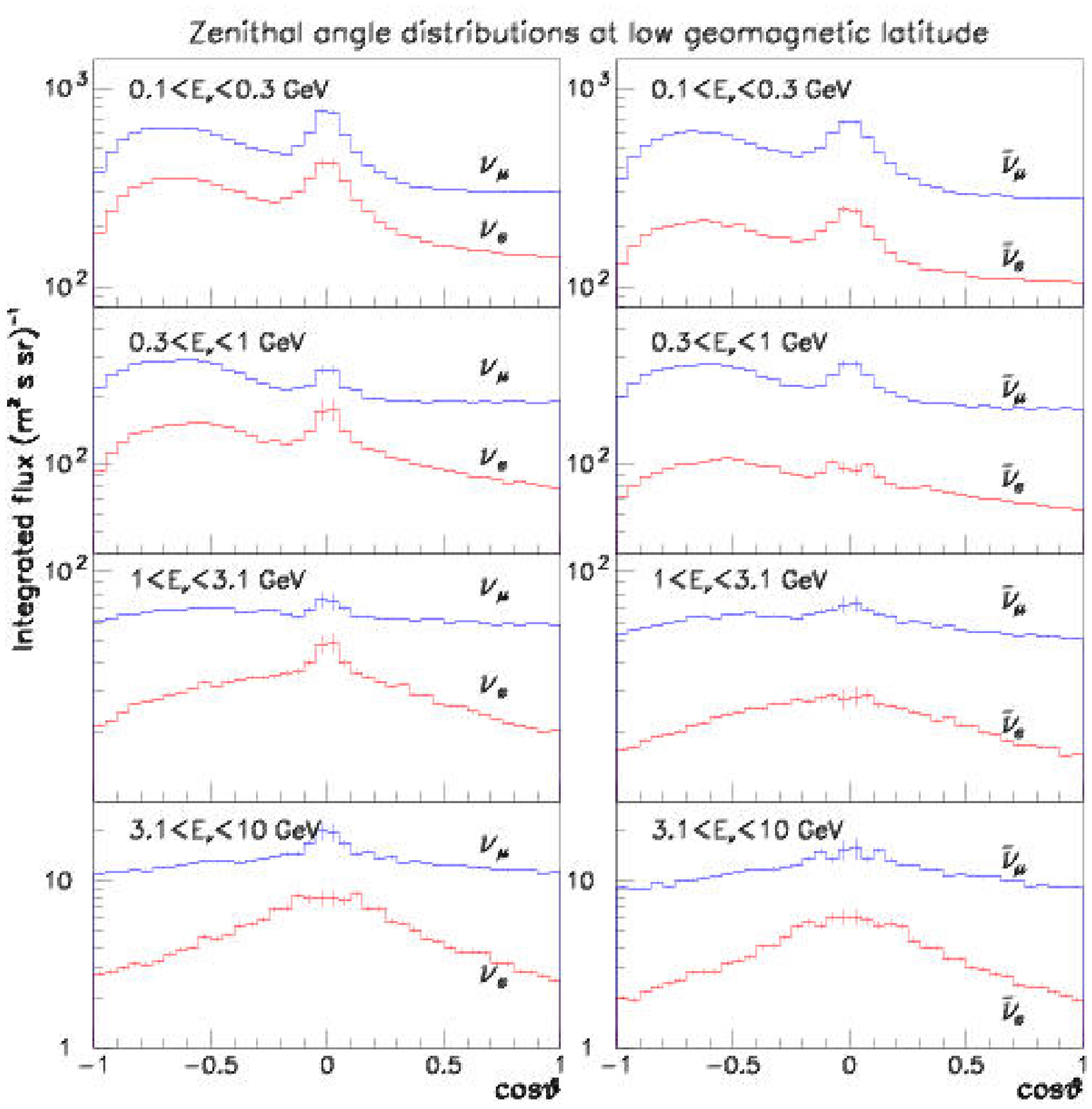}\\
\includegraphics[width=9.5cm]{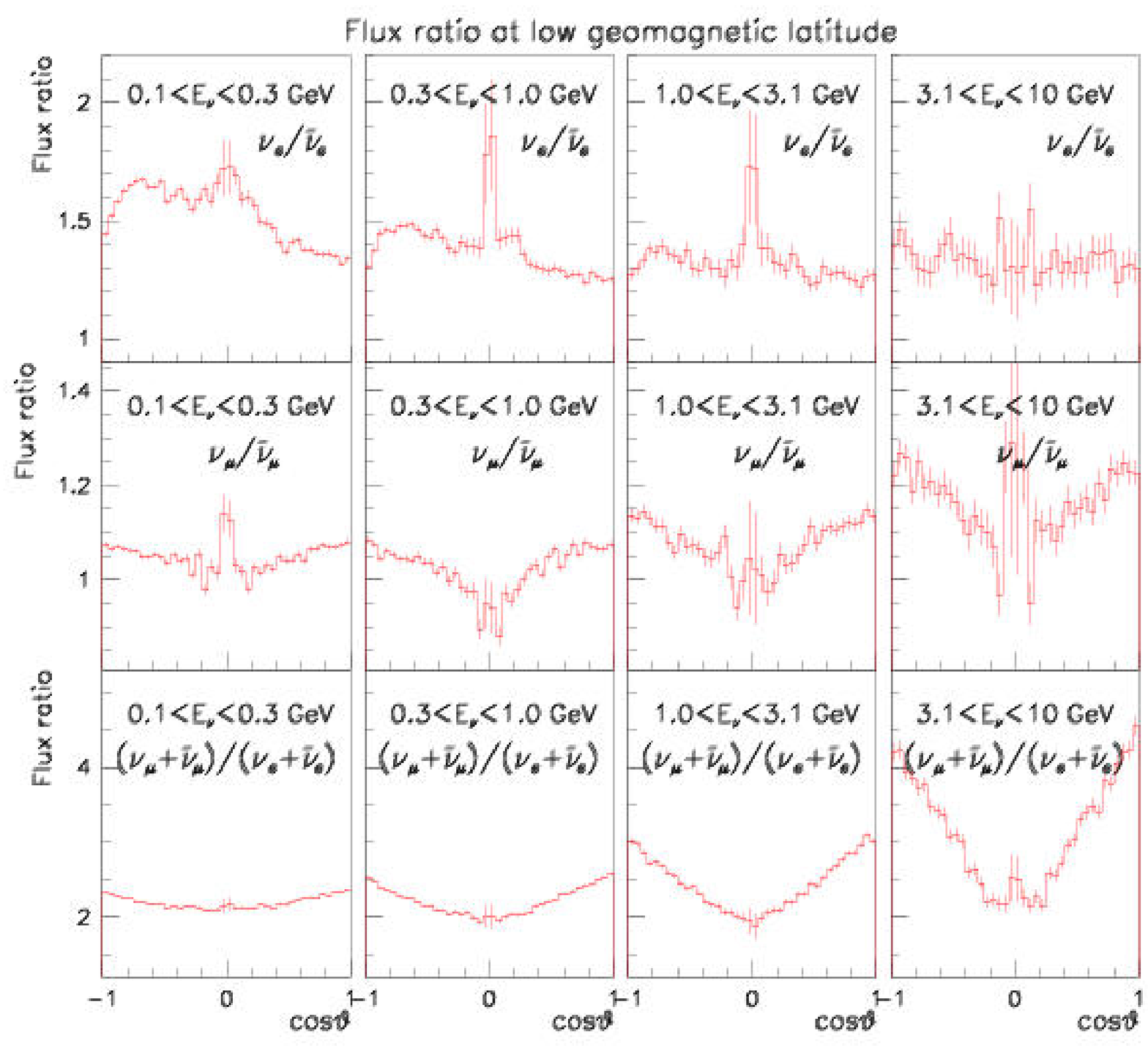}
\end{tabular}
\caption{Zenith angle distribution of the neutrino flux (top) and of the flavor ratios 
(bottom), in the indicated energy bins and for the indicated flavor,
for low geomagnetic latitudes ($\theta_{lat}<0.5$~rad), and averaged
over $4\pi$ solid angle. Here, the zenith angle is defined such that
$\cos\theta = 1$ corresponds to the downward
direction.\label{zenithlow}}
\end{figure}

\begin{figure}[!htbp]
\begin{tabular}{c}
\includegraphics[width=9.5cm]{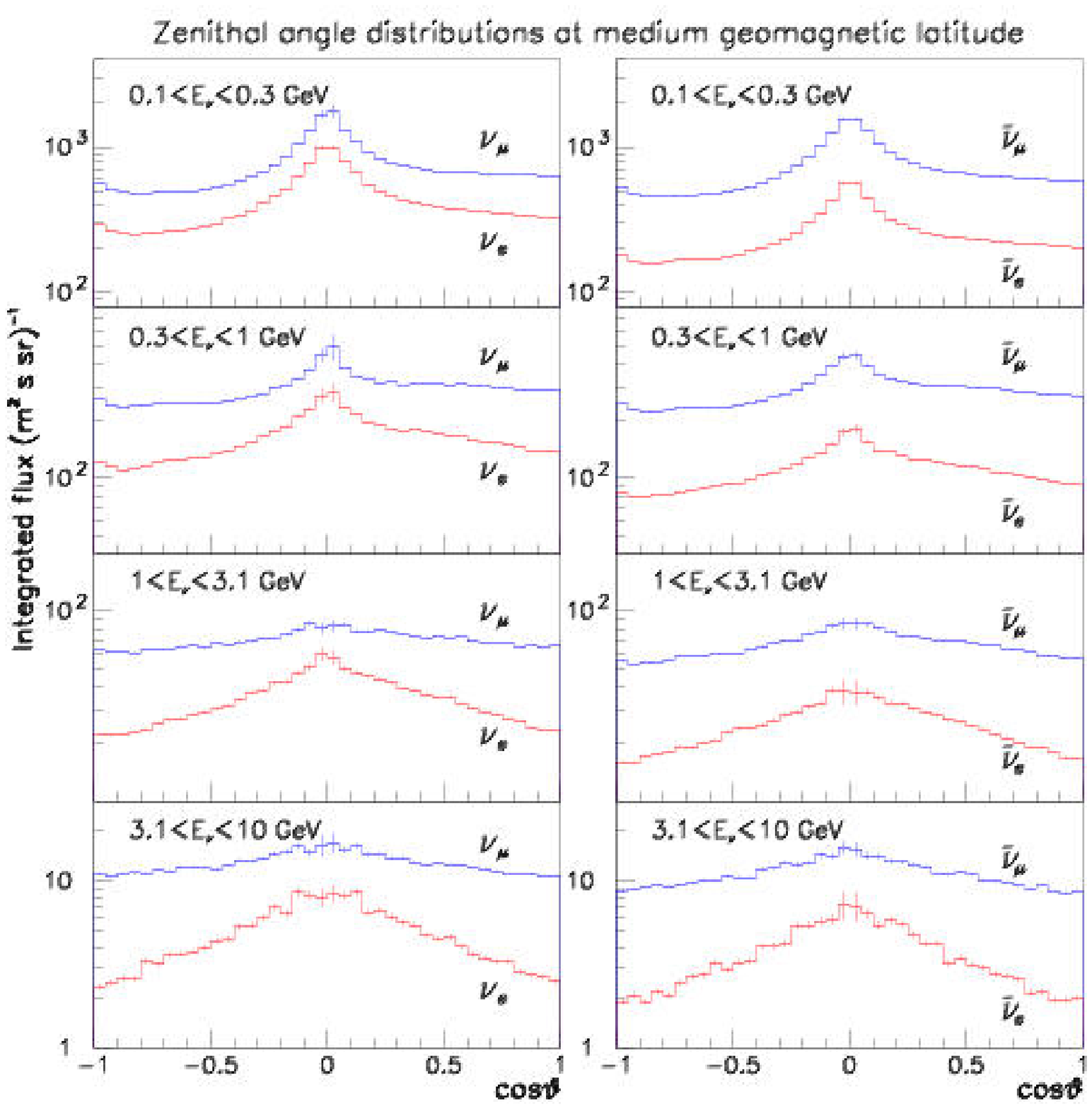}\\
\includegraphics[width=9.5cm]{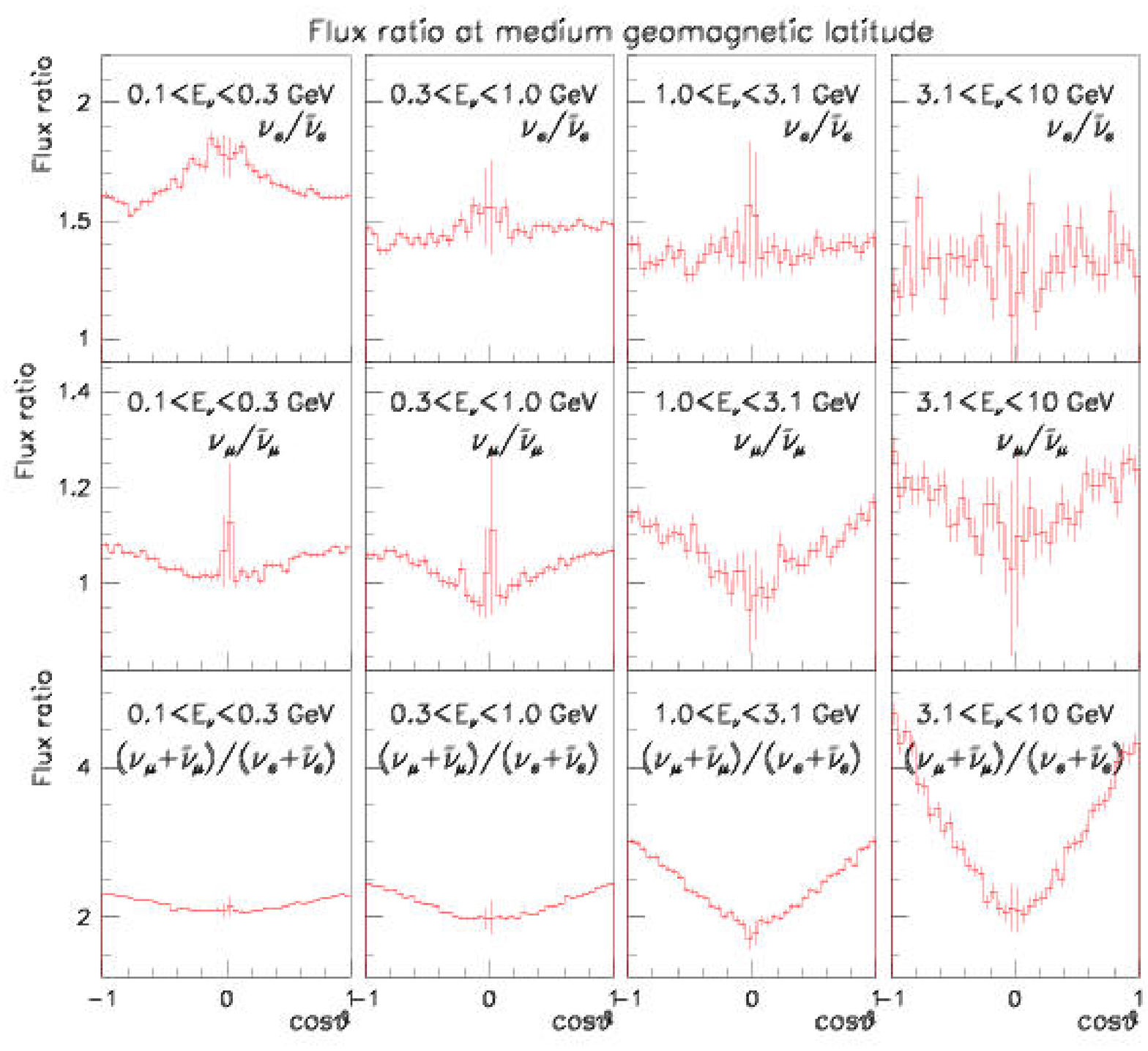}
\end{tabular}
\caption{Same as figure~\ref{zenithlow} for the geomagnetic latitudes \newline
$0.5<\theta_{lat}<1$.\label{zenithmid}}
\end{figure}

\begin{figure}[!htbp]
\begin{tabular}{c}
\includegraphics[width=9.5cm]{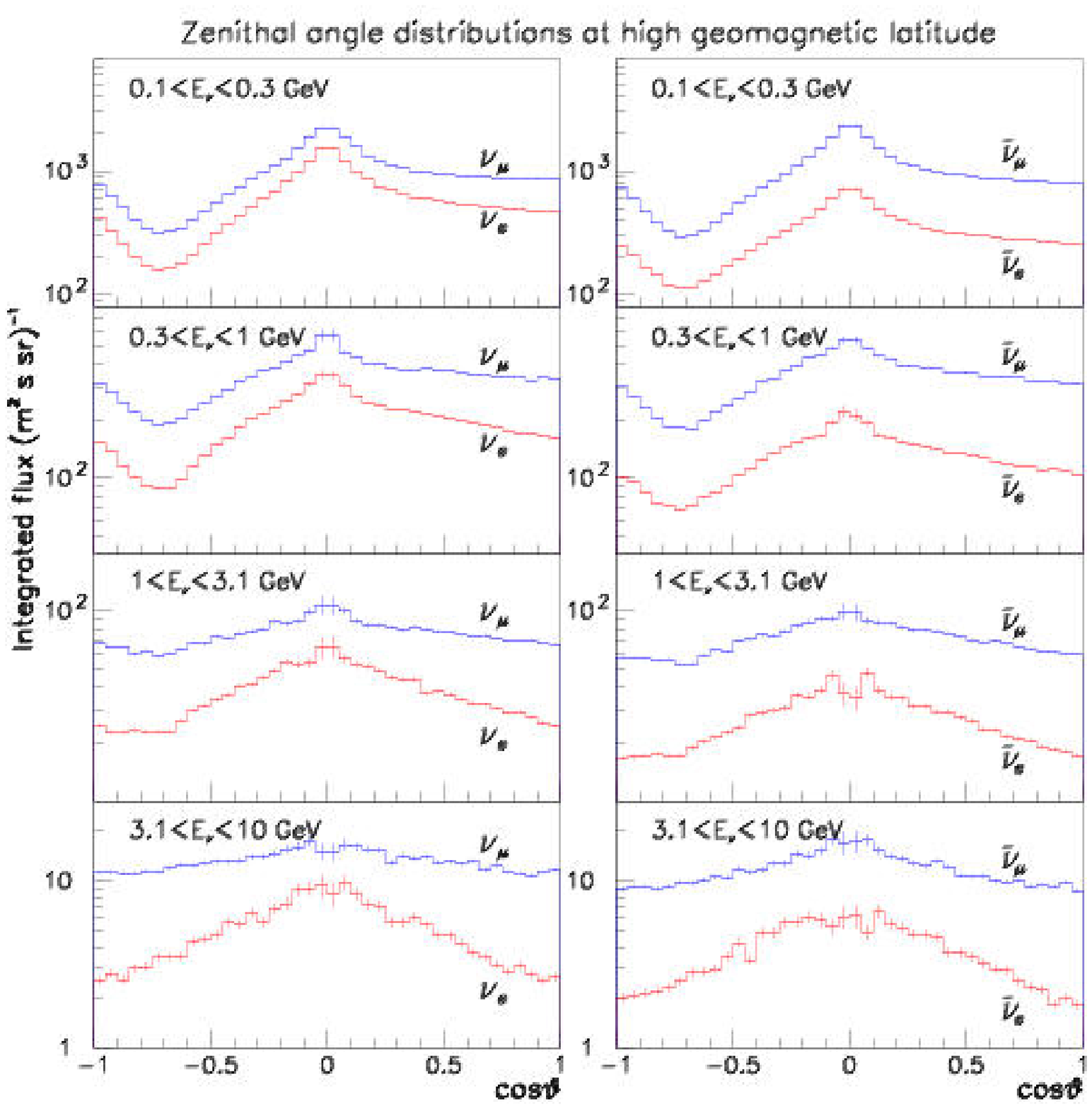}\\
\includegraphics[width=9.5cm]{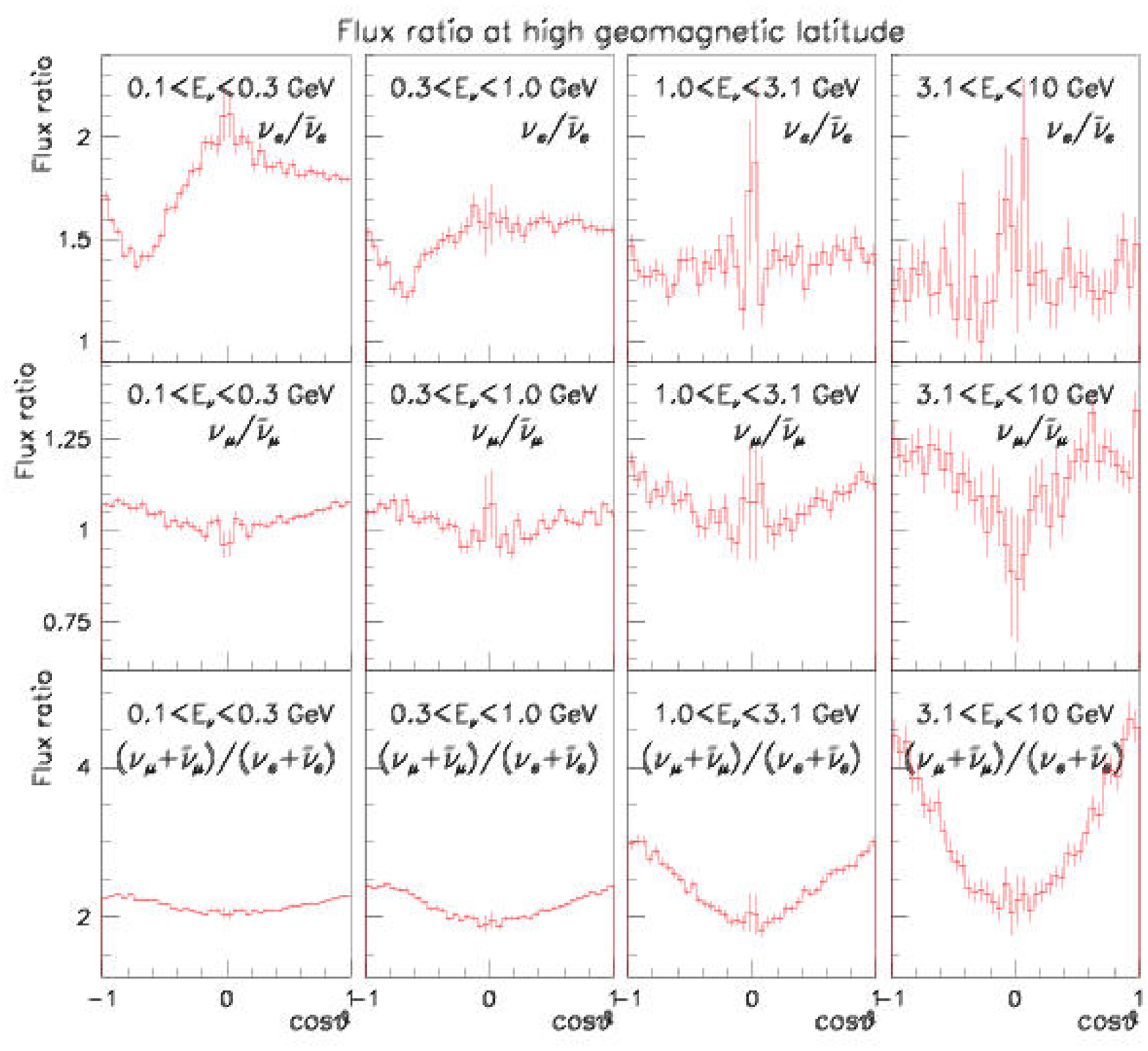}
\end{tabular}
\caption{Same as figure~\ref{zenithlow} for the geomagnetic latitudes \newline
$1<\theta_{lat}<\pi/2$.\label{zenithhig}}
\end{figure}

The zenith angle distributions are shown on figures~\ref{zenithlow},
\ref{zenithmid} and \ref{zenithhig} for the geomagnetic latitude bins
(rad) 0-0.5, 0.5-1 and 1-$\frac{\pi}{2}$, respectively, and for
several energy bins between 0.1 and 10~GeV.

The figures show that the zenith angle distributions for downward
going neutrinos ($\cos\theta>0$) have about the same shape for all the
energy bins, while for upward going neutrinos, they display quite
different shapes for the low energy bins, showing: a) A clear maximum
for low latitudes, b) A flat shape at middle latitude, and c) A deep
minimum at high latitudes. This can be understood by considering the
following two particular cases: 1) For a detection point at the north
pole, the ($\pi/4$) zenith angle $\cos \theta \sim -1/\sqrt{2}$ points
to particles coming from the equatorial region where the geomagnetic
cutoff for CR particles is highest. The neutrino flux is then expected
lower for this angle. This effect is more sensitive for the low energy
region of the neutrino spectrum. 2) Conversely, for a detection point
at the equator, the zenith angle $\cos \theta \sim -\sqrt{2}/2$ points
to particles coming from the poles (although for a narrow azimuthal
angle region only) where the GC is minimum and hence the observed
neutrino flux is expected larger around this angle.

The zenith angle dependence of the flavor ratios is modulated clearly
by the $\pi^+/\pi^-$ production ratio, GC and the muon decay
kinematics. Interestingly, in low energy bins, the $\nu_e/{\bar
\nu_e}$ dependence on the zenith angle displays a similar behavior as
observed for the flux. In addition to its dependence on the GC, this
feature is also related to the pion multiplicity. Because the
production ratio for $\pi^+$ to $\pi^-$ is larger below 10~GeV
\cite{ldl11}, the incident CR flux within this energy range which
contributes a large fraction of low energy neutrinos, is very
sensitive to the GC, and hence the $\nu_e/{\bar{\nu}_e}$ changes
accordingly with the geomagnetic latitude and zenith angle.

The $\nu_\mu/{\bar{\nu}_\mu}$ distribution is flat in low energy bins
but behaves similarly as $(\nu_\mu+{\bar{\nu}_\mu})/(\nu_e+{\bar
\nu_e})$ at high energies. For both the downward and upward flux, it
is expected that the $\nu_\mu/{\bar\nu_\mu}$ ratio approaches
asymptotically the $\pi^+/\pi^-$ production cross section ratio at the
limit of infinite muon lifetime.

The $(\nu_\mu+{\bar{\nu}_\mu})/(\nu_e+{\bar{\nu}_e})$ ratio displays a
clear dependence on the zenith angle, with a symmetric up-down
distribution, increasing from a minimum around the horizontal
direction to a maximum for vertical incidences, the larger the energy
bin, the more dramatic being the variation. This is because muons
coming in horizontal directions have long path lengths in the
atmosphere and then enough time to decay, producing a $\nu_\mu-{\bar
\nu_\mu}$ pair and a $\nu_e ({\bar{\nu}_e})$ for each pion decay. The
situation is different for muons moving downward or upward, since the
flight path available is much shorter. In this case, a fraction of
high energy muons do not decay before reaching the earth and then do
not contribute to the $\nu_e({\bar{\nu}_e})$ flux, wherefrom a larger
$(\nu_\mu+{\bar{\nu}_\mu})/(\nu_e+{\bar{\nu}_e})$ ratio. The larger the
muon energy, the larger the effect.

\subsection{Azimuth angle distributions and geomagnetic latitude dependence}\label{GLAZ}
   
\begin{figure}[!htbp]
\begin{tabular}{c}
\includegraphics[width=9.5cm]{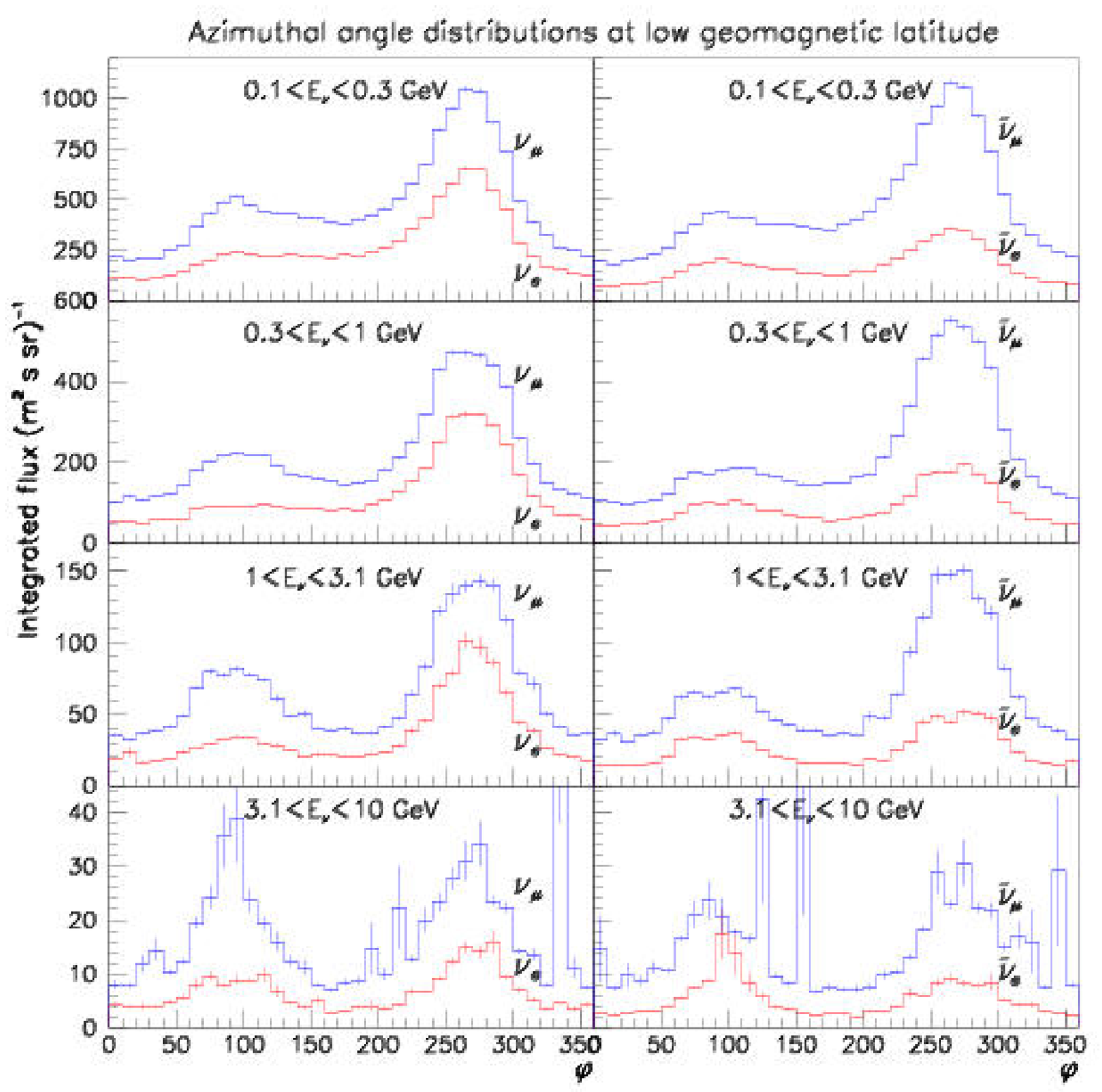}\\
\includegraphics[width=9.5cm]{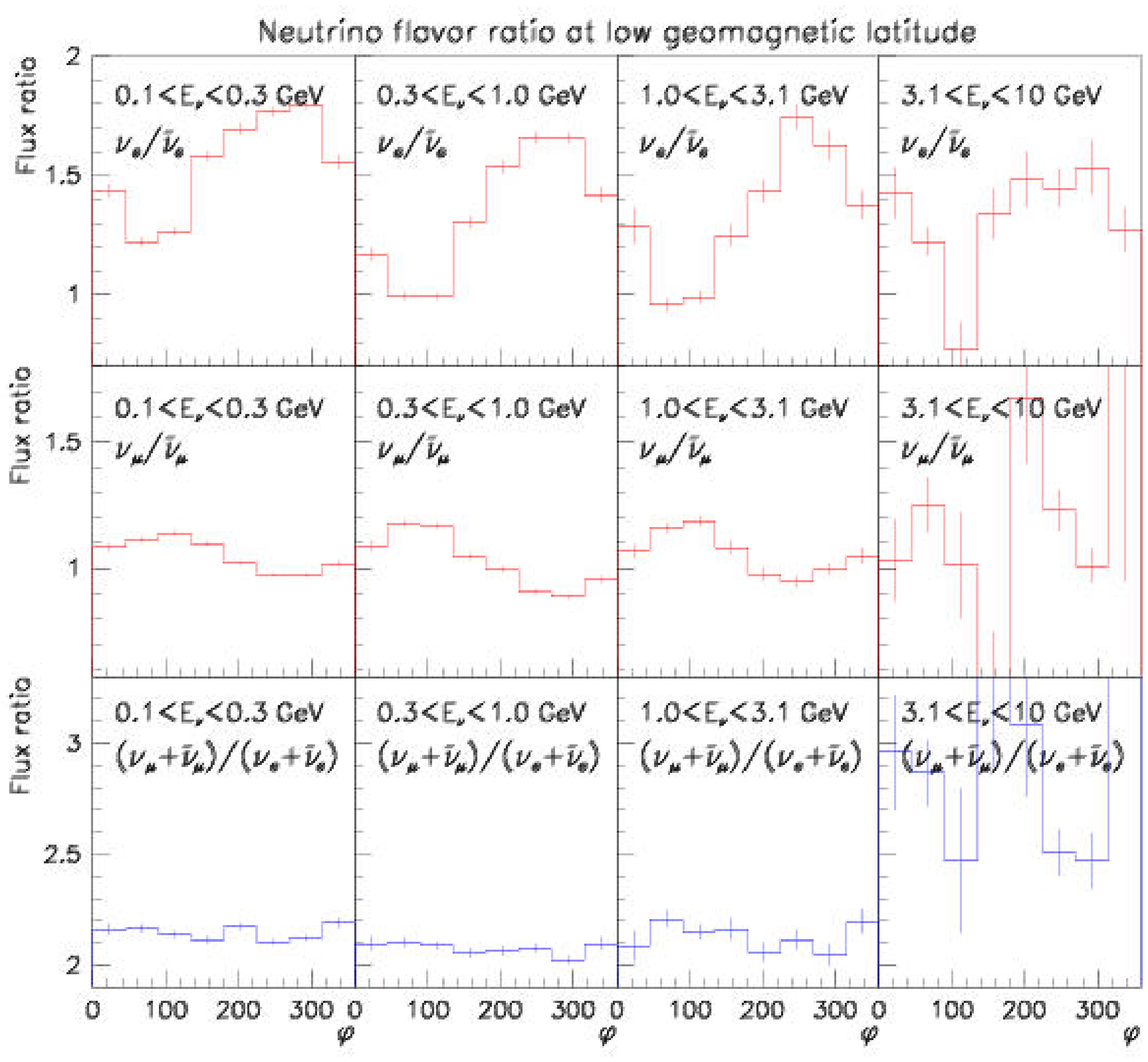}
\end{tabular}
\caption{Azimuth angle distribution of the neutrino flux (top) and of the flavor ratios 
(bottom), in the indicated energy bins and for the indicated flavor,
for low geomagnetic latitudes ($\theta_{lat}<0.5$~rad), and averaged
over $4\pi$ solid angle.  The angle convention is $\phi = 0^o, 90^o$
for geomagnetic south and east, respectively.
\label{azimuthlow}}
\end{figure}

\begin{figure}[!hbp]
\begin{tabular}{c}
\includegraphics[width=9.5cm]{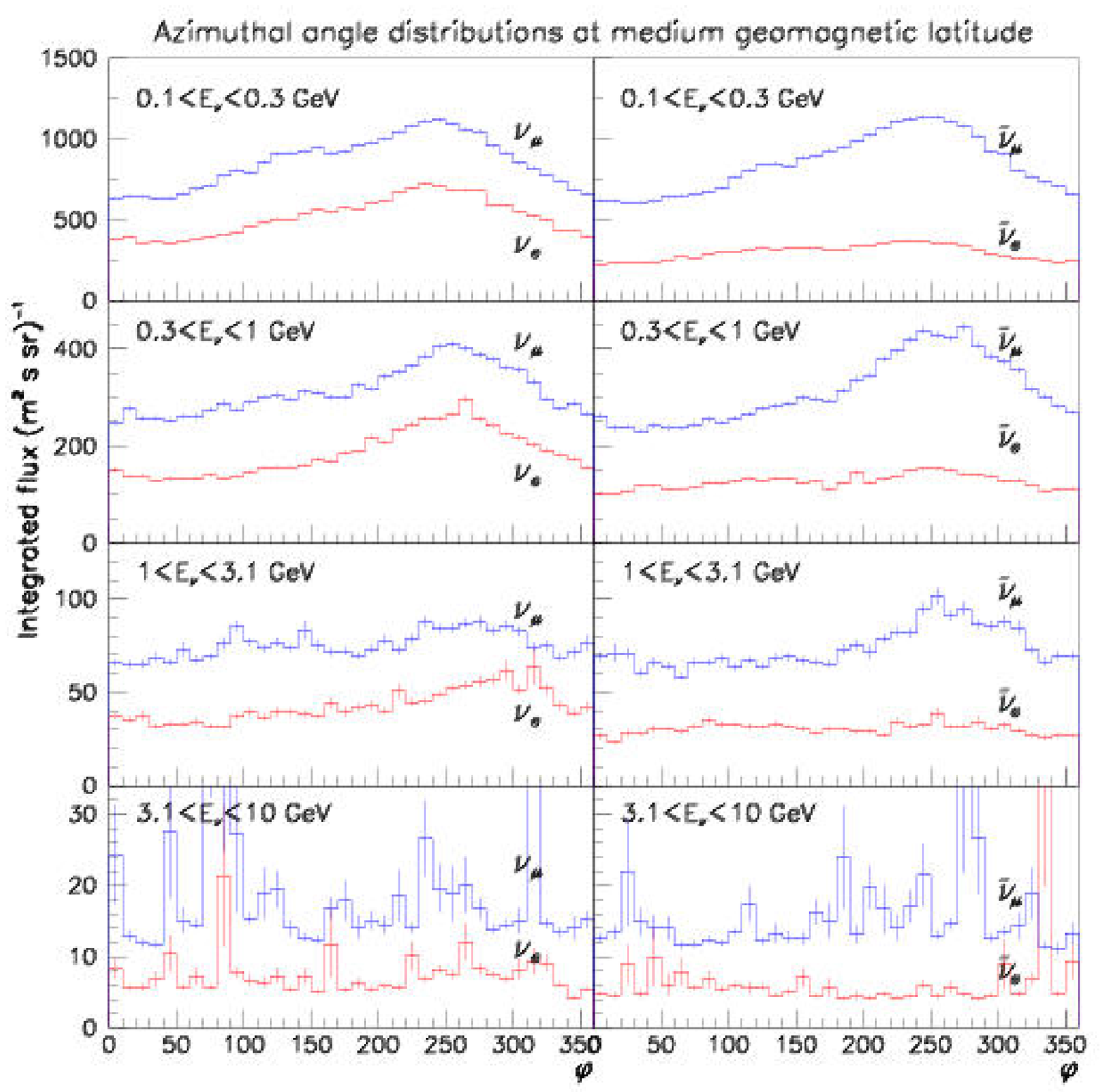}\\
\includegraphics[width=9.5cm]{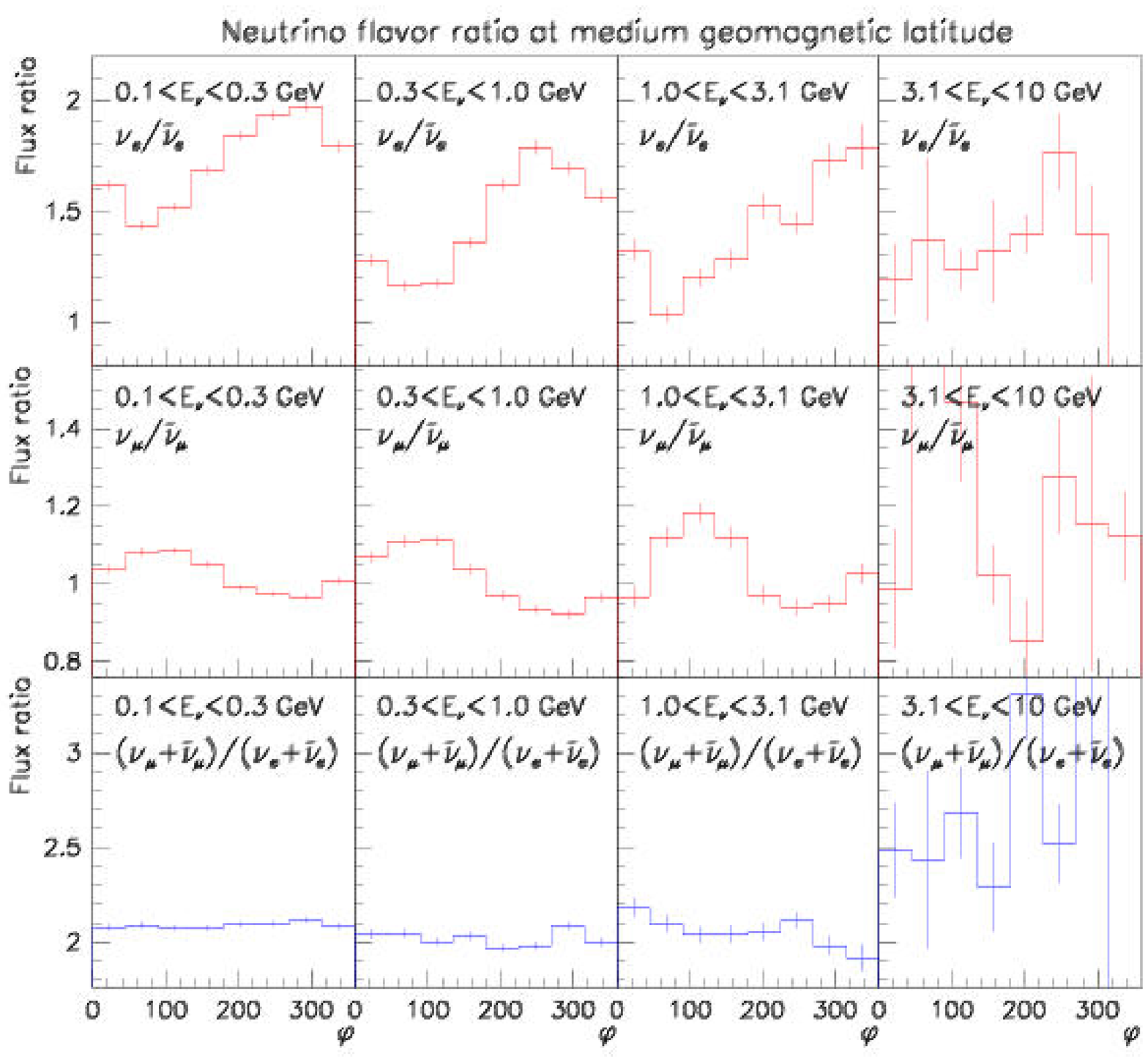}
\end{tabular}
\caption{Same as figure~\ref{azimuthlow} for intermediate geomagnetic latitudes 
($0.5<\theta_{lat}<1$).\label{azimuthmid}}
\end{figure}

\begin{figure}[!htbp]
\begin{tabular}{c}
\includegraphics[width=9.5cm]{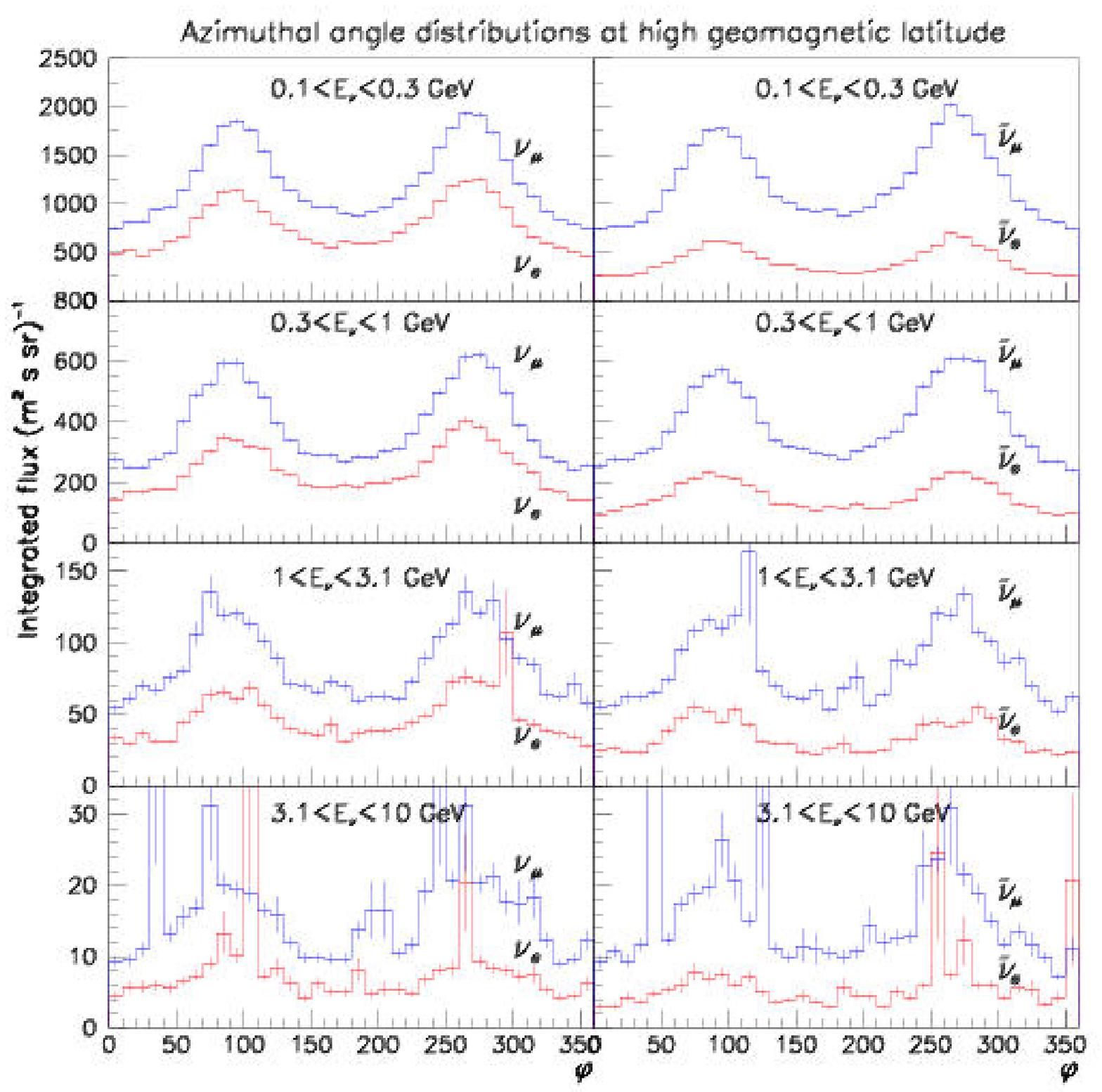}\\
\includegraphics[width=9.5cm]{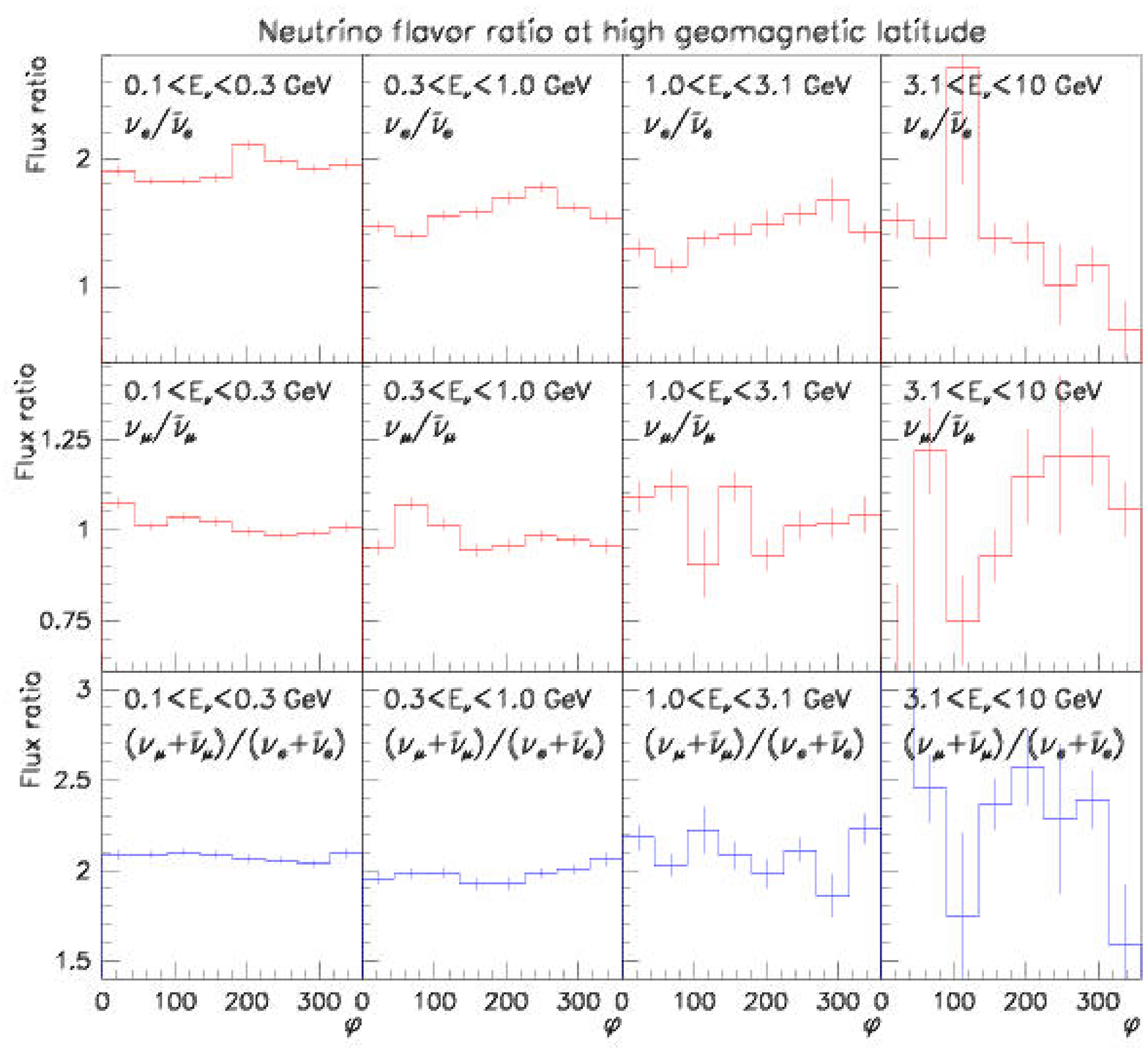}
\end{tabular}
\caption{Same as figure~\ref{azimuthlow} for high geomagnetic latitudes 
($1.<\theta_{lat}<\pi/2$).
\label{azimuthhig}}
\end{figure}

The azimuth angle distributions of the neutrino flux and of the flavor
ratios are shown on figures~\ref{azimuthlow} to \ref{azimuthhig} in
bins of neutrino energy (0.1-0.3, 0.3-1, 1-3.1, 3.1-10~GeV), and for
three bins of geomagnetic latitude (in absolute value): 0-0.5, 0.5-1,
and 1-$\pi/2$ rad, respectively.

It can be observed on these figures that the azimuth angle
distributions display a significant dependence on the latitude
considered. For low geomagnetic latitudes (figure~\ref{azimuthlow}),
the distributions are dominated by two peaks located symmetrically at
the East (E) and West (W) azimuth angles, but with different
heights. These latter features are clearly related to the dipole
nature of the earth magnetic field for the E and W peaking, and to the
GC for the EW asymmetry. It is also seen that the EW asymmetry
decreases as expected with the increasing energy.

For intermediate geomagnetic latitudes (figure~\ref{azimuthmid}), the
West peak appears dampen and wider compared to low latitudes, while
the East peak is even fainter and disappears in the high energy
bins. The two-peaks pattern reappears with nearly equal heights at
high latitude (figure~\ref{azimuthhig}). This latter symmetry was
expected since the GC disappears in the polar region.

For the azimuth angle distributions of the flavor ratios, it can be
noted that the $\nu_e/{\bar{\nu}_e}$ ratio is strongly EW asymmetric
with a broad enhancement on the West side at low and intermediate
latitudes. The opposite trend is observed for the $\nu_\mu/{\bar
\nu_\mu}$ ratio, while the
$(\nu_\mu+\bar{\nu_\mu})/(\nu_e+\bar{\nu_e})$ ratio is found to be
almost structureless at all latitudes.

The reason for these features is similar to that of zenithal angle
dependence. The latitude dependence of the zenith and azimuth angle
distributions reflects the characteristic of the geomagnetic field and
reminds us that an appropriate treatment of the geomagnetic field is
required to account for the upward going muon event data and for the
east-west asymmetry. Besides, detailed simulation for detectors at
different locations is needed to provide a sensitive investigation of
the geomagnetic dependence of the measurements.

\subsection{Cosmic Ray parent proton energy distributions}\label{CRPD}

\begin{figure}[!htbp]
\includegraphics[width=9.5cm]{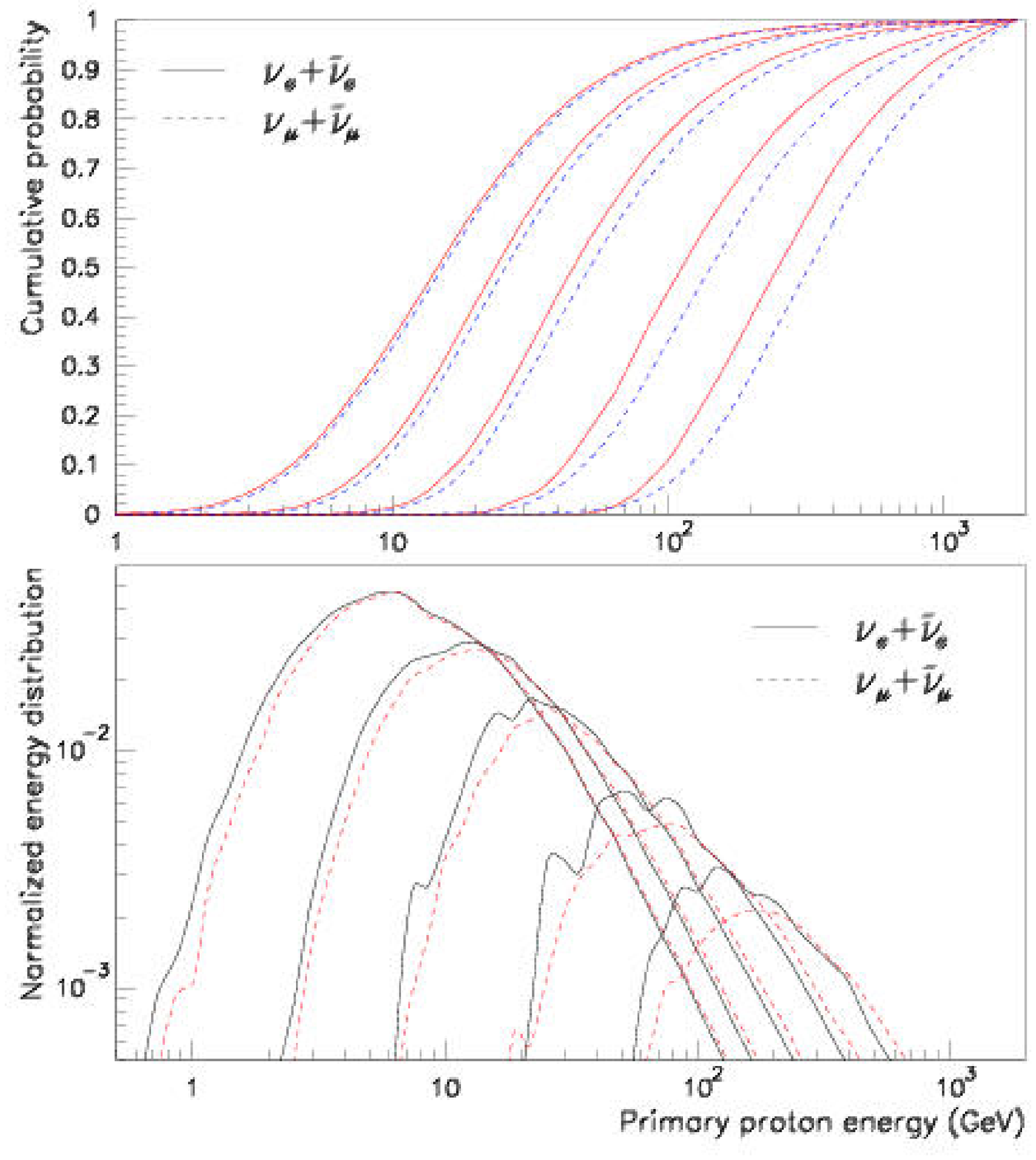}
\caption{The cumulative (top) and normalized (bottom) primary proton energy 
distribution for $\nu_e + \bar{\nu_e}$ (solid lines) and $\nu_\mu +
\bar{\nu_\mu}$ (dashed lines) production in 5 energy bins 0.28-0.32,
0.8-1.2, 2.8-3.2, 8.0-12.0 and 20.0-30.0~GeV from the left to the
right respectively.
\label{proton}}
\end{figure}

The incident CR proton energy distribution inducing the neutrino flux are shown on 
figure~\ref{proton} for neutrino energies around 0.3, 1, 3, and 10~GeV. The 
distributions are given for $\nu_e + \bar{\nu_e}$ and $\nu_\mu + \bar{\nu_\mu}$. They 
are averaged over the whole detection sphere and 4$\pi$ solid angle, and presented as 
cumulative probabilities for neutrinos within the energy bins to be produced by protons 
with an energy below the value given in abscissa, and energy distributions normalized to 1.

At variance with the results of \cite{honda1}, the mean energy of the
primary proton flux producing 0.3~GeV $(\nu_e + {\bar{\nu}_e})$
obtained in these calculations, appears to be hardly smaller than the
for $(\nu_\mu + {\bar{\nu}_\mu})$ in the same energy bin. The
difference becomes larger with the increasing energy however. The mean
values of the proton distributions of the figure are given in
table~\ref{pn}.

\begingroup
\squeezetable
\begin{table}[!htbp]
\begin{ruledtabular}
\begin{tabular}{c|c|c|c|c|c|c|c}
$E_\nu (GeV)$  & 0.1-0.2 & 0.28-0.32 & 0.8-1.2   & 2.8-3.2 & 8.0-12 & 20-30  & 0.1-30 \\ \hline
$<\! E_p\!>_{\nu_e+\bar{\nu}_e}$
               & 30.1    & 37.1      & 57        & 102     & 220    & 405  & 39  \\ 
$<\! E_p\!>_{\nu_\mu+\bar{\nu}_\mu}$
               & 31      & 39        & 64        & 120     & 271    & 498  &43 
\end{tabular}
\end{ruledtabular}

\caption{Mean incident kinetic energy in GeV of the CR protons producing neutrinos in the energy bins 
$E_\nu$.\label{pn}}
\end{table}

\begin{table}[!htbp]
\begin{ruledtabular}\begin{tabular}{c|c|c|c|c|c|c|c}
$ E_p $ (GeV) & 0.2-10 & 10-30 & 30-50 &50-200 & 200-500 & 500-2000 & 0.20-2000\\
\hline
$<\!  E_{\nu_e+\bar{\nu}_e}\!>$ 
                        &  0.25  & 0.40  & 0.52  &  0.67 & 0.93    & 1.14  & 0.42  \\  

$<\! E_{\nu_\mu+\bar{\nu}_\mu} \!>$
                        &  0.25  & 0.40  & 0.56  & 0.81  & 1.36    & 2.19   & 0.48
\end{tabular}\end{ruledtabular}
\caption{Mean neutrino energies $<\!E_{\nu}\!>$ produced by CR protons within kinetic energy 
bins $E_p$. Neutrinos with energy below 0.05~GeV are not included in the average.
\label{tab20}}

\end{table}
\endgroup
Table~\ref{tab20} gives for comparison, the mean energy of the $(\nu_e + {\bar{\nu}_e})$ and 
$(\nu_\mu + {\bar{\nu}_\mu})$ neutrino spectrum produced by protons within the indicated 
energy bins. The mean neutrino energy value is found to be smaller for
$(\nu_e + {\bar{\nu}_e})$ than for $(\nu_\mu + {\bar{\nu}_\mu})$ for the
higher energy bins of the incident protons. This is consistent with
the fact that in the decay chain $\pi^\pm \rightarrow \mu^\pm +
\nu_\mu ({\bar{\nu}_\mu})$, $\mu^\pm \rightarrow e^\pm + \nu_e ({\bar
\nu_e})+{\bar
\nu_\mu}(\nu_\mu)$, the $\nu_e ({\bar{\nu}_e})$ particle shares the
energy of decaying $\mu^\pm$ which momentum is equal to that of the
first $\nu_\mu ({\bar{\nu}_\mu})$ in $\pi^\pm$ center of mass frame
with $e^\pm$ and another ${\bar{\nu}_\mu} (\nu_\mu)$. The
energy-momentum conservation implies that the $\nu_e ({\bar{\nu}_e})$
energy is on the average smaller than the mean energy of the two
${\bar{\nu}_\mu} (\nu_\mu)$.

On figure~\ref{proton}, it is seen that low energy primary protons
contributes more to $(\nu_e + {\bar{\nu}_e})$ than to the same energy
$(\nu_\mu + {\bar{\nu}_\mu})$ production.  This is easy to understand
if we note that higher energy muons which are produced by higher
energy primaries on the average, do not contribute to electron
neutrino, but accompany with them, a muon neutrino has been created in
pion decay. Even at low energy, this process can happen near the
detection surface, and so the mean primary proton energy for muon
neutrino production is raised up.

\subsection{Neutrino production altitudes}\label{NPALT}

\begin{figure*}[!htbp]
\begin{tabular}{cc}
\includegraphics[width=9.0cm,height=12.6cm]{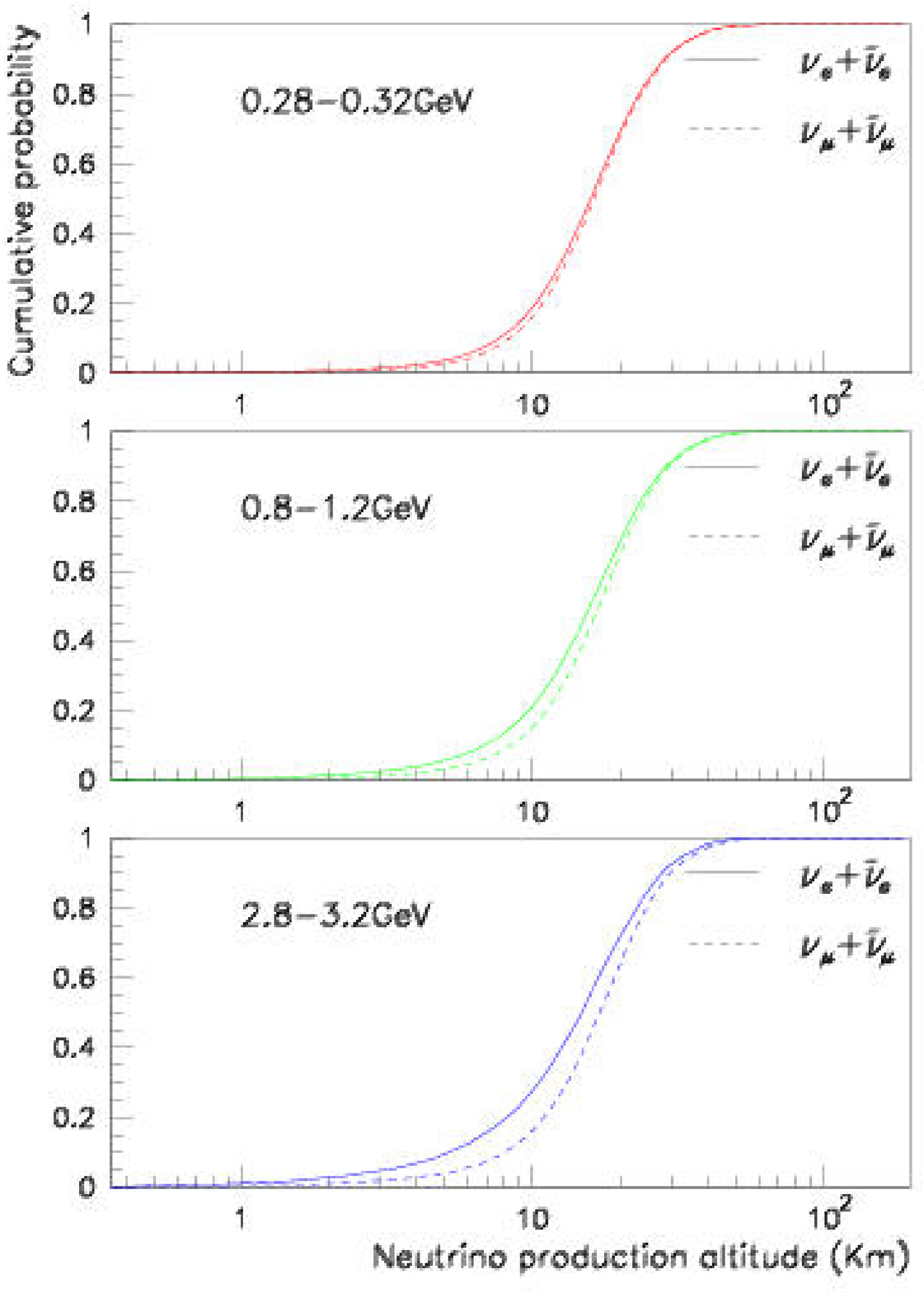}&\hspace{-1cm}
\includegraphics[width=9.0cm,height=12.6cm]{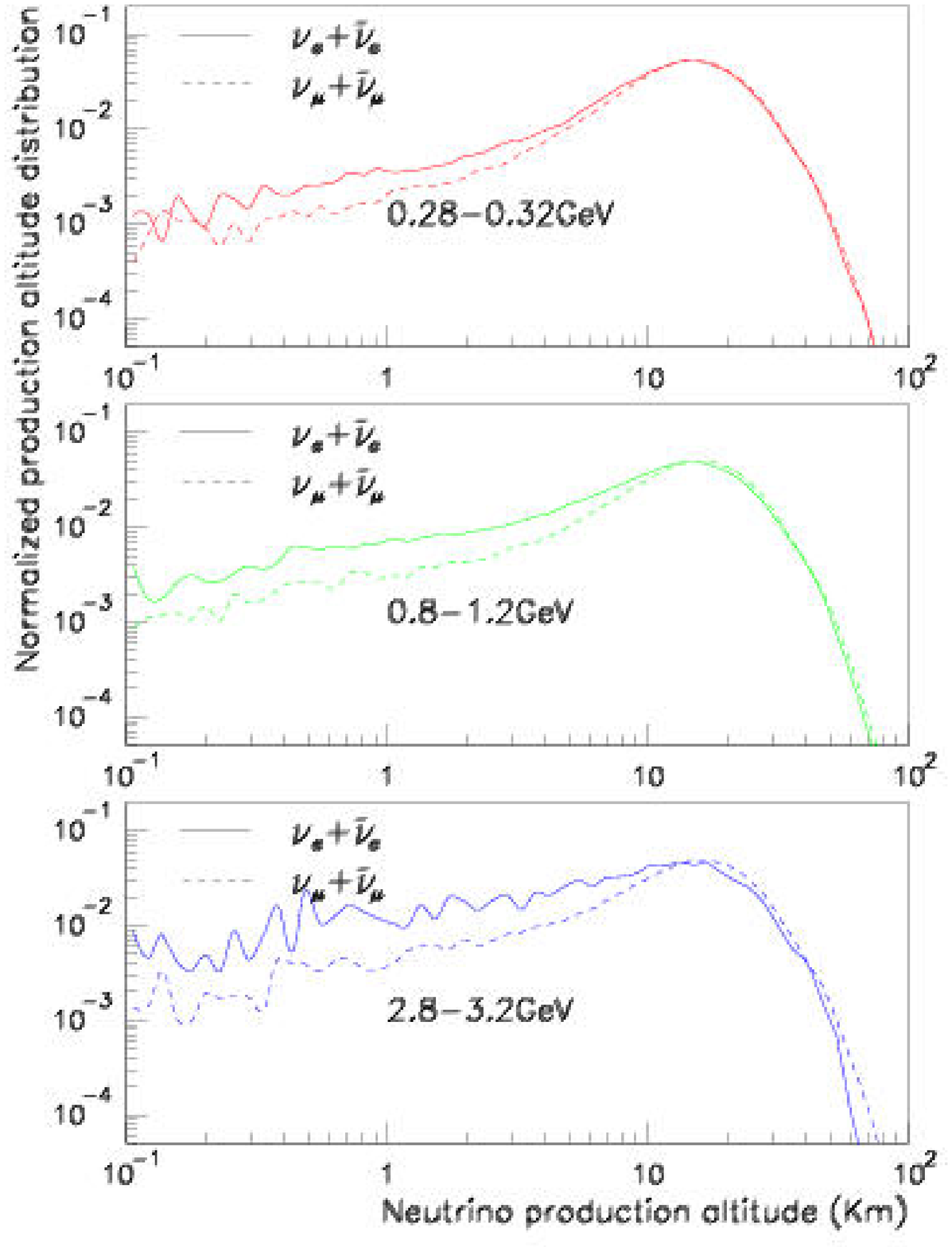}
\end{tabular}
\caption{ Cumulative (left) and normalized (right) $\nu_e + \bar{\nu_e}$, and
$\nu_\mu + \bar{\nu_\mu}$ production altitude distributions, for
different neutrino energy bins, averaged over the whole detection
sphere and 4$\pi$ solid angle.
\label{altitude}}
\end{figure*}

The distribution of the neutrino production altitude is plotted in figure~\ref{altitude}, 
where about $90 \%$ neutrinos appear to be produced between 5 and 40~km altitude, with the 
most probable production altitude being within the 15-20~km range. These numbers are consistent 
with the previous 1-dimensional analysis \cite{tkts}. 
Below a few kilometers, the production distribution tends to flatten, governed by the 
absorption rate, while beyond 40 km, it drops following the atmospheric density profile. 

\begin{table}[htb]
\begin{ruledtabular}
\begin{tabular}{c|c|c|c|c|c|c}

$E_\nu (GeV)$ & 0.1-0.2 & 0.28-0.32 & 0.8-1.2 & 2.8-3.2 & 8.0-12 & 20-30  \\  
\hline
$<\!\textrm{Alt}_{\nu_e+\bar{\nu}_e}\!>$ 
              & 18.     & 18.5      & 18.3     & 17.1   & 16.1   & 15.3 \\  
$<\!\textrm{Alt}_{\nu_\mu+\bar{\nu}_\mu}\!>$ 
              & 18.1    & 19        & 19.6     & 19.6   & 19.3   & 19.1  
\end{tabular}\end{ruledtabular}
\caption{Mean neutrino production altitudes (km) corresponding to different neutrino energy 
bins. \label{tab3}}
\end{table}

Table~\ref{tab3} lists the average production altitudes corresponding
to different neutrino energy bins. The average electron neutrino
production altitude appears to be a little below that of muon
neutrinos, because muons can fly over a certain distance - on the km
scale for the energies considered here - before decaying into
electrons.

The distribution of the altitude where the first interaction of the
incidents CRs with atmosphere takes place was found to have a mean
value of about 26~km on average, 23~km for nearly vertical ($0.9 \leq
\cos \theta \leq 1.0$ at the first collision point) and 38~km for
nearly horizontal ($0.0 \leq \cos \theta~\leq 0.1$ at the first
collision point) CRs, with no other selection cut than a produced
neutrino reaching the detection sphere eventually.

\begin{figure}[!htbp]
\includegraphics[width=8cm]{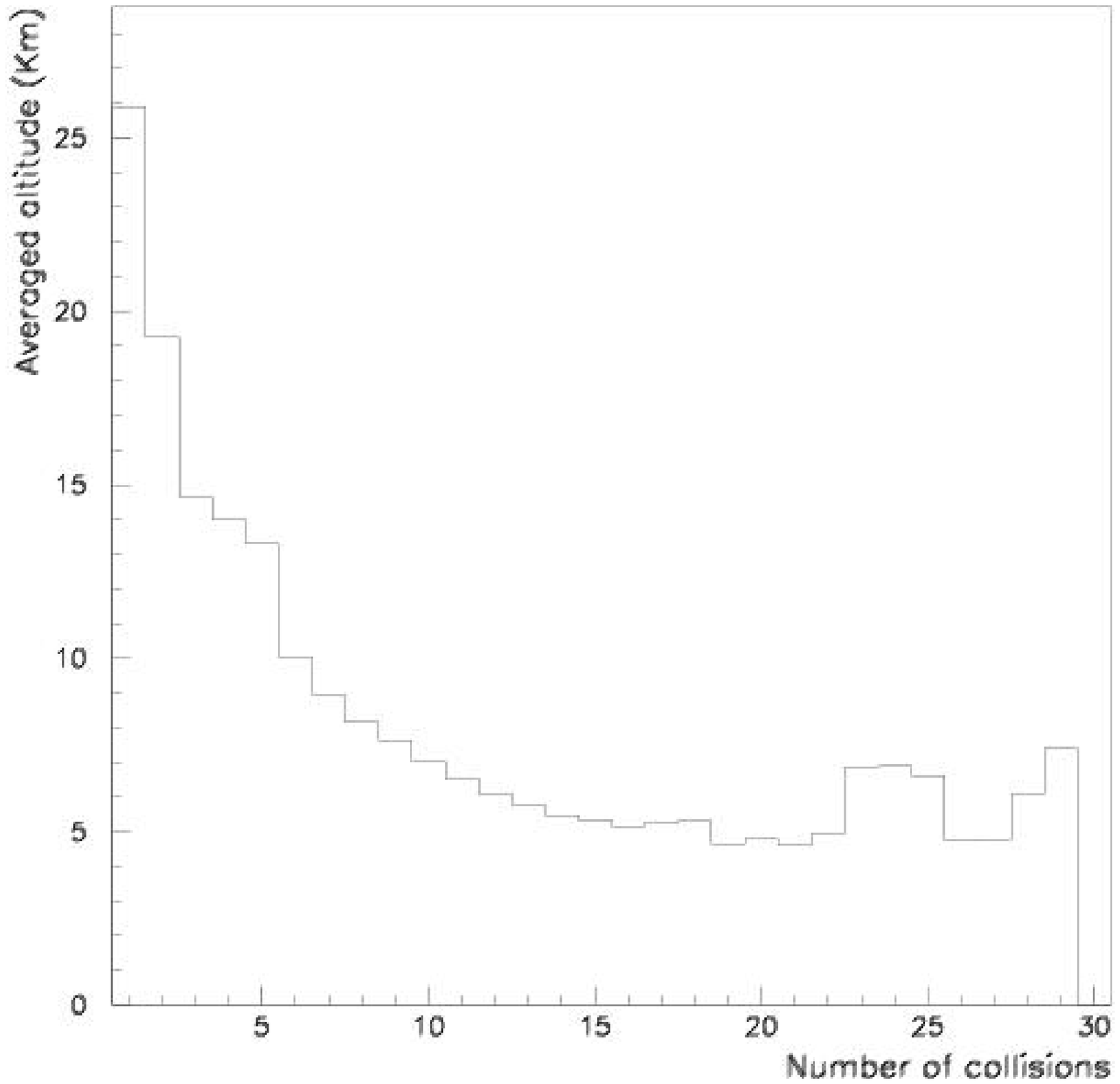}
\caption{Average collision altitude of the $N$th collision versus collision number 
 $N$. \label{highrank}}
\end{figure}

Figure~\ref{highrank} shows the distribution of the averaged altitude
at which the $N$-th collision occurs in the neutrino producing
reaction chain as a function of the number $N$ of the collision.

\begin{figure*}[!htbp]
\begin{tabular}{cc}
\includegraphics[width=9.0cm,height=12.6cm]{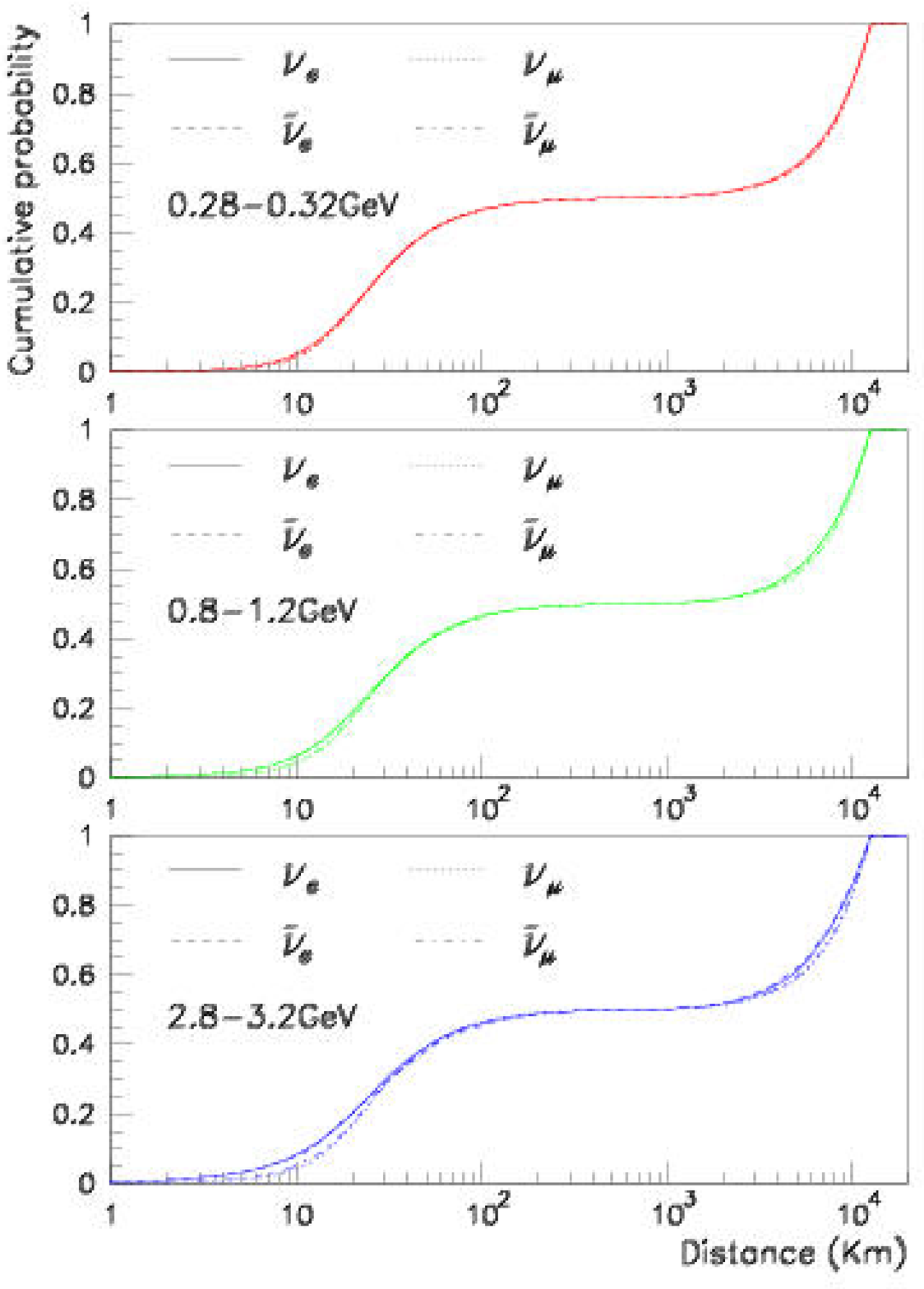}&\hspace{-1cm}
\includegraphics[width=9.0cm,height=12.6cm]{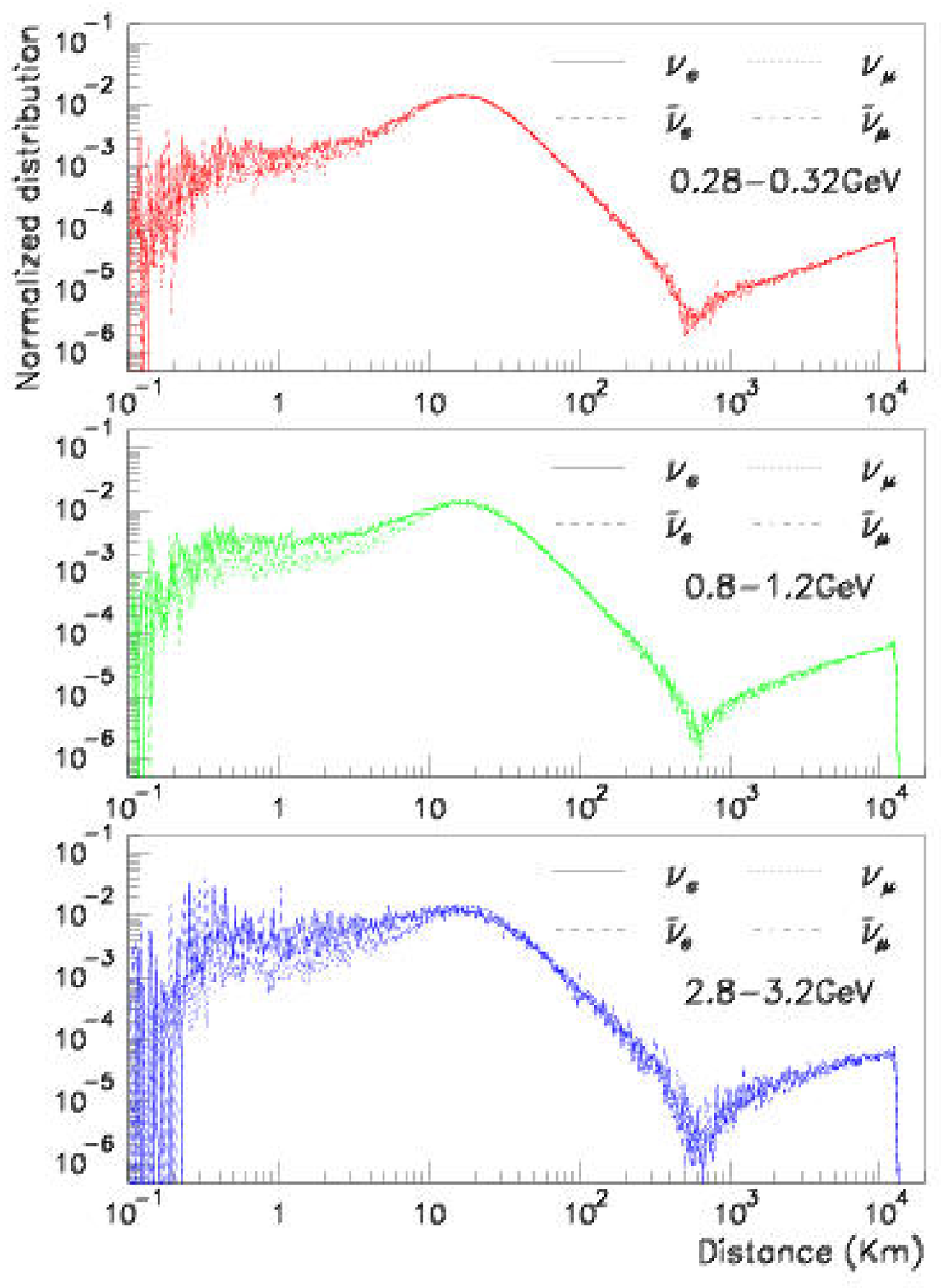}
\end{tabular}
\caption{ Cumulative (left) and normalized (right) $\nu_e$, $\bar{\nu}_e$, 
$\nu_\mu$ and $\bar{\nu}_\mu$ flying distance 
distributions, for different neutrino energy bins, 
averaged over the whole detection sphere and 4$\pi$ solid angle. 
\label{flyd}}
\end{figure*}

\begin{figure}[!htbp]
\includegraphics[width=9.5cm]{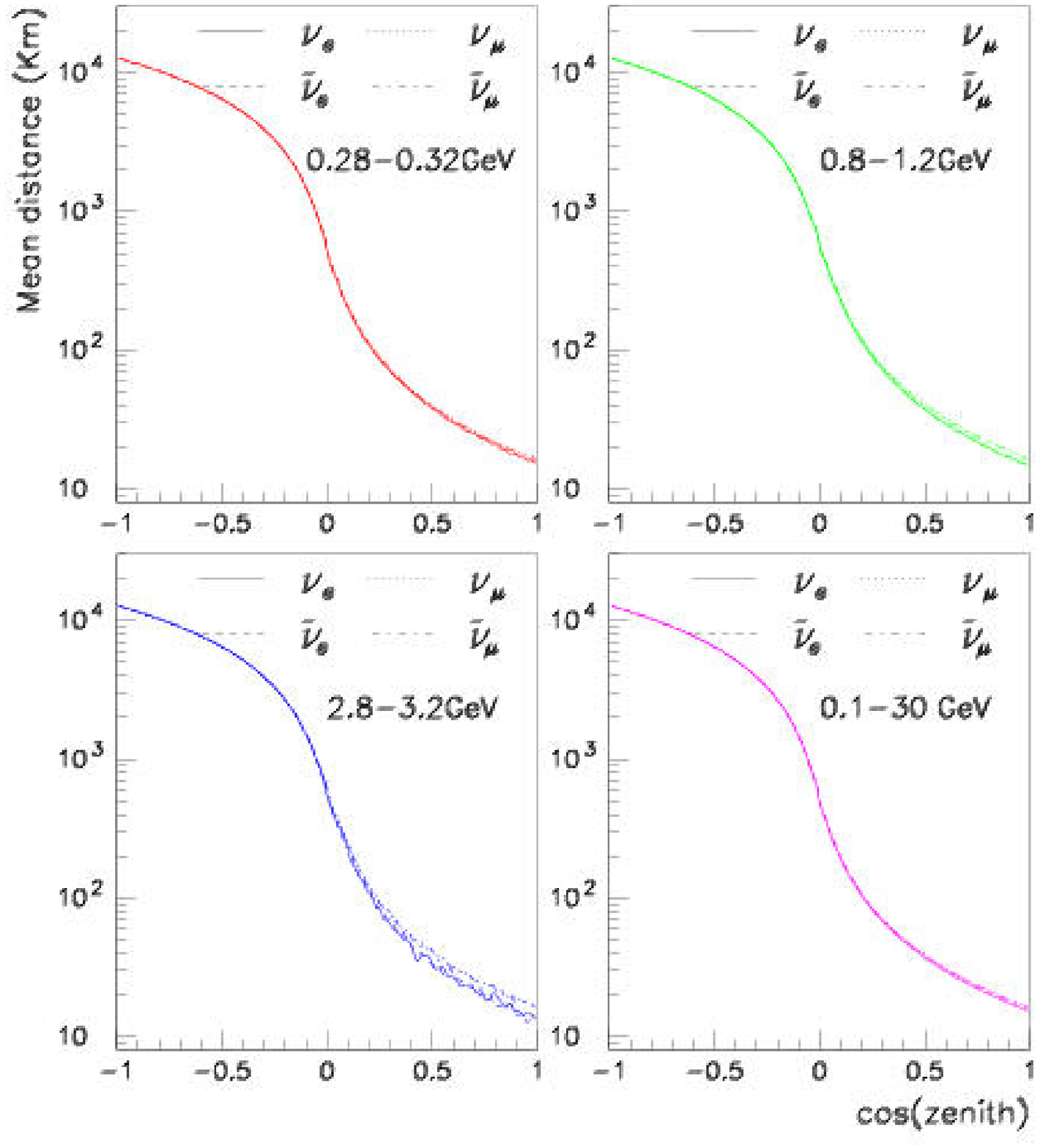}
\caption{ Mean flying distance versus the zenithal angle distribution of $\nu_e$, $\bar{\nu_e}$, 
$\nu_\mu$ and $\bar{\nu_\mu}$, in the indicated energy bins (GeV),
averaged over the whole detection sphere.\label{dcos}}
\end{figure}

It can be observed that, on the average, the second
collision takes place around 19~km, and the third collision 
around 15~km. In addition, below 4~km altitude, there is almost no meson
producing collision because the initial CR energy has been damped
through the cascade and most particles produced below this altitude
are below the pion production threshold.

\subsection{Neutrino flying distance}\label{FLYDIS}

Figure~\ref{flyd} shows the distribution of the neutrino flying
distance between production and detection point. The distributions of
the mean values as a function of the cosine of the zenith angle are
plotted on figure~\ref{dcos} for four neutrino energy bins and for the
four neutrino flavors. These distributions are very similar to those
obtained in \cite{honda6}.

The degeneracy of the curves shown in figure~\ref{dcos} for the 4
flavors can be easily understood by looking at figure~\ref{altitude}
and table~\ref{tab3}, the scales of the peak production altitude and
the difference of the production altitude for different flavors are so
small in comparison with the earth diameter that it can be negligible.

\subsection{3-Dimensional effects }\label{3DEF}

\begin{figure}[!htbp]  
\includegraphics[width=9.5cm]{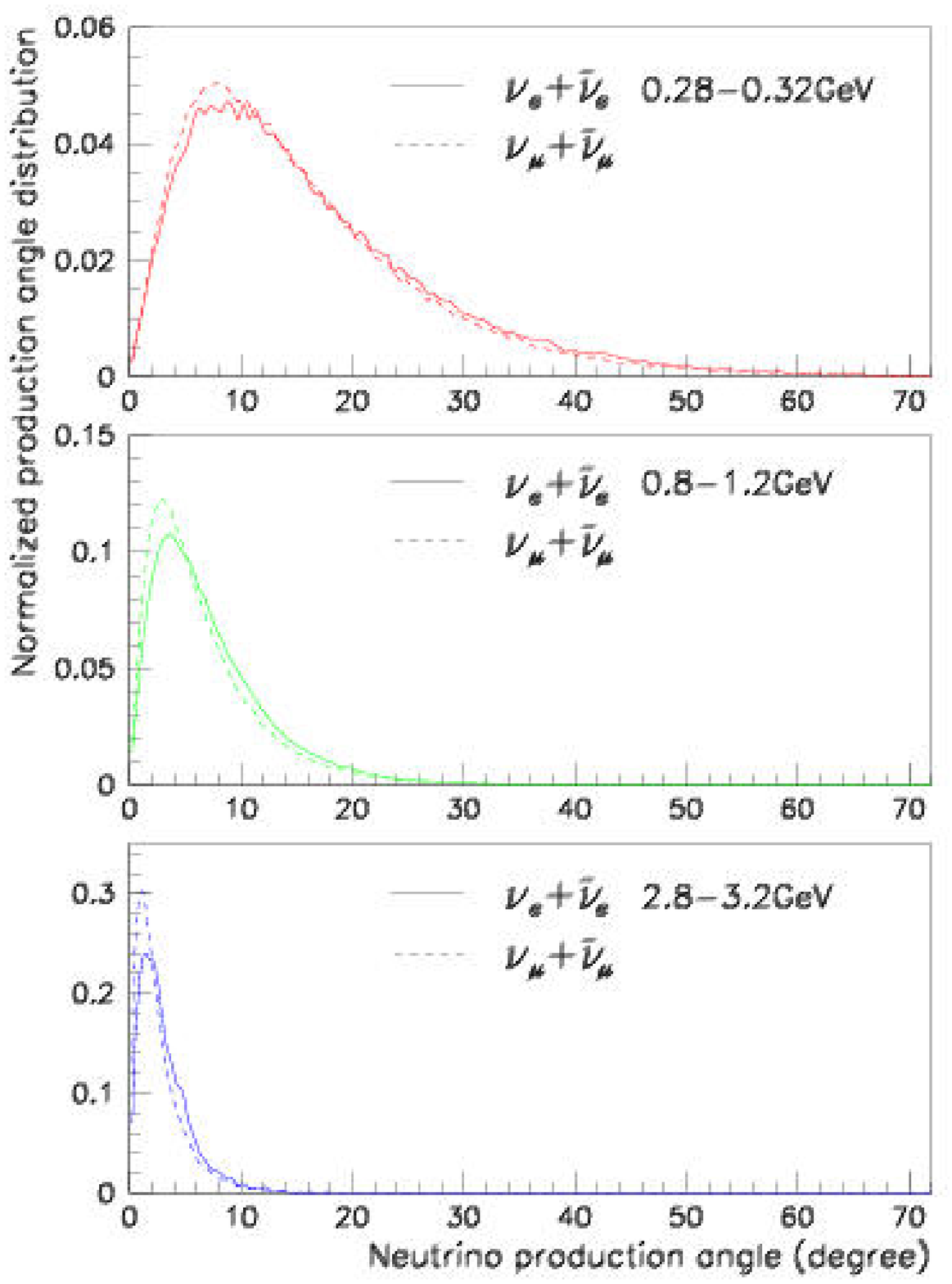}
\caption{ Normalized distributions of neutrino production angle relative to the 
direction of the incident CR proton at the first collision, for three
neutrino energy bins (GeV) : 0.28-0.32, 0.8-1.2, 2.8-3.2. \label{angle}}
\end{figure}

The normalized distributions of angles between the directions of the incident CRs at the first
collision and of the 
neutrinos in the end of the reaction chain, 
averaged over the whole detection sphere and 4 $\pi$ solid angle, are plotted on 
figure~\ref{angle} for three energy bins.  

The mean values corresponding to the different energy bins are listed in table~\ref{tab4}.
These values indicate that the 3-dimensional effects are not important for GeV neutrinos, which 
is consistent with the general expectation and the result reported in \cite{gflm}. 

The mean angles between the momenta of : 1)- the incident CR proton (at the first collision)  
and the pion-producing secondary proton (at the nth collision in the cascade), 
2)- The incident proton (called pion parent) and the produced pion in the production collision 
(production angle), and 3)- The muon parent ($\pi$ or $K$) and the produced muon, are given in 
tables~\ref{tabang1} and \ref{tabang2}, for different neutrino energy bins. From these numbers 
it appears that large deviations of the particle directions occur at the pion production level 
and only at low energy. 

\begingroup
\squeezetable
\begin{table}[!htbp]
\begin{ruledtabular}\begin{tabular}{c|c|c|c|c|c|c|c}
$E_\nu$ (GeV) & 0.1-0.2 & 0.28-0.32 & 0.8-1.2 & 2.8-3.2 & 8.0-12 & 20-30 & 0.1-30 \\  
\hline
$<\!\!\theta_{CR-(\nu_e,\bar{\nu}_e)}\!\!>$ 
              & 28.3    & 16.7      & 7.3     & 3.2     & 1.4    & 0.69  & 20.2  \\  
$<\!\!\theta_{CR-(\nu_\mu,\bar{\nu}_\mu)}\!\!>$ 
              & 27.7    & 15.7      & 6.6     & 2.7     & 1      & 0.46  & 19.1  
\end{tabular}\end{ruledtabular}
\caption{ Mean neutrino production angle (deg) relative to the CR particle direction 
at the first collision, for different neutrino energy bins.
\label{tab4}}
\end{table}

\begin{table}[!htbp]
\begin{ruledtabular}\begin{tabular}{c|c|c|c|c|c|c|c}

$E_{\nu_e({\bar{\nu}_e})}$ (GeV) & 0.1-0.2 &0.28-0.32 & 0.8-1.2 & 2.8-3.2 
&8.0-12 &20-30 & 0.1-30  
\\  \hline
$<\!\! \theta_{CR-\pi^\pm par.}\!\!>$ 
& 4.98 & 2.90 & 1.22   & 0.57  & 0.32   & 0.26  & 3.56   \\  
$<\!\!\theta_{\pi^\pm par.-\pi^\pm}\!\!>$ 
&24.28  & 14.09  & 5.88  & 2.33  & 0.85   &  0.40 &  17.28  \\
$<\!\!\bar \theta_{\mu^\pm par.-\mu^\pm}\!\!>$ 
& 3.73 & 1.95  & 0.72   & 0.30  & 0.26   & 0.25  & 2.55   \\  
\end{tabular}\end{ruledtabular}
\caption{ Mean angles between muons, pions, and their parent particle (p,n,K),
and between the parent particle and the incident CR, for different
energy bins of $\nu_e({\bar{\nu}_e})$~(deg).
\label{tabang1}}

\end{table}
 
\begin{table}[!htbp]
\begin{ruledtabular}\begin{tabular}{c|c|c|c|c|c|c|c}
$E_{\nu_\mu+{\bar{\nu}_\mu}}$ (GeV) & 0.1-0.2 &0.28-0.32 & 0.8-1.2 & 2.8-3.2 
&8.0-12 &20-30 & 0.1-30  
\\  \hline
$<\!\!\theta_{CR-\pi^\pm par.}\!\!>$ 
& 5.33 & 2.91  & 1.18 & 0.54 & 0.30   & 0.26  & 3.89   \\  
$<\!\!\theta_{\pi^\pm par.-\pi^\pm}\!\!>$ 
& 23.96 & 13.41  & 5.36   & 1.99  &  0.66  & 0.32 & 16.48  \\
$<\!\!\theta_{\mu^\pm par.-\mu^\pm}\!\!>$ 
& 4.00 & 2.12   & 0.79   & 0.30  & 0.26 & 0.25  & 2.73    \\ 
\end{tabular}\end{ruledtabular}
\caption{ Same as table \ref{tabang1} for different energy bins of 
 $\nu_\mu({\bar\nu_\mu}$) .\label{tabang2}}
\end{table}
\endgroup

The deflection of the charged particles by the geomagnetic field has also been examined. 
The largest deflection is undergone by the primary proton between the generation point 
and the first collision point (which depends on the generation radius). The pion deflection 
angle is negligible in account of its short lifetime, while for muons it is less than 
4$^\circ$.

Hence, it is confirmed that the 1-dimensional approximation used for GeV neutrinos in the
previous calculations/simulations \cite{bn89,bs89,hk90,HO95,lk90} was an acceptable approximation.

\section{Neutrino flux around the Super-Kamiokande detector\label{SK}}

In this section, the features of the atmospheric neutrino flux at the geographical position of 
the Super-Kamiokande (SuperK) detector are investigated. The detection sphere was defined 
 at 0.372 km altitude, and within  
$(36^o 25' 33'' \pm 7.5^o) N$ and $(137^o 18' 37'' \pm 15^o) E$ latitude and longitude bins 
respectively. 
 
\subsection{Flux and flavor ratio\label{FSK}}

\begin{figure*}[!htbp]
\begin{tabular}{cc}
\includegraphics[width=9.5cm]{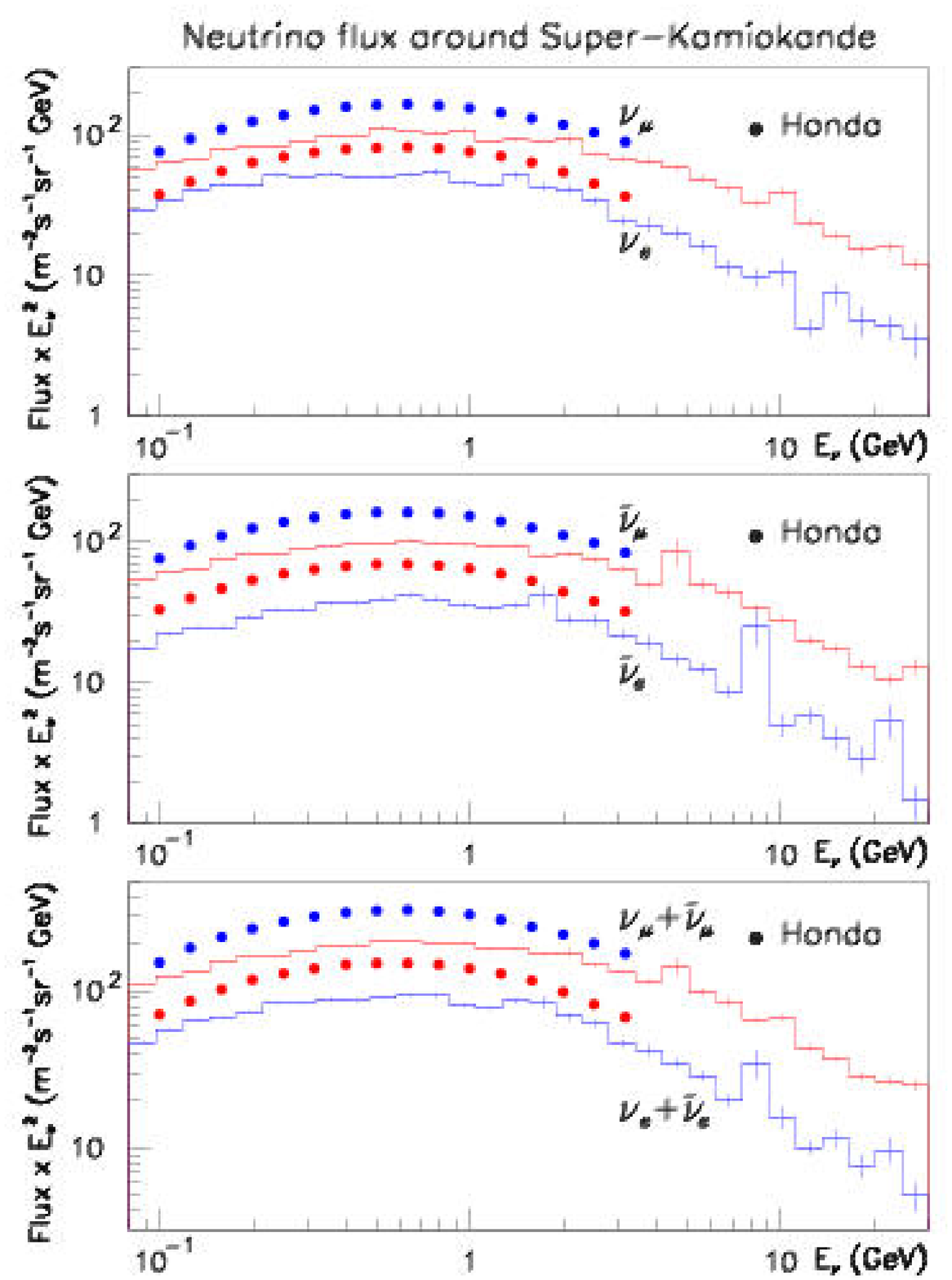}&\hspace{-1cm}
\includegraphics[width=9.5cm]{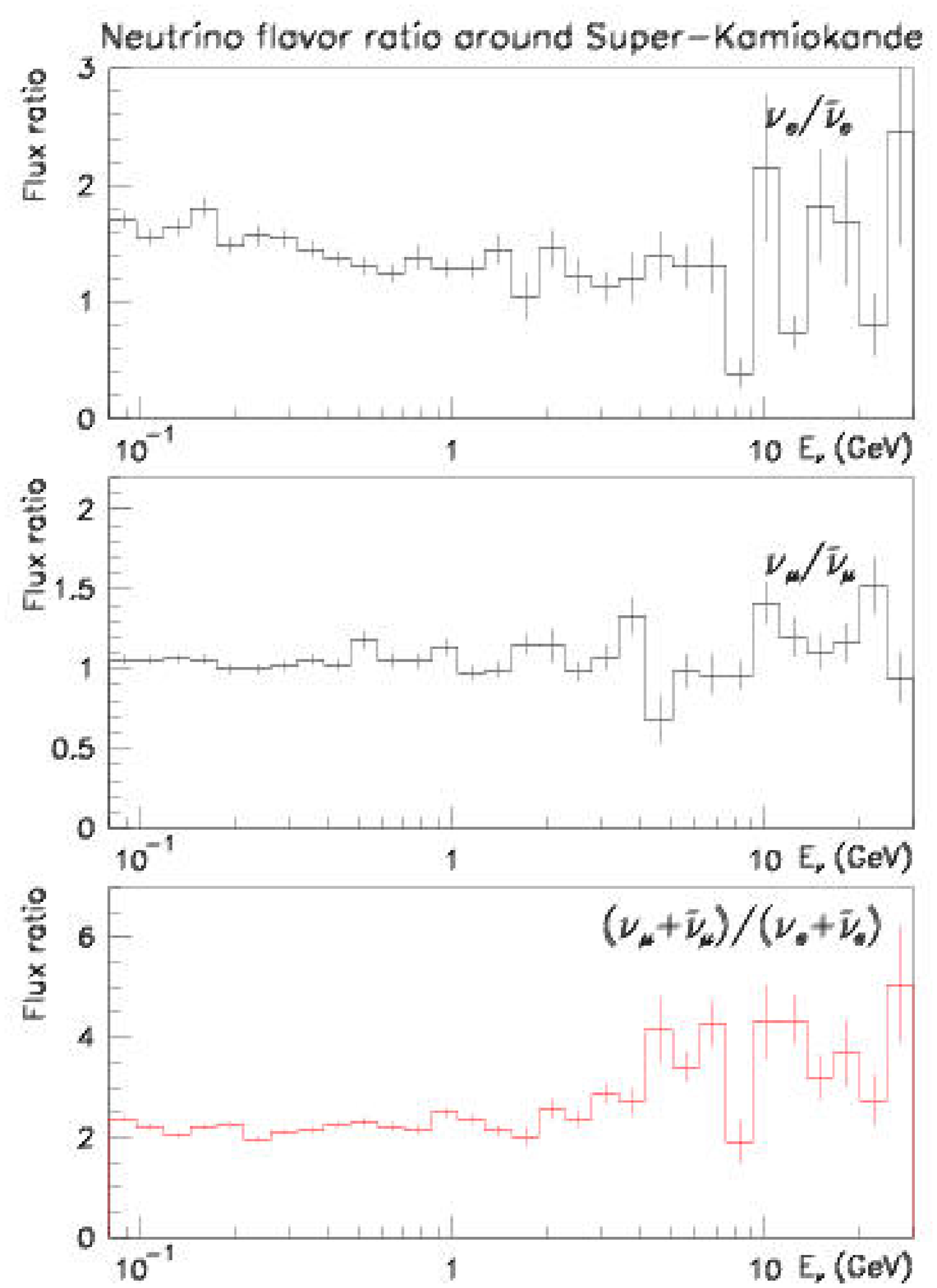}
\end{tabular}
\caption{ Simulated atmospheric neutrino spectra and flavor ratio around 
Super-Kamiokande detector.\label{skflux}}
\end{figure*}

The calculated energy distributions of the atmospheric neutrino flux and the $\nu/\bar{\nu}$ 
flux ratios around the SuperK detector, averaged over 4$\pi$ solid angle, are compared with the 
results of Honda et al. \cite{hk90,HO95} on figure~\ref{skflux}, for the various flavors. 
The flux obtained in the present work appear to be significantly smaller than in the 
1-Dimensional calculations for low neutrino energies.

\begin{table*}[!htbp]
\begin{ruledtabular}\begin{tabular}{ccccc}
 \multicolumn{1}{c}{ } & \multicolumn{4}{c}{FLUX
 ($\textrm{m}^{-2}\textrm{s}^{-1}\textrm{sr}^{-1}\textrm{GeV}^{-1}$) } \\ \cline{2-5}
$E_\nu$ (GeV) & $\nu_e$ & ${\bar{\nu}_e}$ & 
$\nu_\mu$ & ${\bar{\nu}_\mu}$ \\  \hline 
  0.100-0.124 &  2948 $\pm$ 84 & 1780 $\pm$ 59 & 5229 $\pm$ 103 & 4915 $\pm$ 99 \\  
  0.124-0.153 &  2078 $\pm$ 53 & 1324 $\pm$ 45 & 3677 $\pm$ 68 & 3629 $\pm$ 80 \\  
  0.153-0.189 &  1595 $\pm$ 50 & 892 $\pm$ 29 & 2771 $\pm$ 63 & 2569 $\pm$ 48 \\  
  0.189-0.233 &  1023 $\pm$ 30 & 702 $\pm$ 29 & 1875 $\pm$ 40 & 1853 $\pm$ 40 \\  
  0.233-0.289 &  774 $\pm$ 28 & 485 $\pm$ 19 & 1260 $\pm$ 27 & 1266 $\pm$ 27 \\  
  0.289-0.357 &  503 $\pm$ 15 & 335 $\pm$ 10 & 909 $\pm$ 20 & 901 $\pm$ 21 \\  
  0.357-0.441 &  325 $\pm$ 12 & 238 $\pm$ 8 & 649 $\pm$ 17 & 610 $\pm$ 14 \\  
  0.441-0.545 &  218 $\pm$ 8 & 162 $\pm$ 7 & 450 $\pm$ 15 & 389 $\pm$ 9\\  
  0.545-0.674 &  140 $\pm$ 5 & 113 $\pm$ 5 & 291 $\pm$ 9 & 268 $\pm$ 7 \\  
  0.674-0.833 &  99 $\pm$ 6 & 70 $\pm$ 3 & 185 $\pm$ 5 & 183 $\pm$ 5 \\  
  0.833-1.029 &  54 $\pm$ 2 & 42 $\pm$ 2 & 125 $\pm$ 5 & 114 $\pm$ 4 \\  
  1.029-1.272 &  33 $\pm$ 1 & 26 $\pm$ 1 & 72 $\pm$ 2 & 70 $\pm$ 2 \\  
  1.272-1.572 &  27 $\pm$ 2 & 18 $\pm$ 1 & 46 $\pm$ 2 & 46 $\pm$ 2 \\  
  1.572-1.943 &  13.1 $\pm$ 0.8 & 13 $\pm$ 2 & 29 $\pm$ 1 & 26 $\pm$ 1 \\  
  1.943-2.402 &  8.8 $\pm$ 0.6 & 5.7 $\pm$ 0.5 & 20 $\pm$ 2 & 17.4 $\pm$ 0.8 \\  
  2.402-2.969 &  4.1 $\pm$ 0.3 & 3.9 $\pm$ 0.4 & 9.7 $\pm$ 0.4 & 10.1 $\pm$ 0.4 \\  
  2.969-3.670 &  2.4 $\pm$ 0.2 & 1.8 $\pm$ 0.2 & 6.4 $\pm$ 0.4 & 5.5 $\pm$ 0.3 \\  
  3.670-4.537 &  1.2 $\pm$ 0.2 & 1.1 $\pm$ 0.1 & 3.8 $\pm$ 0.2 & 3.1 $\pm$ 0.2 \\  
  4.537-5.608 &  0.79 $\pm$ 0.08 & 0.54 $\pm$ 0.05 & 2.2 $\pm$ 0.1 & 3.1 $\pm$ 0.7 \\  
  5.608-6.931 &  0.35 $\pm$ 0.04 & 0.27 $\pm$ 0.03 & 1.14 $\pm$ 0.07 & 1.2 $\pm$ 0.1 \\  
  6.931-8.568 &  0.19 $\pm$ 0.02 & 0.19 $\pm$ 0.04 & 0.55 $\pm$ 0.04 & 0.72 $\pm$ 0.06 \\  
  8.568-10.59 &  0.12 $\pm$ 0.03 & 0.22 $\pm$ 0.07 & 0.47 $\pm$ 0.03 & 0.32 $\pm$ 0.02 \\  
  10.59-13.09 &  0.033 $\pm$ 0.005 & 0.036 $\pm$ 0.005 & 0.17 $\pm$ 0.01 & 0.158 $\pm$ 0.009 \\  
  13.09-16.18 &  0.035 $\pm$ 0.006 & 0.022 $\pm$ 0.004 & 0.103 $\pm$ 0.007 & 0.086 $\pm$ 0.007 \\  
  16.18-20.00 &  0.013 $\pm$ 0.003 & 0.009 $\pm$ 0.002 & 0.046 $\pm$ 0.004 & 0.040 $\pm$ 0.003
\end{tabular}\end{ruledtabular}
\caption{ Neutrino flux calculated for Super-Kamiokande experiment location.\label{tab1}}
\end{table*}

The values of the simulated 0.1-20~GeV neutrino flux for the location of the 
Super-Kamiokande experiment are tabulated in table~\ref{tab1}. These spectra  
can be fit by means of the function form:

\begin{equation}
\label{func}
f(E_\nu) = c_1 \ ( 1.0 \ + \ c_2 \ {\rm Exp}( \ - c_3 \ E_\nu \ )) \ E_\nu^{c_4} 
\end{equation}
very nicely,
where $f(E_\nu)$ is the flux, $E_\nu$ the neutrino energy, and $c_1, c_2, c_3$ and 
$c_4$ are fitting parameters given in table~\ref{tab30}.

\begingroup
\squeezetable
\begin{table}[!htbp]
\begin{ruledtabular}\begin{tabular}{ccccc}
 Particle & $c_1$ &  $c_2$  &  $c_3$  &  $c_4$  \\  \hline 
$\nu_e$           & 89.92 & -1.043 & 0.776 & -2.99 \\  
${\bar{\nu}_e}$    & 60.47  & -1.056 & 0.935 & -2.88 \\ 
$\nu_\mu$         & 198.4  & -1.029 & 0.744 & -2.82 \\ 
${\bar{\nu}_\mu}$  & 201.3  & -1.031 & 0.674 & -2.88 \\  
\end{tabular}\end{ruledtabular}
\caption{ Parameters used to fit the simulated neutrino spectra around the Super-Kamiokande
detector.\label{tab30}}
\end{table}
\endgroup
The functional form is inspired from \cite{kmnn}, carrying some
features of the hadroproduction of pions and kaons.

For the flavor ratios, our $(\nu_\mu + {\bar{\nu}_\mu})/(\nu_e + {\bar
\nu_e})$ is similar to that reported in \cite{hk90,HO95}, but
$\nu_e/{\bar{\nu}_e}$ differ evidently from all the previous
1-dimensional calculation summarized in figure 12 of \cite{HO95}.

From figures~\ref{muonfr} and \ref{muonpfr}, it can be observed that
the positive muon flux is overestimated in our simulation, which
subsequently raises the $\nu_e/\bar{\nu_e}$ ratio. This observation
can explain partially the large difference observed.

However, as explained at the end of section \ref{GLFL}, it is believed
that the tendency of decreasing $\nu_e/{\bar{\nu}_e}$ with the
increasing energy should be expected because of the constraint on the
$\pi^+/\pi^- (K^+/K^-)$ ratio imposed by the relevant data.

\subsection{Zenith angle distribution }\label{ZSK}

\begin{figure}[!htbp]
\begin{tabular}{c}
\includegraphics[width=9.5cm]{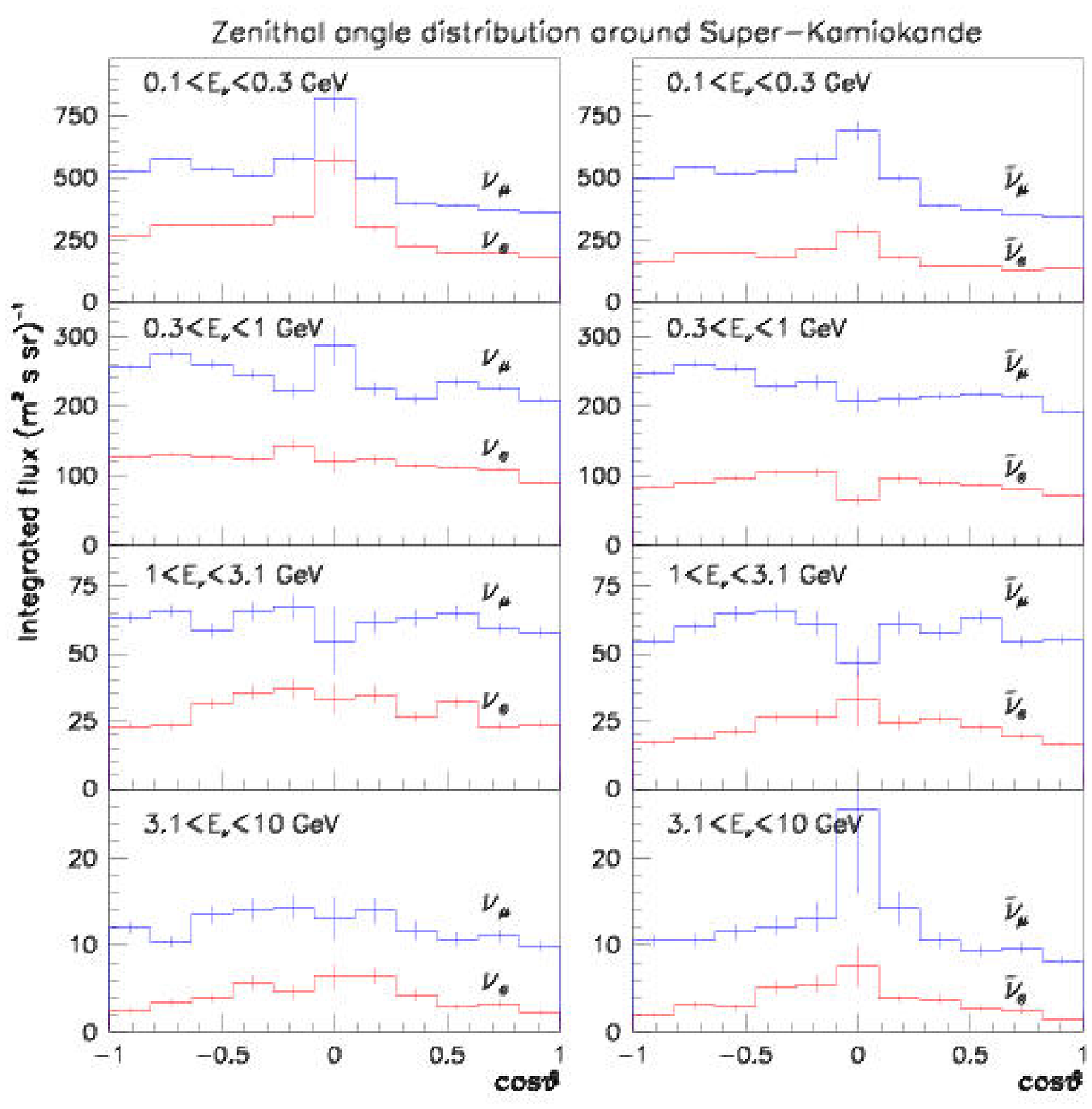}\\
\includegraphics[width=9.5cm]{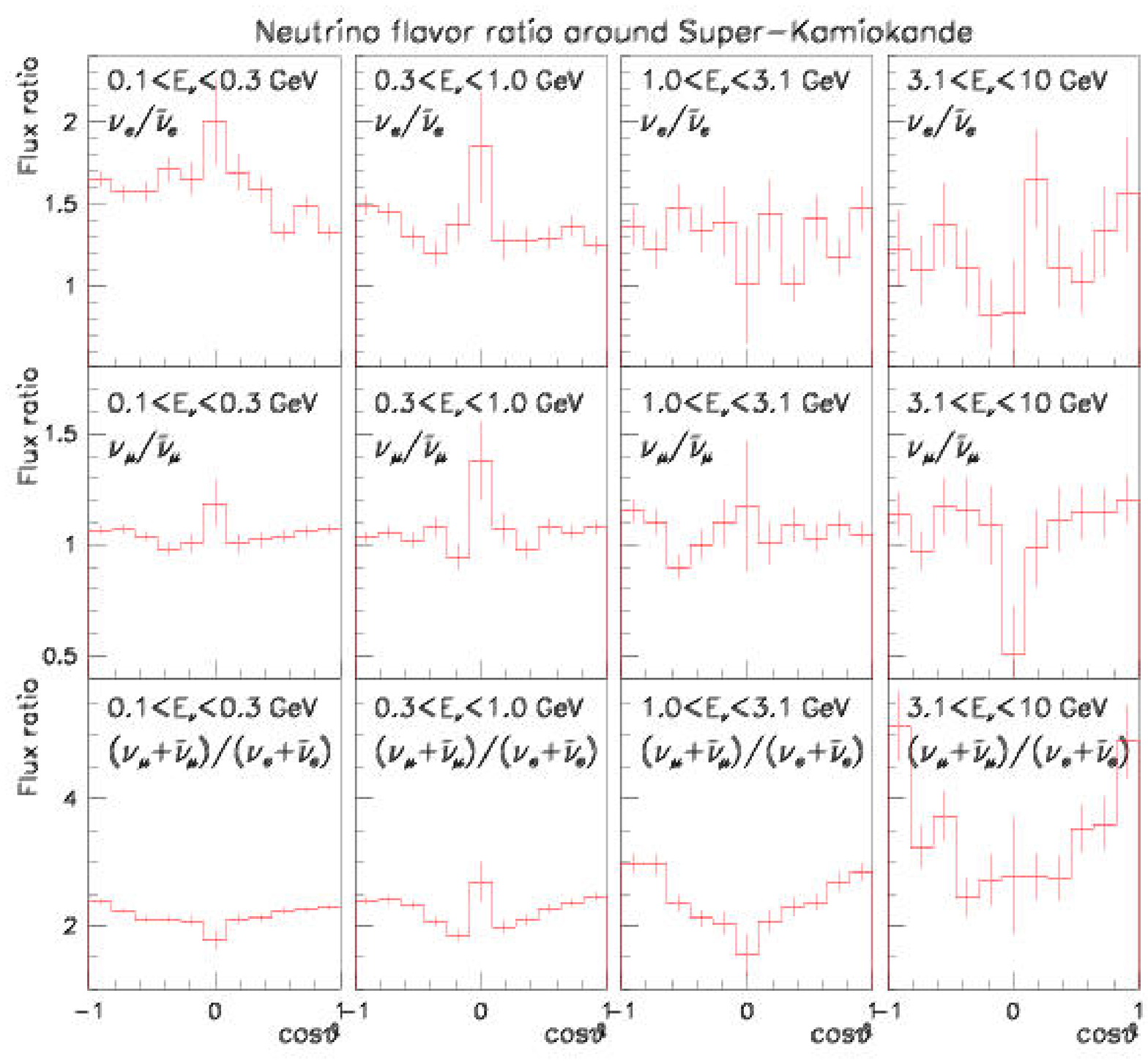}
\end{tabular}
\caption{ Zenith angle distribution of the calculated neutrino flux, and of the 
corresponding flavor ratios around the Super-Kamiokande detector. \label{skzenith}}
\end{figure}

The zenith angle distribution of the flux and of the flavor ratios are
shown on figure~\ref{skzenith}, where the enhancement of the flux in
horizontal directions discussed in section \ref{GLOBZEN} is observed,
in qualitative agreement with \cite{lap1}. In the low energy bin, the
maximum of this enhancement is about twice the downward or upward
going flux out of the horizontal plane in the low energy bin. Again,
it can be seen on the figure (right panel), that the flavor ratio
$(\nu_\mu+{\bar{\nu}_\mu})/(\nu_e+{\bar{\nu}_e})$ also depends on the
zenith angle, especially for high energy neutrinos, for the same
reason as given previously.

\subsection{Azimuthal angle distribution \label{SKZ}}

\begin{figure}[!htbp]
\begin{tabular}{c}
\includegraphics[width=9.5cm]{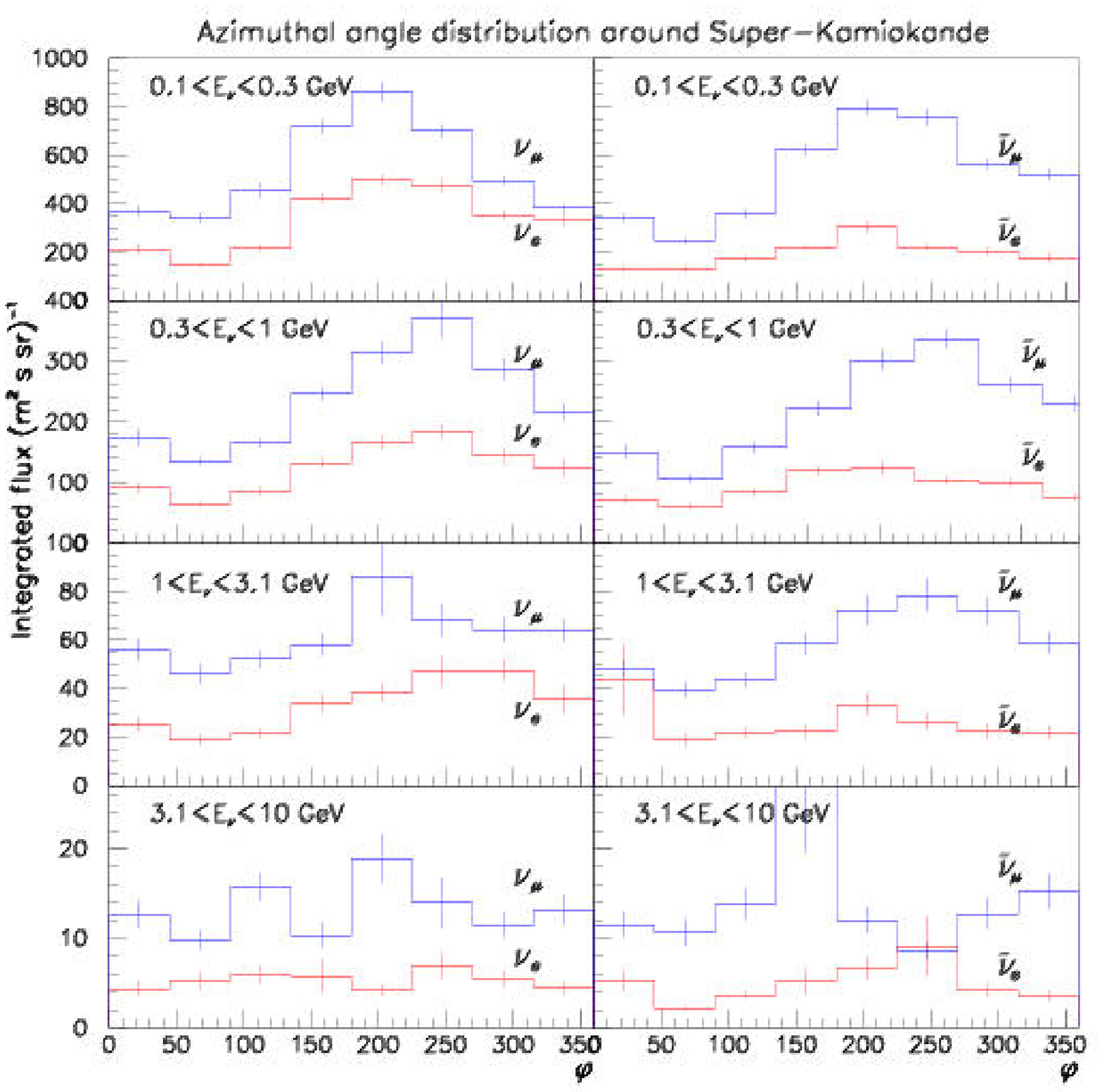}\\
\includegraphics[width=9.5cm]{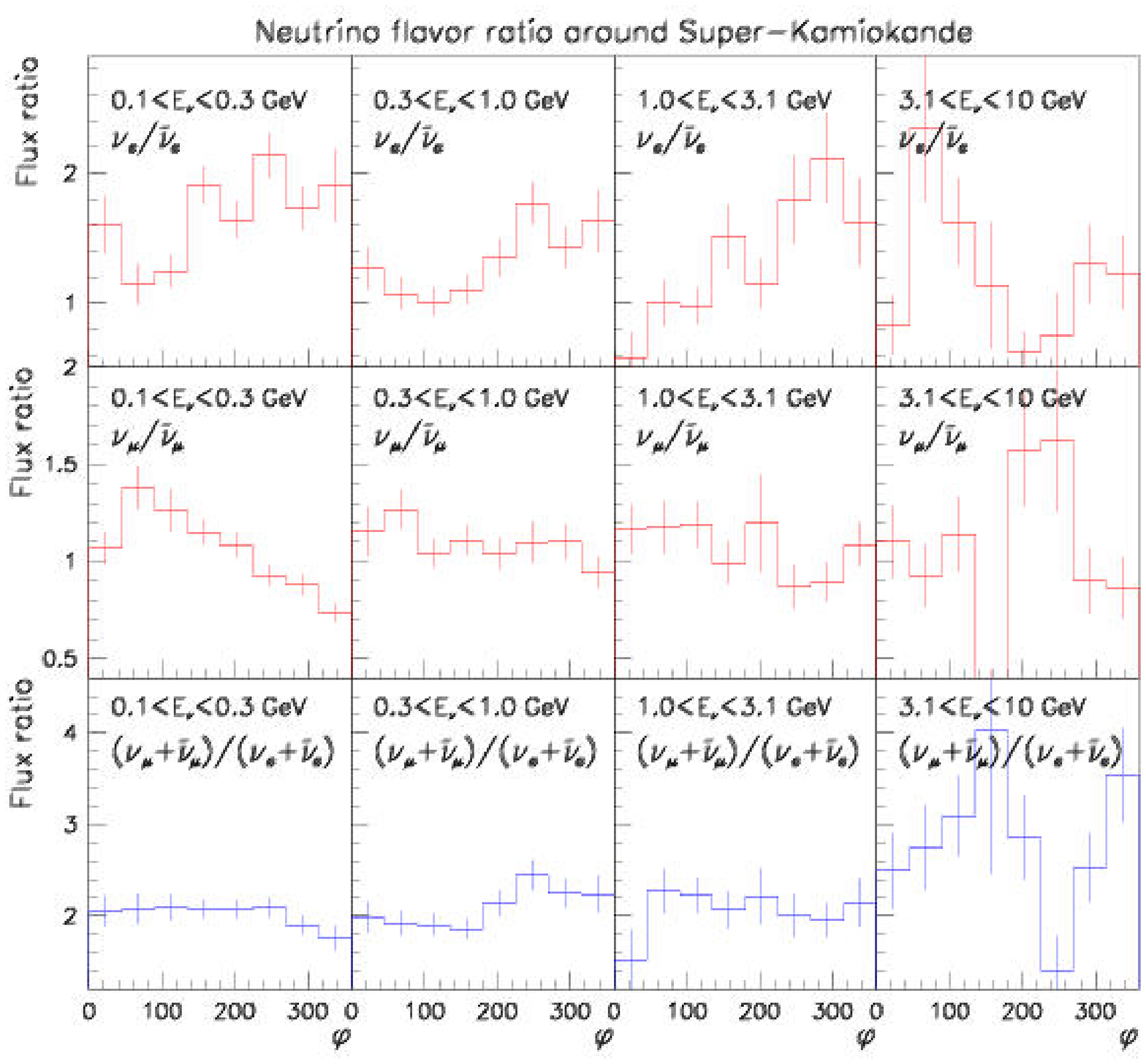}
\end{tabular}
\caption{ Simulated azimuth angle distribution of the neutrino flux and of the flavor ratios 
around the SuperK detector. 
\label{skazimuth}}
\end{figure}

The azimuth angle distribution of the flux and of the flavor ratios
are shown on figure~\ref{skazimuth} for the four neutrino species. The
large angular binning used in the figure was dictated by the low
counting statistics obtained, even for very long computer time used
for running the simulation program.
The East-West effect discussed previously in section \ref{GLAZ} is
clearly observed here, more prominent for low energy neutrinos. The EW
asymmetry discussed below has been estimated on the basis of these
distributions.

\subsection{Estimate of the East-West asymmetry} 

In addition to the atmospheric muon flux, the EW asymmetry
measurements provide an important global test of the reliability of
the overall approach. Since muons are charged particles, having finite
lifetime, and losing energy during their flying in the atmosphere,
while neutrinos have none of these features, these differences make
the relation between muon and neutrino flux a complicated issue
\cite{tk20}. However, the EW asymmetry is independent on the physics
beyond the standard model and is not expected to depend on the
neutrino mass or oscillations. A reliable simulation should thus be
able to reproduce the experimental EW asymmetry data quantitatively.

For the asymmetry parameter $A_{EW}$ defined as:

\begin{equation*}A_{EW} = \frac{N_E-N_W}{N_E+N_W} \end{equation*}

$N_E (N_W)$ being the number of eastward (westward) lepton events, the
SuperK Collaboration reported the following measured values:
$$A_{EW}^{e-like} = 0.21 \pm 0.04 \hspace{0.5cm} A_{EW}^{\mu-like} =
0.08 \pm 0.04 $$ for e-like and $\mu$-like events respectively, for a
selection of single-ring events with momentum between $0.4-3$~GeV and
$-0.5 \leq \cos \tau_{zenith} \leq 0.5$, with $\tau_{zenith}$
being the zenith angle.

In order to obtain the $A_{EW}$ value precisely, the 3-Dimension
differential neutrino reaction cross section \cite{plea,nrcs} needs to
be known.  The asymmetry $A_{EW}$ however could be estimated by using
the following procedures:
\begin{enumerate}
\item Evaluate the energy of the neutrinos which can induce lepton events
over the same range as selected by superK, i.e., 0.4 to 3~GeV. The
leptons with momentum $p_{\it l} \sim 100$~MeV/c, appear to carry 65\%
of the incident neutrino energy, this proportion increases to $\sim
85\%$ for $p_{\it l}=1$~GeV/c \cite{superk}. It can be assumed that
the fraction keeps on increasing and asymptotically tends to 100\%.
Based on this supposition, the corresponding neutrino energies of about 0.55 and 3.1~GeV associated to 
the 0.4 and 3~GeV lepton events are obtained respectively. 

\item Estimate the production angle of the neutrino induced lepton.
This can be done following \cite{ew122}, by noting that the interested neutrino energy range 
is above 0.5~GeV within which there is no large difference between the production angles of 
$\nu_e$ and $\nu_\mu$ events. Another way is by using the values $55^o$ at $p_{\it l}=0.4$~GeV/c 
and $20^o$ at $p_{\it l}=1.5$~GeV/c as reported in \cite{superk}, and fit them as a function 
of the neutrino energy, assuming that the angle smoothly tends to zero as expected from basic 
kinematics.

\item For each neutrino with energy between 0.55 and 3.1~GeV, lepton events are randomly 
generated with the scattering angle relative to the neutrino direction estimated as above, 
and with a uniform distribution for the azimuth angle. 
After weighting by the energy dependent cross section \cite{oerr,pdb92,plya} and 
application of a normalization 
procedure, the events with $\mid \cos(\theta_{\it l}) \mid \le 0.5$ are selected 
to calculate the asymmetry parameter.
\end{enumerate}

Using the above procedure, the following values are obtained for the
EW asymmetry parameters for e-like and $\mu$-like events around the
Super-Kamiokande detector: $$A_{E-W}^{e-like}=0.12\pm 0.03
\hspace{0.5cm} A_{E-W}^{\mu-like}= 0.13\pm 0.02$$

In contrast, the EW asymmetries obtained for the two neutrino flux
with the same cut over the zenithal angle and within the same energy
bin of $[0.55, 3.1]$~GeV, are: $$A_{E-W}^{\nu_e+{\bar{\nu}_e}}= 0.17\pm
0.03\hspace{0.5cm} A_{E-W}^{\nu_\mu+{\bar{\nu}_\mu}}= 0.22\pm 0.02$$
which shows that the predicted neutrino asymmetry is largely washed
out by the angular distribution of the neutrino induced lepton
production process.

The larger EW flux asymmetry obtained for $\nu_\mu (\bar{\nu_\mu})$
than for $\nu_e (\bar{\nu_e})$ can be qualitatively understood: The
production of $\nu_e (\bar{\nu_e})$ takes place in the second step of
the pion decay chain. The decay kinematics should thus wash out the EW
asymmetry of neutrinos more than for muons which are produced
dominantly in the first stage of the decay sequence.

 
In comparison with the data \cite{ew111} and other simulation result \cite{ew122}, the present 
calculations departs further from the data. It can be noted that, the present overestimated ratio 
of $\pi^+/\pi^-$ may affect this prediction. More works is still 
needed to clear this problem.

\section{Neutrino flux around the Soudan experiment\label{SOUD}}

The same calculations for the simulated atmospheric neutrino flux
around the SOUDAN detector as for superK experiment in the previous
section, are reported in this section.  The normalization area used in
the calculations corresponds to the geographical latitude and
longitude bin of $(48^o \pm 5^o) N$ and $(98^o \pm 10^o) W$
respectively. The altitude is taken the same as for Super-K, as we
have seen from figure~\ref{altitude}, the error resulting from the
altitude difference between Super-Kamiokande and Soudan detector
should be negligible in evaluating the flux and relevant
distributions.

\subsection{Flux and flavor ratio } 

\begin{figure*}[!htbp]
\begin{tabular}{cc}
\includegraphics[width=9.5cm]{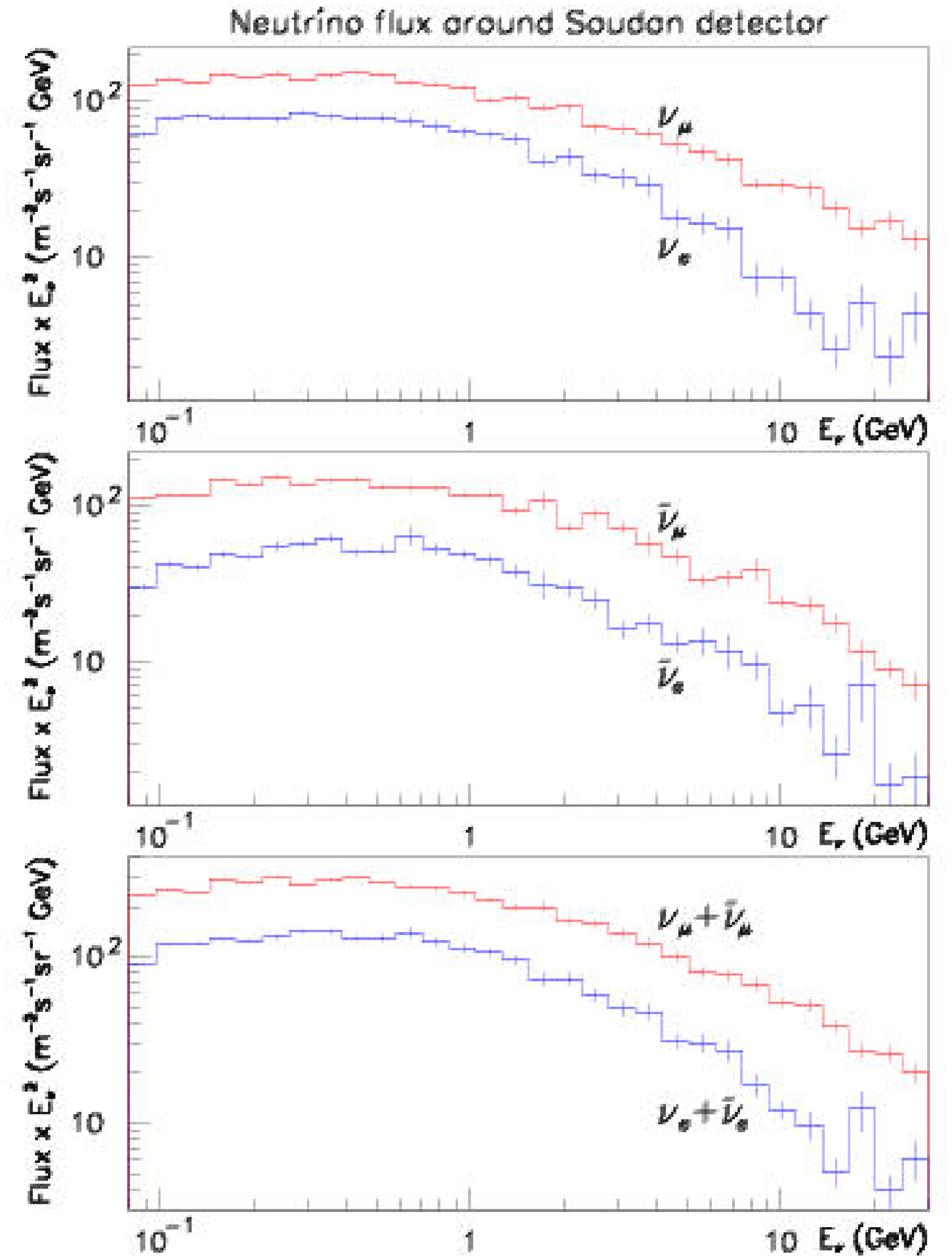} &\hspace{-1cm}
\includegraphics[width=9.5cm]{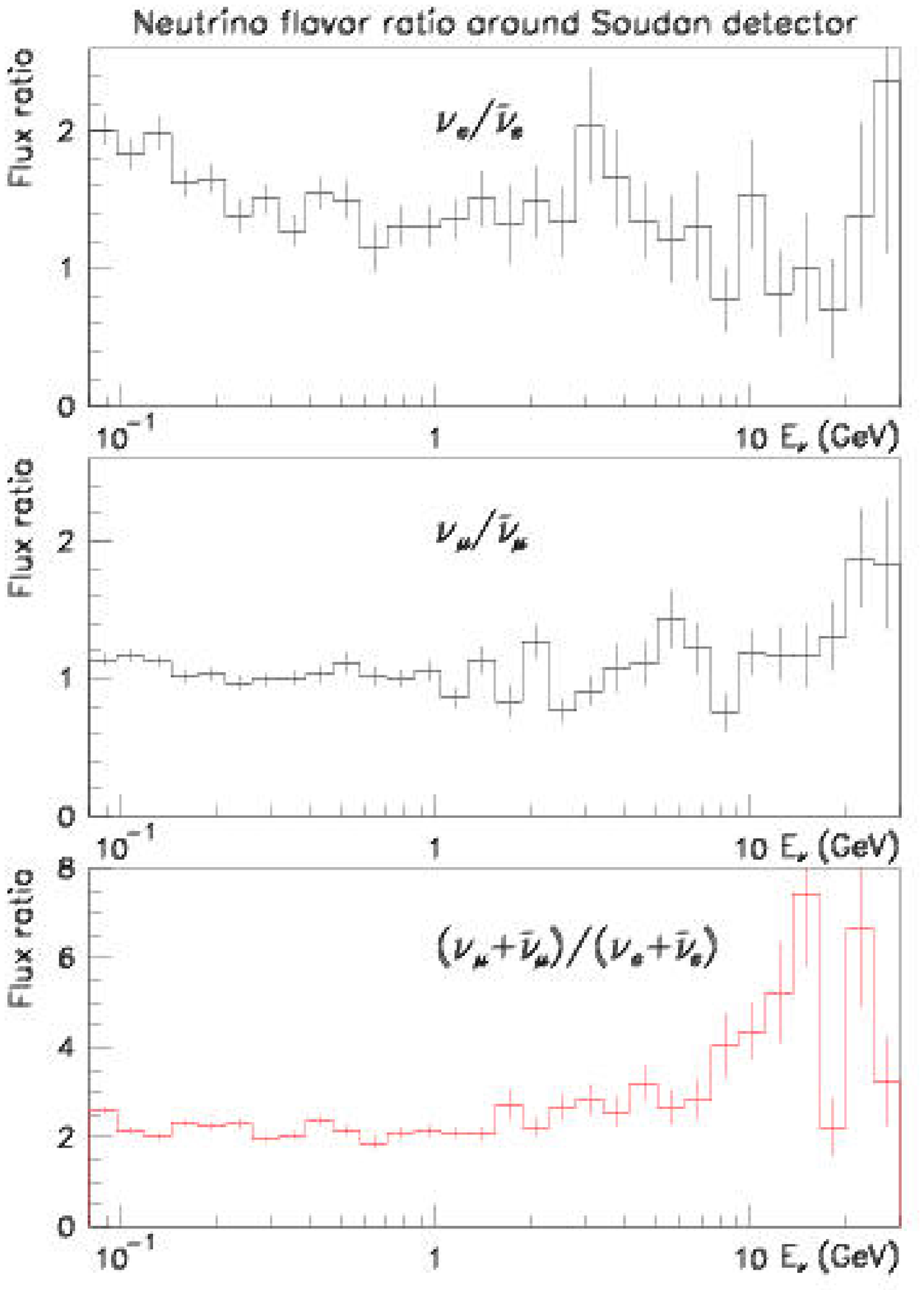}
\end{tabular}
\caption{ Simulated atmospheric neutrino spectra and flavor ratios around the
SOUDAN detector. \label{soudanflux}}
\end{figure*}

The energy spectra and energy dependence of flavor ratios, averaged
over 4$\pi$ solid angle are shown in figure~\ref{soudanflux}. Due to
the relatively high geomagnetic latitude of the the Soudan detector
location, and hence to the lower geomagnetic cutoff, the calculated
neutrino flux are much higher than found for Super-K.

\begin{table*}[!htbp]
\begin{ruledtabular}\begin{tabular}{ccccc}
 \multicolumn{1}{c}{ } & \multicolumn{4}{c}{FLUX
 ($\textrm{m}^{-2}\textrm{s}^{-1}\textrm{sr}^{-1}\textrm{GeV}^{-1}$) } \\ \cline{2-5}
$E_\nu$ (GeV) & $\nu_e$ & ${\bar{\nu}_e}$ & 
$\nu_\mu$ & ${\bar{\nu}_\mu}$ \\  \hline 
  0.100-0.124 & 6222 $\pm$ 240  & 3344 $\pm$ 163  & 10967 $\pm$ 313  & 9453  $\pm$ 257 \\  
  0.124-0.153 & 4338 $\pm$ 175  & 2283 $\pm$ 97   & 7154 $\pm$ 194   & 6604  $\pm$ 188 \\  
  0.153-0.189 & 2698 $\pm$ 100  & 1674 $\pm$ 83   & 5078 $\pm$ 159   & 4872  $\pm$ 154 \\  
  0.189-0.233 & 1738 $\pm$ 71   & 997 $\pm$ 49    & 3344 $\pm$ 110   & 3254  $\pm$ 110 \\  
  0.233-0.289 & 1200 $\pm$ 47   & 860 $\pm$ 62    & 2062 $\pm$ 60    & 2124  $\pm$ 78 \\  
  0.289-0.357 & 771 $\pm$ 32    & 623 $\pm$ 45    & 1396 $\pm$ 53    & 1397  $\pm$ 52 \\  
  0.357-0.441 & 488 $\pm$ 26    & 361 $\pm$ 19    & 936 $\pm$ 32     & 876   $\pm$ 36 \\  
  0.441-0.545 & 333 $\pm$ 20    & 214 $\pm$ 13    & 628 $\pm$ 28     & 595   $\pm$ 31 \\  
  0.545-0.674 & 212 $\pm$ 15    & 163 $\pm$ 21    & 359 $\pm$ 15     & 352   $\pm$ 15 \\  
  0.674-0.833 & 122 $\pm$ 9     & 92 $\pm$ 7      & 231 $\pm$ 10     & 234   $\pm$ 12 \\  
  0.833-1.029 & 73 $\pm$ 6      & 60 $\pm$ 4      & 145 $\pm$ 6      & 138   $\pm$ 7 \\  
  1.029-1.272 & 48 $\pm$ 3      & 34 $\pm$ 3      & 80 $\pm$ 4       & 92    $\pm$ 6 \\  
  1.272-1.572 & 28 $\pm$ 3      & 19 $\pm$ 2      & 52 $\pm$ 3       & 49    $\pm$ 3 \\  
  1.572-1.943 & 13 $\pm$ 1      & 10 $\pm$ 2     & 30 $\pm$ 2       & 32    $\pm$ 4 \\  
  1.943-2.402 & 9 $\pm$ 1       & 6.1 $\pm$ 0.7      & 19 $\pm$ 1       & 17    $\pm$ 1 \\  
  2.402-2.969 & 4.4 $\pm$ 0.5     & 3.5 $\pm$ 0.5     & 9.9 $\pm$ 0.7      & 11.1   $\pm$ 0.9 \\  
  2.969-3.670 & 3.4 $\pm$ 0.5     & 1.5 $\pm$ 0.2     & 5.5 $\pm$ 0.5      & 6.4    $\pm$ 0.6 \\  
  3.670-4.537 & 1.6 $\pm$ 0.3     & 0.9 $\pm$ 0.1     & 3.6 $\pm$ 0.3      & 3.3    $\pm$ 0.4 \\  
  4.537-5.608 & 0.53 $\pm$ 0.08   & 0.49 $\pm$ 0.08   & 2.0 $\pm$ 0.2      & 1.5    $\pm$ 0.2 \\  
  5.608-6.931 & 0.44 $\pm$ 0.07   & 0.43 $\pm$ 0.09   & 1.1 $\pm$ 0.1      & 0.9    $\pm$ 0.1 \\  
  6.931-8.568 & 0.16 $\pm$ 0.03   & 0.12 $\pm$ 0.02   & 0.58 $\pm$ 0.05    & 0.6    $\pm$ 0.1 \\  
  8.568-10.59 & 0.10 $\pm$ 0.02   & 0.08 $\pm$ 0.02   & 0.32 $\pm$ 0.03    & 0.29   $\pm$ 0.03 \\  
  10.59-13.09 & 0.031 $\pm$ 0.007 & 0.04 $\pm$ 0.01   & 0.20 $\pm$ 0.02    & 0.16   $\pm$ 0.02 \\  
  13.09-16.18 & 0.017 $\pm$ 0.004 & 0.012 $\pm$ 0.004 & 0.10 $\pm$ 0.01    & 0.09   $\pm$ 0.01 \\  
  16.18-20.00 & 0.016 $\pm$ 0.005 & 0.022 $\pm$ 0.009 & 0.053 $\pm$ 0.006  & 0.037  $\pm$ 0.005 \\    
\end{tabular}\end{ruledtabular}
\caption{ Neutrino flux calculated for Soudan experiment location.\label{tab2}}
\end{table*}

Also, the 0.1-20.0~GeV neutrino energy spectra simulated for the Soudan experiment
site are tabulated in table~\ref{tab2}. They can be fit with the same functional 
form (relation~\ref{func}). The coefficients are listed in table~\ref{tab40}. 

\begingroup
\squeezetable
\begin{table}[!htbp]
\begin{ruledtabular}\begin{tabular}{ccccc}
Particle  & $c_1$ &  $c_2$  &  $c_3$  &  $c_4$  \\  \hline 
$\nu_e$           & 105.0  & -1.034 & 0.869 & -3.12 \\ 
${\bar{\nu}_e}$    & 63.9  & -1.069  & 1.235 & -2.99 \\  
$\nu_\mu$         & 143.0  & -0.970 & 1.859 & -2.67 \\  
${\bar{\nu}_\mu}$  & 362.2  & -1.021 & 0.416 & -3.16 \\  
\end{tabular}\end{ruledtabular}
\caption{ Fitting parameters for simulated neutrino spectra around Soudan
detector. \label{tab40}}
\end{table}
 \endgroup

\subsection{Zenith angle distributions} 

\begin{figure}[!htbp]
\begin{tabular}{c}
\includegraphics[width=9.5cm]{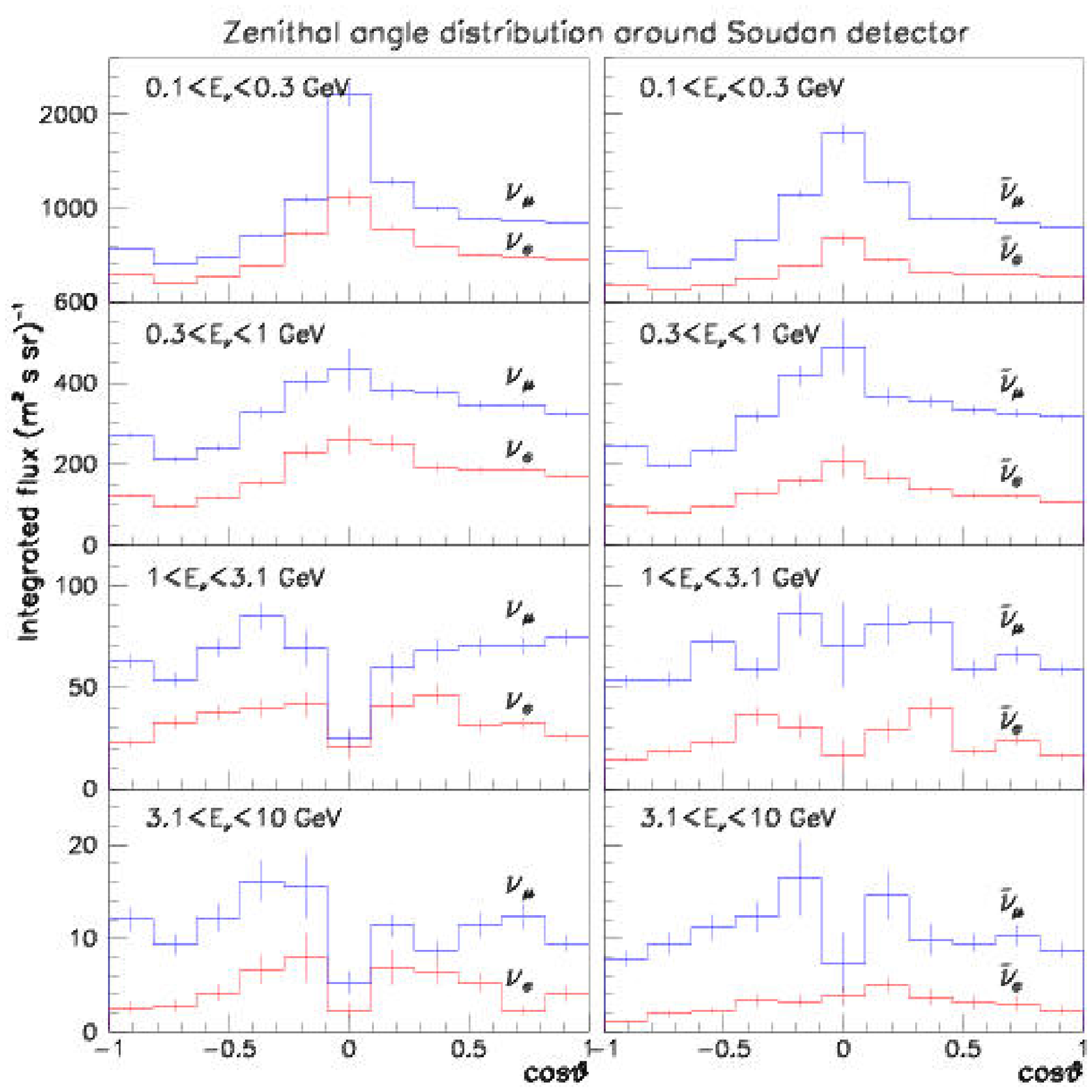} \\
\includegraphics[width=9.5cm]{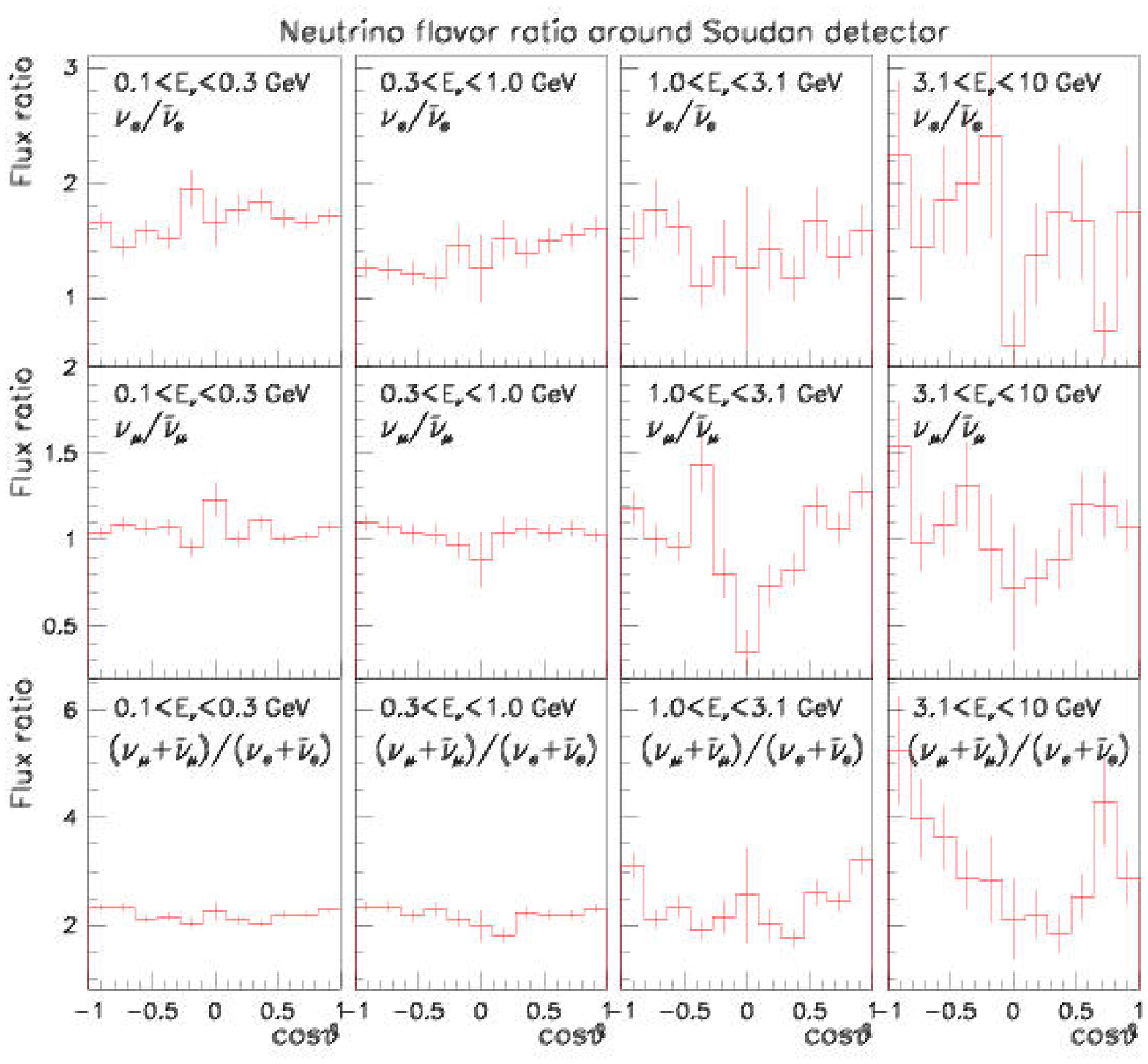}
\end{tabular}
\caption{Zenith angle distributions of the simulated neutrino flux and 
flavor ratios around the SOUDAN detector. \label{soudanzenith}}
\end{figure}

The calculated zenithal angle distributions are shown in
figure~\ref{soudanzenith}, for the same energy bins as previously,
where the high geomagnetic latitude feature of such distribution
illustrated in Section~\ref{GLOBZEN} is evident.

\subsection{Azimuth angle distributions }

\begin{figure}[!htbp]
\begin{tabular}{c}
\includegraphics[width=9.5cm]{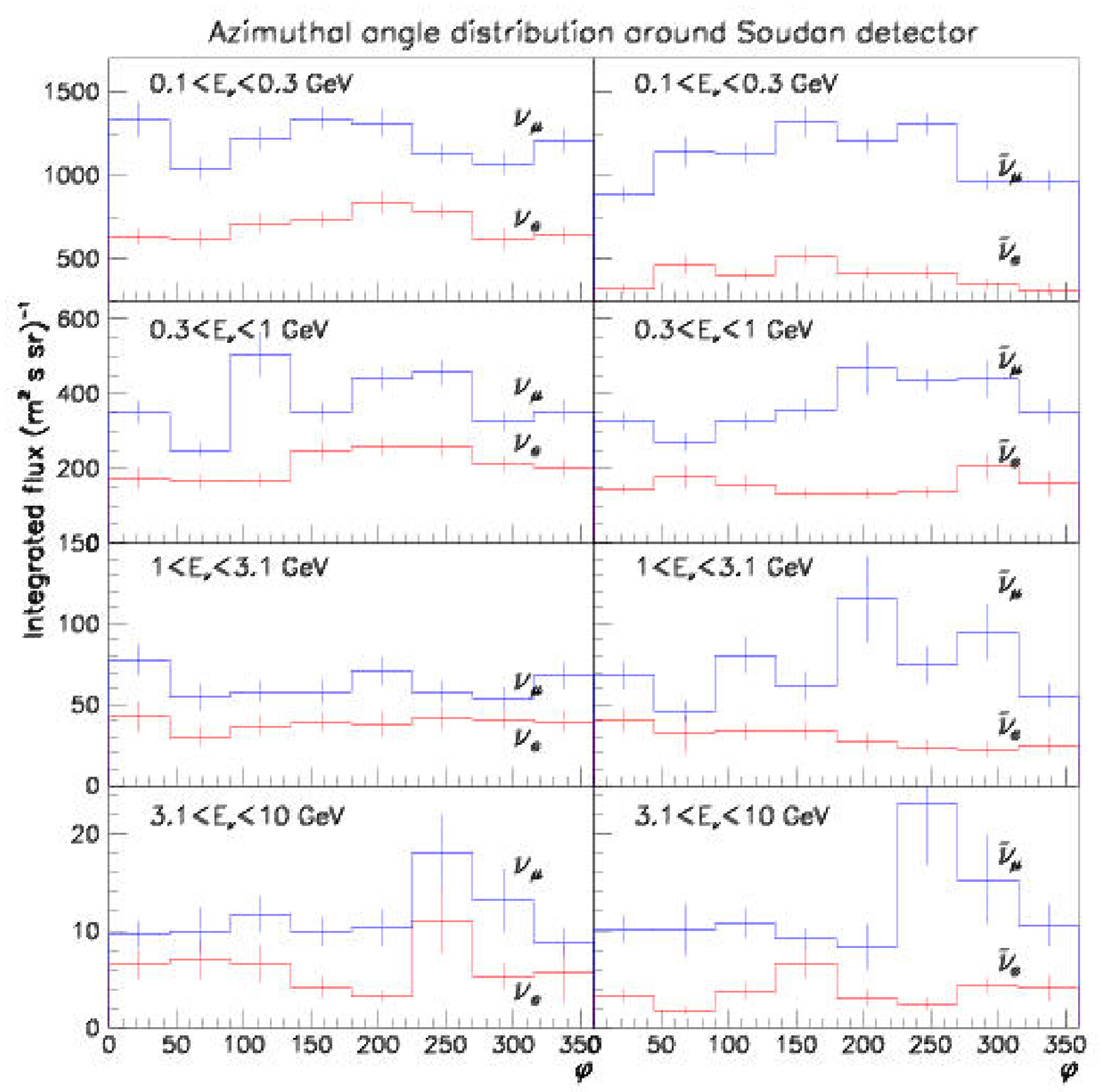}\\
\includegraphics[width=9.5cm]{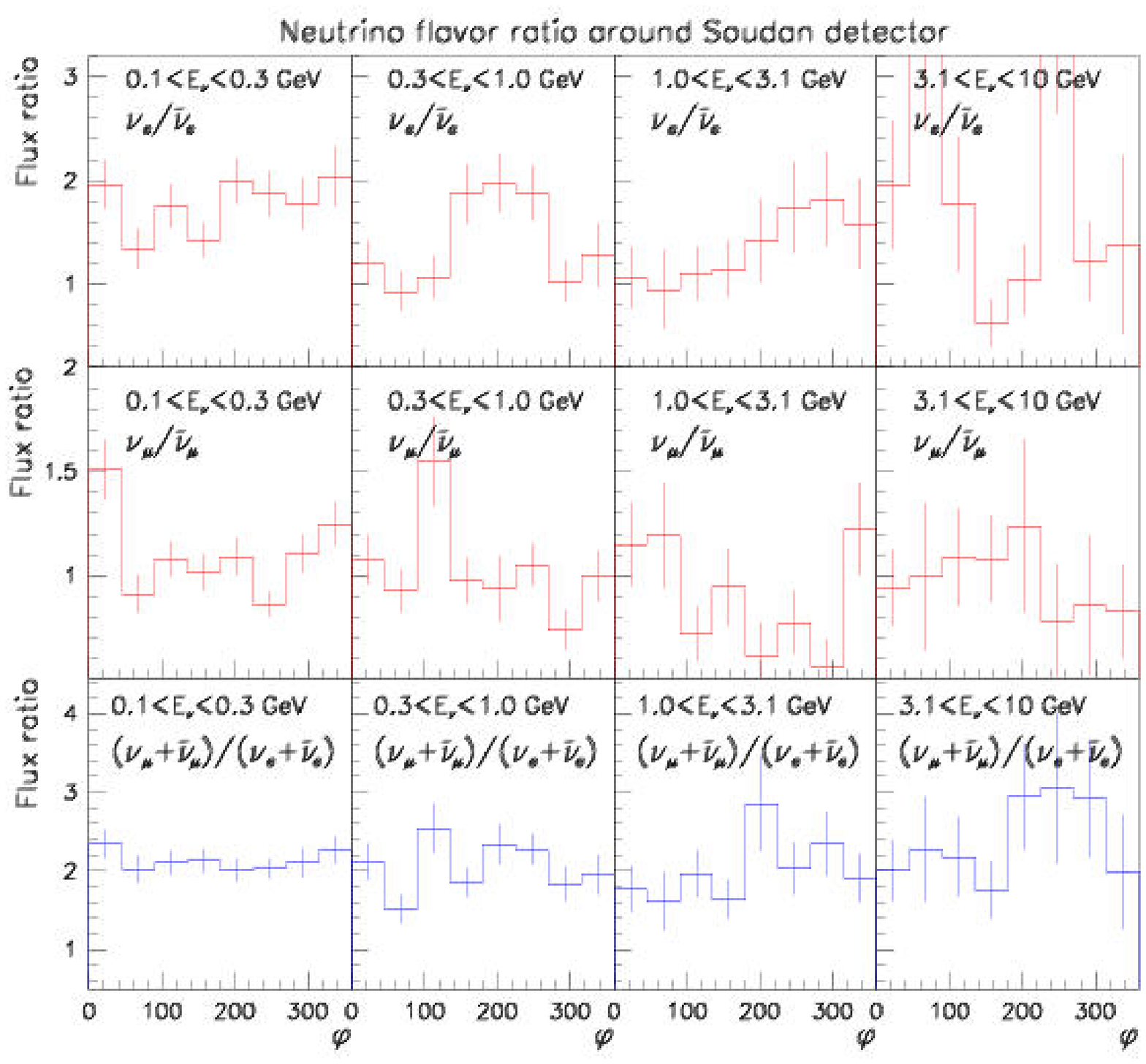}
\end{tabular}
\caption{Azimuth angle distributions of the simulated neutrino flux and of the flavor 
ratios around the SOUDAN detector.\label{soudanazimuth}}
\end{figure}

Figure~\ref{soudanazimuth} shows the azimuth angle distributions of the atmospheric
neutrino flux around the SOUDAN detector, based on which the EW asymmetry parameter for 
neutrino flux and lepton events in the detector are also estimated along the same lines 
as for the superK experiment. 

The EW asymmetry for the lepton events is predicted as
$$A_{E-W}^{e-like}=0.006 \pm 0.037 \hspace{0.5cm} A_{E-W}^{\mu-like}=
0.026\pm 0.024 $$ while for the neutrino flux the following values are
obtained: $$A_{E-W}^{\nu_e+{\bar\nu_e}} = 0.068 \pm 0.043
\hspace{0.5cm} A_{E-W}^{\nu_\mu+{\bar\nu_\mu}}= 0.071 \pm 0.025 $$
with the same selection criteria as used for SuperK in the previous
section.  It would therefore be of great interest to have some
experimental values for this experiment, to provide a further test the
present approach, and a step forward to a better global understanding
of the problem.

\section{Conclusion and discussion\label{CONC}}

In summary, the atmospheric neutrino flux has been simulated by means
of an event generator dedicated to the general Cosmic Ray - Atmosphere
interactions, developed to account for the AMS01 results,
originally. The successful calculations reported previously on the
hadron and lepton flux close to earth, strongly support the method and
the models used.

In this work, the CAPRICE and HEAT muon flux data measured at various
altitudes in the atmosphere, from about sea level to 38km have been
successfully reproduced, providing further grounds to the correctness
of the approach.

A significantly lower absolute value of the flux than found in
1-dimensional calculations \cite{hk90,HO95} in the low energy range
has been reported here. Also, the $\nu_e/\bar{\nu}_e$ ratio is
largely different from all the previous calculations. In addition, the
present results are in agreement with similar results reported in
\cite{naumov01}.

Detailed features of the neutrino flux: zenithal and azimuthal angle
distribution and flavor ratios at different latitudes, corresponding
energy distributions, production altitude and production angle
distributions, have been calculated and discussed. The components
originating from Cosmic Rays and from the atmospheric cascade have
been distinguished and discussed separately. All the observed features
of the calculated flux could be traced back to the primary spectra,
$\pi^+/\pi^-$($K^+/K^-$) ratio, muon kinetics, geomagnetic effect, and
geometry.

Specific 3-dimensional effect seen in the simulation results are very
small, which indicates that the 1-dimensional approximation provides
reliable results for GeV neutrino flux calculations.

The calculations fail to reproduce quantitatively the East-West
asymmetry of the atmospheric neutrino-induced events in the Super-K
detector. More investigations are needed to understand this
disagreement.

A complete comparison of the parametrizations used to describe the
various hadronic production channels with the available data will be
reported in details later.

\begin{acknowledgments}

The authors are grateful to N. Mokhov, and V. Naumov for helpful discussion 
during the course of this work, and to M. Boezio, M. Circella, and T. Sanuki for making their data 
available and for providing experimental details.

\end{acknowledgments}


\begin{thebibliography}{30}
\bibitem{superk} Super-Kamiokande Collaboration, Y. Fukuda, et al., Phys. Rev. Lett. 
                {\bf 81}, 1562(1998); {\bf B 433}, 9(1998); {\bf B 436}, 33(1998).  
\bibitem{soudan} The Soudan 2 Collaboration, W. W. M. Allison, et al., Phys. Lett. {\bf B 
                449}, 137(1999). W. W. M. Allison, et al., Phys. Lett. {\bf B 391}, 
                491(1997). 
\bibitem{frati} W. Frati, T. K. Gaisser, A. K. Mann and T. Stanev, Phys. Rev. {\bf D 48}, 
                1140(1993).
\bibitem{fogli} Fogli G. L., Lisi E., Marrone A., and Scioscia G., Phys. Rev. {\bf D 59}, 
                033001(1998).
\bibitem{bn89} E. V. Bugaev and V. A. Naumov, Phys. Lett. {\bf B 232}, 391(1989).
\bibitem{pl93} P. Lipari, Astropart. Phys. {\bf 1}, 195(1993).
\bibitem{pk94} D. H. Perkins,  Astropart. Phys. {\bf 2}, 249(1994). 
\bibitem{bs89} G. Barr, T. K. Gaisser, and T. Stanev, Phys. Rev. {\bf D 39}, 3532(1989); 
               V. Agrawal, T. K. Gaisser, P. Lipari, and T. Stanev, Phys. Rev. {\bf D 53}, 
               1314(1996).
\bibitem{hk90} M. Honda, K. Kasahara, K. Hikada, and S. Midorikawa, Phys. Lett. {\bf B 248}, 
               193(1990). 
\bibitem{vv80} L. V. Volkova, Yad. Fiz. {\bf 31}, 1510(1980); Sov. J. Nucl. Phys. {\bf 31}, 
               784(1980).  
\bibitem{HO95} M. Honda, T. Kajita, K. Kasahara and S. Midorikawa; Phys. Rev. {\bf D 52}, 4985(1995).
\bibitem{lk90} H. Lee and Y. S. Koh, Nuovo Cimento {\bf B 105}, 883(1990).
\bibitem{tk20} T. K. Gaisser, Nucl. Phys. {\bf B 87} (Proc. Suppl.), 145(2000).
\bibitem{re20} R. Engel, T. K. Gaisser and T. Stanev, Phys. Lett. {\bf B 472}, 113(2000). 
\bibitem{GA96} T. K. Gaisser et al., Phys. Rev. {\bf D 54}, 5578(1996).
\bibitem{ams00} AMS Collaboration, Alcaraz J. et al., Phys. Lett. {\bf B 472}, 215(2000); 
                Phys. Lett. {\bf B 484}, 10(2000).
\bibitem{bess00} BESS Collaboration, T. Sanuki et al., ApJ {\bf 545}, 1135(2000).
\bibitem{honda1} M. Honda et al., {\it 27th ICRC, Aug. 7-14, 2001, Hamburg}, P. 1162.
\bibitem{ms98} Mokhov N. V. and Striganov S. I., {\it Workshop on the Front End of a Muon 
               Collider}, pp. 453-459(1998). 
\bibitem{ew111} Super-Kamiokande Collaboration, T.Futagami, et al., Phys. Rev. Lett. {\bf 82}, 5194(1999). 
\bibitem{ew122} T. Kajita, Y. Totsuka, Rev. Mod. Phys. {\bf 73}, 85(2001).
\bibitem{honda2} Honda M., Kajita T., Kasahara K., and Midorikawa S., Phys.Rev. {\bf D64}:053011(2001) 
\bibitem{naumov01} G. Fiorentini, V. A. Naumov, and F. L. Villante, Phys. Lett. {\bf B 510},  
                 173(2001); G. Fiorentini, V. A. Naumov, and F. L. Villante, hep-ph/0106014;
                {\it Proc. of 27th ICRC, Aug. 7-14, 2001, Hamburg}.
\bibitem{lap1} P. Lipari, Astropart. Phys. {\bf 14}, 153(2000); P. Lipari, hep-ph/0003013.
\bibitem{gai01} T. K. Gaisser, astro-ph/0104327.
\bibitem{gflm} G. Battistoni et al., Astropart. Phys. {\bf 12}, 315(2000); G. Battistoni, 
               A. Ferrari, T. Montaruli, and P. R. Sala, {\it Proc. of 27th ICRC, Aug. 7-14, 
               2001, Hamburg}. G. Battistoni et al., hep-ph/0207035.
\bibitem{tknw} Y. Tserkovnyak, R. Komar, C. Nally, and C. Waltham, {\it Proc. of27th ICRC, 
                Aug. 7-14, 2001, Hamburg }, hep-ph/9907450.
\bibitem{plya} V. Plyaskin, Phys. Lett. {\bf B 516}, 213(2001).
\bibitem{ldbd} Yong Liu, L. Derome, and M. Bu\'enerd,  {\it Proc. of 27th ICRC, Aug. 7-14, 
               2001, Hamburg}
\bibitem{wentz} J. Wentz et al., {\it Proc. of 27th ICRC, Aug. 7-14, 2001, Hamburg}.
\bibitem{cbdv} HEAT Collaboration, S. Coutu, et al., Phys. Rev. {\bf D 62}, 032001(2000).
\bibitem{capri} CAPRICE Collaboration, J. Kremer, et al., Phys. Rev. Lett. {\bf 83}, 4241(1999); 
                M. Boezio, et al., Phys. Rev. Lett. {\bf 82}, 4757(1999).
\bibitem{mass99} MASS Collaboration, R. Bellotti, et al., Phys. Rev. {\bf D 60}, 
                052002(1999);
\bibitem{ams02} AMS Collaboration, Alcaraz  J., et al., Phys. Lett. {\bf B 494}, 193(2000).
\bibitem{DERP} L. Derome et al., Phys. Lett. {\bf B 489}, 1(2000); 
               L. Derome, and M. Bu\'enerd, Nucl. Phys. {\bf A 688}, 66c(2001).
\bibitem{DERL} L. Derome, M. Bu\'enerd, and Yong Liu,  Phys. Lett. {\bf B 515}, 1(2001).
\bibitem{DERH} L. Derome, and M. Bu\'enerd, Phys. Lett. {\bf B 521}, 139(2001); See also
               {\it Proc. of ICRC 2001, Hamburg, Aug. 7-14, 2001}.
\bibitem{hdb03} C.Y. Huang, L. Derome, and M. Bu\'enerd, In preparation.
\bibitem{VA61} M.S. Vallarta, Handbuch der Physik, vol 61-1, Springer Verlag, 
               S. Fl\"ugge ed., Berlin-G\"ottingen-Heidelberg, 1961.
\bibitem{SOLMOD} J.S. Perko, Astro. Astrophys. 184(1984)119; see M.S. Potgieter, Proc. ICRC
                 Calgary, 1993, p213, for a recent review of the subject.
\bibitem{SI83}   J.A. Simpson, Ann. Rev. Nucl. Sci. {\bf 33}, 323(1983).
\bibitem{WI98} B. Wiebel-Sooth, P.L. Biermann, and H. Meyer, A. \& A. {\bf 330}, 389(1998).
\bibitem{AG81} G.N. Agakishiev et al., Sov. J. Nucl. Phys. {\bf 33}, 552(1981)
\bibitem{AG84} G. N. Agakishiev, et. al., Sov. J. Nucl. Phys. {\bf 40}, 767(1984)
               SKM-200 Collaboration, A. Kh. Abdurakhimov et. al., Nucl. Phys. {\bf A 362},
               376(1981); G. N. Agakishiev, et. al., Z. Phys. {\bf C 27}, 177(1985); 
               R.N. Bekmirzaev, E. N. Kladnitskaya, and S. A. Sharipova, Phys. Atomic Nuclei 
               {\bf 58}, 58(1995) [Yad. Fiz. {\bf 58}, 1(1995)].
\bibitem{agfm1} A. E. Hedin, J. Geophys. Res. {\bf 96}, 1159(1991).
\bibitem{kmnn}  A. N. Kalinovsky, N. V. Mokhov, and Y. P. Nikitin, {\it "Passage of High 
                Energy Particles Through Matter"}, New-York, AIP ed., 1989.
\bibitem{ldl11} Yong Liu, L. Derome, and M. Bu\'enerd, {\it Parametrization for the 
                Inclusive Production Cross Section of proton induced $\pi^\pm$ particles on 
                nuclei}, ISN Internal Report 01-012, February 2001.
\bibitem{pdg00} D. E. Groom, et al., (Particle Data Group), Eur. Phys. J. {\bf C15}, 1(2000).
\bibitem{sh57}  S. Hayakawa, Phys. Rev. {\bf D 15}, 1533(1957); G. Barr, T.K. Gaisser, and 
                T. Stanev, Phys. Rev. {\bf D 39}, 3532(1989).
\bibitem{DERLICRC} L. Derome, Yong Liu, and M. Bu\'enerd, {\it Proc. of ICRC 2001, Hamburg, 
                   Aug. 7-14, 2001, OG1.018}.
\bibitem{phan}   P. Hansen, for The WIZARD/CAPRICE Collaboration, in  {\it Proc. of27th ICRC, 
                 Aug. 7-14, 2001, Hamburg}.
\bibitem{bess02} BESS Collaboration, T. Sanuki et al., Phys. Lett. {\bf B 541}, 234(2002).
                 M. Motoki et al., astro-ph/0205344. 
\bibitem{coch} D. R. F. Cochran, {\it et al}, Phys. Rev. {\bf D 6}, 3085(1972). 
\bibitem{bogg} H. Boggild, {\it et al}, Phys. Rev. {\bf C 59}, 328(1999). 	
\bibitem{acfs} D. Antreasyan et al., Phys. Rev. {\bf D 19}, 764(1979); 
               S. Fredriksson, Phys. Rev. {\bf D 18}, 4144(1978).
\bibitem{tkts} T. K. Gaisser, and T. Stanev,  Phys. Rev. {\bf D 57}, 1977(1998).
\bibitem{honda6} T. K. Gaisser and M. Honda, hep-ph/0203272, 
                 to appear in the Annual Review of Nuclear \& Particle Science Vol. 52.	       
\bibitem{plea}  P. Lipari, M. Lusignoli, and F. Sartogo, Phys. Rev. Lett. {\bf 74},
                4384(1995); E. A. Paschos, and J. Y. Yu, hep-ph/0107261.
\bibitem{nrcs}  Kamiokande Collaboration, M. Nakahata et al., J. Phys. Soc. Jap. {\bf 55}, 
                3786(1986). 
\bibitem{oerr}  O. Erriquez et al. Phys. Lett. {\bf B 80}, 309(1979). 
\bibitem{pdb92} Particle Data Group, Phys. Rev. {\bf D 45}, No. 11, III. 82(1992) 
                and references therein.

\end{thebibliography}
\end{document}